\documentclass[twocolumn,tighten]{aastex62}

\usepackage{amsmath}

\newcommand{\keyw}[1]{\textcolor{gray}{#1}}

\usepackage{longtable}
 
\begin{document}

\title{The Disk Substructures at High Angular Resolution Project (DSHARP):  \\
VII. The Planet-Disk Interactions Interpretation}

\correspondingauthor{Zhaohuan Zhu}
\email{zhaohuan.zhu@unlv.edu}

\author[0000-0002-8537-9114]{Shangjia Zhang}
\affiliation{Department of Physics and Astronomy, University of Nevada, Las Vegas, 
                4505 S. Maryland Pkwy, Las Vegas, NV, 89154, USA}

\author[0000-0003-3616-6822]{Zhaohuan Zhu}
\affiliation{Department of Physics and Astronomy, University of Nevada, Las Vegas, 
                4505 S. Maryland Pkwy, Las Vegas, NV, 89154, USA}

\author[0000-0001-6947-6072]{Jane Huang}
\affiliation{Harvard-Smithsonian Center for Astrophysics, 60 Garden Street, Cambridge, MA 02138, USA}

\author[0000-0003-4784-3040]{Viviana V. Guzm\'an}
\affiliation{Joint ALMA Observatory, Avenida Alonso de Córdova 3107, Vitacura, Santiago, Chile}
\affiliation{Instituto de Astrof\'isica, Pontificia Universidad Cat\'olica de Chile, Av. Vicu\~na Mackenna 4860, 7820436 Macul, Santiago, Chile}

\author{Sean M. Andrews}
\affiliation{Harvard-Smithsonian Center for Astrophysics, 60 Garden Street, Cambridge, MA 02138, USA}

\author[0000-0002-1899-8783]{Tilman Birnstiel}
\affiliation{University Observatory, Faculty of Physics, Ludwig-Maximilians-Universit\"at M\"unchen, Scheinerstr. 1, 81679 Munich, Germany}

\author[0000-0002-7078-5910]{Cornelis P. Dullemond}
\affiliation{Zentrum für Astronomie, Heidelberg University, Albert Ueberle Str. 2, 69120 Heidelberg, Germany}

\author[0000-0003-2251-0602]{John M. Carpenter}
\affiliation{Joint ALMA Observatory, Avenida Alonso de Córdova 3107, Vitacura, Santiago, Chile}

\author[0000-0001-8061-2207]{Andrea Isella}
\affiliation{Department of Physics and Astronomy, Rice University, 6100 Main Street, Houston, TX 77005, USA}

\author[0000-0002-1199-9564]{Laura M. P\'erez}
\affiliation{Departamento de Astronom\'ia, Universidad de Chile, Camino El Observatorio 1515, Las Condes, Santiago, Chile}

\author[0000-0002-7695-7605]{Myriam Benisty}
\affiliation{Unidad Mixta Internacional Franco-Chilena de Astronom\'{i}a (CNRS, UMI 3386), Departamento de Astronom\'{i}a, Universidad de
Chile, Camino El Observatorio 1515, Las Condes, Santiago, Chile}
\affiliation{Univ. Grenoble Alpes, CNRS, IPAG, 38000 Grenoble, France}

\author[0000-0003-1526-7587]{David J. Wilner}
\affiliation{Harvard-Smithsonian Center for Astrophysics, 60 Garden Street, Cambridge, MA 02138, USA}

\author{Cl{\'e}ment Baruteau}
\affiliation{CNRS / Institut de Recherche en Astrophysique et Plan{\'e}tologie, 14 avenue Edouard Belin, 31400 Toulouse, France}

\author[0000-0003-1172-3039]{Xue-Ning Bai}
\affiliation{Institute for Advanced Study and Tsinghua Center for Astrophysics, Tsinghua University, Beijing 100084, China}

\author{Luca Ricci}
\affiliation{Department of Physics and Astronomy, California State University Northridge, 18111 Nordhoff Street, Northridge, CA 91130, USA}

\begin{abstract}
The Disk Substructures at High Angular Resolution Project (DSHARP) provides a large sample of protoplanetary disks having substructures which could be induced by young forming planets.  To explore the properties of planets that may be responsible for these substructures,  we systematically carry out a grid of 2-D hydrodynamical simulations including both gas and dust components. We present the resulting gas structures, including the relationship between the planet mass and 1) the gaseous gap depth/width, and 2) the sub/super-Keplerian motion across the gap.  We then compute dust continuum intensity maps at the frequency of the DSHARP observations. We provide the relationship between the planet mass and 1) the depth/width of the gaps at millimeter intensity maps, 2) the gap edge ellipticity and asymmetry, and 3) the position of secondary  gaps induced by the planet. With these relationships, we lay out the procedure to constrain the planet mass using gap properties, and study the potential planets in the DSHARP disks. We highlight the excellent agreement between observations and simulations for AS 209 and the detectability of the young Solar System analog. Finally, under the assumption that the detected gaps are induced by young planets, we characterize the young planet population in the planet mass-semimajor axis diagram. We find that the occurrence rate for $>$ 5 $M_J$ planets beyond 5-10 au is consistent with direct imaging constraints. Disk substructures allow us probe a wide-orbit planet population (Neptune to Jupiter mass planets beyond 10 au) that is not accessible to other planet searching techniques.
\end{abstract}

\section{Introduction \label{sec:intro}}
Discoveries over the past few decades show that planets are common.
The demographics of exoplanets have put constraints on planet formation theory 
(e.g. review by \citealt{johansen14,  raymond2014, chabrier2014}).
Unfortunately, most discovered exoplanets are billions of years old and have therefore been subject to significant orbital dynamical
alteration after their formation (e.g., review by \citealt{Davies2014}). 
To test planet formation theory, it is crucial to constrain the young planet population right after they are born in protoplanetary disks. 
However, the planet search techniques that have discovered thousands of exoplanets around mature stars are not efficient
at finding planets around young stars ($<$10 Myrs old) mainly due to their stellar variablity and the presence of the protoplanetary disks.
Fewer than 10 young planet candidates in systems $<$10 Myrs have been detected so far (e.g. CI Tau b, \citealt{JohnsKrull2016}; V 830 Tau b, \citealt{Donati2016}; Tap 26 b, \citealt{Yu2017}; PDS 70 b, \citealt{Keppler2018}; LkCa 15 b, \citealt{sallum2015}). 

On the other hand, 
recent high resolution imaging at near-IR wavelengths (with the new adaptive optics systems on 10-meter class telescopes)  and  interferometry at radio wavelengths (especially the ALMA and the VLA)
can directly probe the protoplanetary disks down to au-scales, and a variety of disk features (such as gaps, rings, spirals, and large-scale asymmetries) have been revealed (e.g.\citealt{casassus13, vandermarel13, ALMA2015, andrews16, garufi2017}). 
Despite that there are other possibilities for producing these features,
they may be induced by young planets in these disks, and we can use these features to probe
the unseen young planet population. 

Planet-disk interactions have been studied over the past three decades with both analytical approaches \citep{goldreich80, tanaka2002} and numerical simulations \citep{kley12,baruteau2014}. 
While the earlier work focused on planet migration and gap opening, more recently efforts have been dedicated to studying observable disk features induced by planets \citep{wolf2005, dodson-robinson11, zhu11, gonzalez2012,  pinilla12a, ataiee2013,  bae2016, kanagawa_width, rosotti2016, isella2018}, including the
observational signatures in near-IR scattered light images (e.g. \citealt{dong2015, zhu2015, fung2015}), (sub-)mm dust thermal continuum images \citep{dipierro15, picogna2015, DongFung, dong2018}, and (sub-)mm molecular line channel maps that trace the gas kinematics at the gap edges or around the planet \citep{perez15, pinte2018, teague2018}.

Among all these indirect methods for probing young planets at various wavelengths, only
dust thermal emission at (sub-)mm wavelengths allows us to probe low mass planets,
since a small change in the
gas surface density due to the low mass planet can cause dramatic changes in the dust surface density \citep{paardekooper06, Zhu14}. 
However, this also means that hydrodynamical simulations with both gas and dust components are needed to study the expected
disk features at (sub-)mm wavelengths. Such simulations are more complicated due to
the uncertainties about the dust size distribution in protoplanetary disks. 
Previously, hydrodynamical simulations have been carried out to explain features in individual sources (e.g. \citealt{jin2016,dipierro2018,fedele2018}).
With many disk features revealed by DSHARP \citep{andrews18b}, a
systematic study of how the dust features relate to the planet properties is desirable.
By conducting an extensive series of disk models spanning a substantial range in disk and planet properties, we can enable a broad exploration of parameter space which can then be used to rapidly infer young planet populations from the observations, and
we will also be more confident 
that we are not missing possible parameter space for each potential planet.

In this work, we carry out a grid of hydrodynamical simulations including both gas
and dust components. Then, assuming different
dust size distributions, we generate intensity maps at the observation wavelength of DSHARP. In \S \ref{sec:method},
we describe our methods. The results are presented in \S \ref{sec:results}. The derived young planet properties for the DSHARP disks are
given in \S \ref{sec:flowchart}. After a short discussion in \S \ref{sec:discussion}, we 
conclude the paper in \S \ref{sec:conclusion}.

\section{Method \label{sec:method}}
We carry out 2-D hydrodynamical planet-disk simulations using the modified version of the grid-based code FARGO \citep{masset2000} called Dusty FARGO-ADSG \citep{baruteau2008a, baruteau2008b, baruteau2016}. The gas component is simulated
using finite difference methods \citep{stone1992}, while the dust component is modelled as Lagrangian particles. 
To allow our simulations to be as scale-free as possible, we do not include disk self-gravity, radiative cooling, or dust feedback. These simplifications are suitable for most disks observed in DSHARP. Most of the features in these disks lie beyond 10 au where the irradiation
from the central star dominates the disk heating such that the disk is nearly vertically isothermal close to the midplane \citep{dalessio98}.  Although the dust dynamical feedback to the gas is important when a significant amount of dust  accumulates at  gap edges
or within vortices \citep{fu2014, Crnkovic-Rubsamen2015}, simulations that have dust particles but do not include dust feedback to the gas (so-called "passive dust" models) serve as reference models and allow us to  scale our simulations freely to disks with different
dust-to-gas mass ratios and dust size distributions. As shown in \S \ref{sec:flowchart}, passive dust models are also adequate in
most of our cases (especially when the dust couples with the gas relatively well).
Simulations with dust feedback will be presented in Yang \& Zhu (2018). 

\subsection{Setup: Gas and Dust}
We adopt polar coordinates ($r$, $\theta$) centered on the star and fix the planet on a circular orbit at $r=1$. Since the star is wobbling around the center of mass due to the perturbation by the planet,
indirect forces are applied to this non-inertial coordinate frame.

We initialize the gas surface density as
\begin{equation}
\Sigma_g(r)=\Sigma_{g,0}(r/r_{0})^{-1}\,,
\end{equation}
where $r_{0}$  is also the position of the planet and we set $r_0=r_p=1$. For studying gaps of individual sources in \S \ref{sec:flowchart}, we scale $\Sigma_{g,0}$ to be consistent with the DSHARP observations.
We assume locally isothermal equation of state, and the temperature at radius $r$ follows $T(r)=T_{0}(r/r_{0})^{-1/2}$.
$T$ is related to the disk scale height $h$ as $h/r=c_{s}/v_{\phi}$ where $c_{s}^2=RT/\mu=P/\Sigma$ and $\mu$ = 2.35. With our setup, $h/r$ changes as $r^{1/4}$. In the rest of the text, when we give a value of $h/r$, we are referring to $h/r$ at $r_0$.

 Our numerical grid extends from 0.1 $r_0$ to 10 $r_0$ in the radial direction and 0 to 2$\pi$ in the $\theta$ direction. For low viscosity cases ($\alpha$ = $10^{-4}$ and $10^{-3}$), there are 750 grid points in the radial direction and 1024 grid points in the $\theta$ direction. This is equivalent to 16 grid points per scale height at $r_0$ if $h/r=0.1$. For high viscosity cases ($\alpha$=0.01), less resolution is needed so there are 375 and 512 grid points in the radial and $\theta$ direction. For simulations to fit AS 209 in \S \ref{sec:as209}, the resolution is 1500 and 2048 grid points in the radial and $\theta$ direction to capture additional gaps at the inner disk. We use the evanescent boundary condition, which relaxes the fluid variables to the initial state at $r<$0.12$r_0$ and  $r>$8$r_0$. 
A smoothing length of 0.6 disk scale height at $r_0$ is used to smooth the planet's potential \citep{muller2012}. 

We assume that the dust surface density is
1/100 of the gas surface density initially. 
The open boundary condition is applied for dust particles, so that
the dust-to-gas mass ratio for the whole disk can change with time. 

The dust particles experience both  gravitational forces and  aerodynamic drag forces. The particles are pushed at every timestep with the orbital integrator. When the particle's stopping time is smaller than the numerical timestep, we use the short friction time approximation to push the particle. 
Since we are interested in disk regions beyond 10s of au, the disk density is low enough that
the molecular mean-free path is larger than
the size of dust particles. In this case,
the drag force experienced by the particles is in the Epstein regime. 
The Stokes number $St$ for particles (also called particles' dimensionless stopping time) is 
\begin{equation}
St=t_{stop}\Omega=\frac{\pi s \rho_{p}}{2 \Sigma_{gas}}=1.57\times10^{-3}\frac{\rho_{p}}{1 \mathrm{g \,cm^{-3}}}\frac{s}{1 \mathrm{mm}}\frac{100 \mathrm{g\, cm^{-2}}}{\Sigma_{g}}\,.\label{eq:stokes}
\end{equation}
where $\rho_p$ is the density of the dust particle, $s$ is the radius of the dust particle, 
and $\Sigma_{g}$ is the gas surface density. We assume $\rho_p$=1 g cm$^{-3}$ in our simulations.
We use 200,000 and 100,000 particles for high and low resolution runs, respectively. Each particle is a super particle representing a group of real dust particles
having the same size. The super particles in our simulations 
have Stokes numbers ranging from 1.57$\times10^{-5}$ to 1.57, or physical
radii ranging from 1 $\mu$m to 10 cm if $\Sigma_{g,0}$=10 $\mathrm{g\,cm^{-2}}$ and $\rho_p=1 \mathrm{g\, cm^{-3}}$. We distribute super particles uniformly in $log(s)$ space, which means that we have the same number of super particles per decade in size.  Since dust-to-gas back reaction is not included, 
we can scale the dust size distribution in our simulations to any desired distribution. 

During the simulation, we keep the size of the super-particle the same no matter where it drifts to. Thus,
the super-particle's Stokes number changes when this particle drifts in the disk, because the particle's Stokes number also depends on the local disk surface density (Equation \ref{eq:stokes}). More specifically, during the simulation, the Stokes number of the every particle varies as being inverse proportional to the local gas surface density.

Turbulent diffusion for dust particles is included as random kicks to the particles \citep{Charnoz2011, Fuente2017}. The diffusion coefficient
is related to the $\alpha$ parameter as in \cite{youdin2007} through the so-called Schmidt number $Sc$. In this work, $Sc$ is defined as the ratio between the angular momentum transport coefficient ($\nu$) and the gas diffusion coefficient ($D_{g}$). We set $Sc=1$ which serves as a good first order approximation, although that $Sc$ can take on different values and its value can differ between the radial and vertical directions \citep{Zhu2015b, yang2018},

\subsection{Grid of Models \label{sec:gridofmodels}}
To explore the full parameter space, 
we choose three values for $(h/r)_{r_0}$ $(0.05,  0.07, 0.1)$, five values for the planet-star mass ratio (q$\equiv M_{p}/M_*$ = 3.3$\times$10$^{-5}$, 10$^{-4}$, 3.3$\times$10$^{-4}$, 10$^{-3}$, 3.3$\times$10$^{-3}$ M$_{
*}$, or roughly $M_p$ = 11 $M_{\Earth}$, 33 $M_{\Earth}$, 0.35 $M_{J}$, 1 $M_{J}$, 3.5 $M_{J}$ if $M_{*}=M_{\odot}$), and three values for the disk turbulent viscosity coefficient 
($\alpha=0.01\,,0.001\,,0.0001$). Thus, we have 45 simulations 
in total. We label each simulation in the following manner: 
\texttt{h5am3p1} means $h/r$=0.05, $\alpha=10^{-3}$ (\texttt{m3} in \texttt{h5am3p1} means minus 3), $M_{p}/M_{*}=3.3\times 10^{-5} M_{*}$ (\texttt{p1} refers to the lowest planet mass case). 
We also run some additional simulations for individual sources (e.g. AS 209, Elias 24) which will be presented in \S \ref{sec:individuals} and \citet{guzman18}. 

This parameter space represents typical disk conditions. Protoplanetary disks normally have $h/r$ between 0.05 and 0.1 at $r>10\, \mathrm{au}$ \citep{dalessio98}.
While a moderate $\alpha\sim10^{-2}$ is preferred to explain
the disk accretion \citep{hartmann98}, recent works suggest that a low turbulence level ($\alpha<10^{-2}$) is needed to explain molecular line widths in TW Hya \citep{flaherty2018} and dust settling in HL Tau \citep{pinte2016}. When $\alpha$ is smaller than $10^{-4}$, the viscous timescale over the disk scale height at the planet position ($H_p^2/\nu$) is longer than $10^4/\Omega_{p}$ or 1.6 million years at 100 au, so that the viscosity will not affect the disk evolution significantly. In \S \ref{sec:individuals}, we carry out several simulations with different $\alpha$ values to extend the parameter space for some sources in the DSHARP sample. As shown below, when the planet mass is less than 11 $M_{\Earth}$, the disk features are not
detectable with ALMA. When the planet mass is larger than 3.5 $M_{J}$, the disk features have strong asymmetries, and we should be able to detect the planet directly though
direct imaging techniques.

We run the simulations for 1000 planetary orbits (1000 $T_{p}$), which is equivalent to 1 Myr for a planet at 100 au or 0.1 Myr for a planet at 20 au. These timescales are comparable to the disk ages of the DSHARP sources.

\subsection{Calculating mm Continuum Intensity Maps \label{sec:scaling}}
For each simulation, we calculate the mm continuum intensity maps assuming different disk surface densities and dust size distributions. Since dust-to-gas feedback is neglected, we can freely scale the initial disk surface density and dust size distribution in simulations to match realistic disks. 

Both the disk surface density and dust size distribution have large impacts on the mm intensity maps. If the dust thermal continuum is mainly from micron sized particles and the disk surface density is high, these dust particles have small Stokes numbers (Equation \ref{eq:stokes}). Consequently, they couple to the gas almost perfectly and the gaps revealed in mm are very similar to the gaps in the gas. If the mm emission is dominated by mm sized particles and the disk surface density is low, the dust particles can have Stokes numbers close to 1 and they drift very fast in the disk. In this case, they can be trapped at the gap edges, producing deep and wide gaps. 
To explore how different dust size distributions can affect the mm intensity maps, we choose two very different dust size
distributions to generate intensity maps. For the distribution referred to as DSD1, we assume $n(s)\propto s^{-3.5}$ with a maximum grain size of 0.1 
mm in the initial condition ($p=-3.5$ and $s_{max}$=0.1 mm. This is motivated by recent (sub-)mm polarization measurements \citep{kataoka2017, hull2018}, which indicate that the maximum grain size in a variety of disks is around 0.1 mm. 
In the other case referred to as DSD2, we assume $n(s)\propto s^{-2.5}$ with the maximum grain size of 1 cm ($p=-2.5$ and $s_{max}$=1 cm). This shallower dust
size distribution is expected from dust growth models \citep{birnstiel12} and consistent with SED constraints \citep{dalessio01} and the spectral index at mm/cm wavelengths \citep{ricci10a,ricci10b,lperez15b}. Both cases assume a minimum grain size of 0.005 $\mu$m. We find that the minimum grain size has no effect on the dust intensity maps since most dust mass is in larger particles. 
Coincidentally, these two size distributions
lead to the same opacity at 1.27 mm (1.27 mm is the closest wavelength to 1.25 mm in the table of \citealt{birnstiel18}) in the initial condition (the absorption opacity for the $s_{max}=$0.1 mm case is 0.43 $\mathrm{cm^2\,g^{-1}}$, while for the $s_{max}=$1 cm case it is 0.46 $\mathrm{cm^2\,g^{-1}}$ based on  \citealt{birnstiel18}). More discussion on how to generalize our results to disks with other dust size distributions can be found in \S \ref{sec:gapring}. 

For each simulation, we scale the simulation to different disk surface densities. Then for each surface density, we calculate the 1.27 mm intensity maps using DSD1 or DSD2 dust size distributions.
For the $s_{max}=$ 0.1 mm dust size distribution (DSD1), we calculate the 1.27 mm intensity maps for disks with $\Sigma_{g,0}$ = 0.1 \,$\mathrm{g\, cm^{-2}}$, 0.3 \,$\mathrm{g\,cm^{-2}}$, 1 $\mathrm{g\,cm^{-2}}$, 3 $\mathrm{g\,cm^{-2}}$, 10 $\mathrm{g\,cm^{-2}}$, 30 $\mathrm{g\,cm^{-2}}$, and 100 $\mathrm{g\,cm^{-2}}$ (seven groups of models). The maximum-size particle in these disks (0.1 mm), which dominates the total dust mass, corresponds to $St$= 1.57$\times10^{-1}$, 5.23$\times10^{-2}$, 1.57$\times10^{-2}$, 5.23$\times10^{-3}$, 1.57$\times10^{-3}$, 5.23$\times10^{-4}$, and  1.57$\times10^{-4}$ at $r=r_p$. 
For the $s_{max}=$1 cm cases (DSD2), we vary $\Sigma_{g,0}$ as 1 $\mathrm{g\, cm^{-2}}$, 3 $\mathrm{g\, cm^{-2}}$, 10 $\mathrm{g\, cm^{-2}}$, 30 $\mathrm{g\, cm^{-2}}$, and 100 $\mathrm{g\, cm^{-2}}$ (five groups of models), and the corresponding $St$ for 1 cm particles at $r=r_p$ is  1.57, 5.23$\times10^{-1}$, 1.57$\times10^{-1}$, 5.23$\times10^{-2}$, and 1.57$\times10^{-2}$. For each given surface density above, we only select particles with  Stokes numbers smaller than the corresponding $St$  in our simulations   and use the distribution of these particles to calculate the 1.27 mm intensity maps.
For the $s_{max}=$1 cm dust distribution (DSD2), we do not have $\Sigma_{g,0}$ = 0.1 $\mathrm{g\, cm^{-2}}$, 0.3 $\mathrm{g\, cm^{-2}}$ cases since 1 cm particles in these disks have Stokes numbers larger than the largest Stokes number (1.57) in our simulations. 

Here, we lay out the detailed steps to scale each simulation to the  disks that have surface densities of
$\Sigma_{g,0}$ listed above, and then calculate the mm intensity maps for these disks.

1) First, given a $\Sigma_{g,0}$, we find the relationship between the particle size in this disk and the Stokes number of super-particles in simulations. 
For each particle in the simulation, we use its Stokes number in the initial condition to calculate the corresponding particle size $s$  (Equation \ref{eq:stokes} with known $\Sigma_g$).  
The Stokes number of test particles at $r=r_p$ in the initial condition ranges from $St_{min}$ = 1.57 $\times$ $10^{-5}$ to $St_{max}$ = 1.57, or in terms of grain size,  $s^{code}_{min}$ = $St_{min}\times 2\Sigma_{gas}/(\pi\rho_p)$ and $s^{code}_{max}$ = $St_{max} \times 2\Sigma_{gas}/(\pi\rho_p)$ from Equation \ref{eq:stokes}.
For instance, a 1 $\mu$m particle 
in a disk with $\Sigma_g$ = $10 \,\mathrm{g \,cm^{-2}}$ at the planet position corresponds to the particle with
$St$ = 1.57 $\times$ $10^{-5}$ at $r=r_p$ in the initial setup of the simulation. For dust grains with $St$ $<$ $St_{min}$ = 1.57 $\times$ $10^{-5}$, we use the gas surface density $\Sigma_g(r, \theta)$ in our simulations to represent the dust, assuming small dust grains are well coupled with the gas. 

\begin{figure*}[t!]
\includegraphics[width=\linewidth]{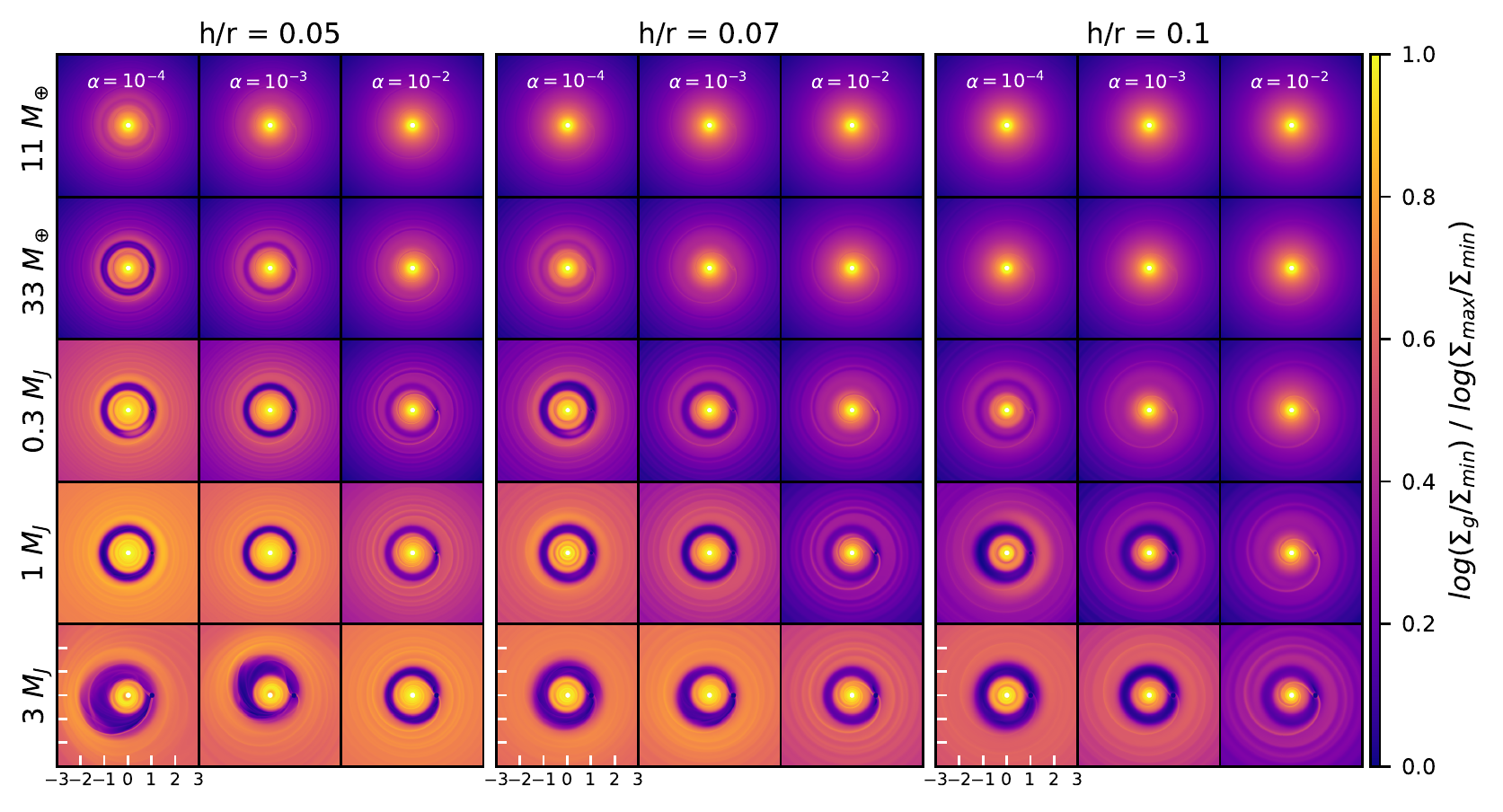} 
\figcaption{The 2-D gas surface density in log scale for $h/r$=0.05, 0.07 and 0.1 from left to right panel blocks.  In each block, the models for $\alpha=10^{-4}\,,10^{-3}\,,10^{-2}$ are shown from left to right. The planet mass increases from top to bottom, namely $M_p$ = 11 $M_\oplus$, 33 $M_\oplus$, 0.3 $M_J$, 1 $M_J$ and 3 $M_J$, if $M_*$ = $M_\odot$. In each panel, the star is located at the center, and the plotting region is 3 $\times$ 3 in units of $r_p$, where $r_p$ is the distance between the star and the planet. The planet is located at (x,y) =  (1,0) and orbits counterclockwise around the star. $\Sigma_{max}$ and $\Sigma_{min}$ are chosen to highlight the structures in each panel. 
\label{fig:2Dgas}}
\end{figure*}

\begin{figure*}[t!]
\includegraphics[width=\linewidth]{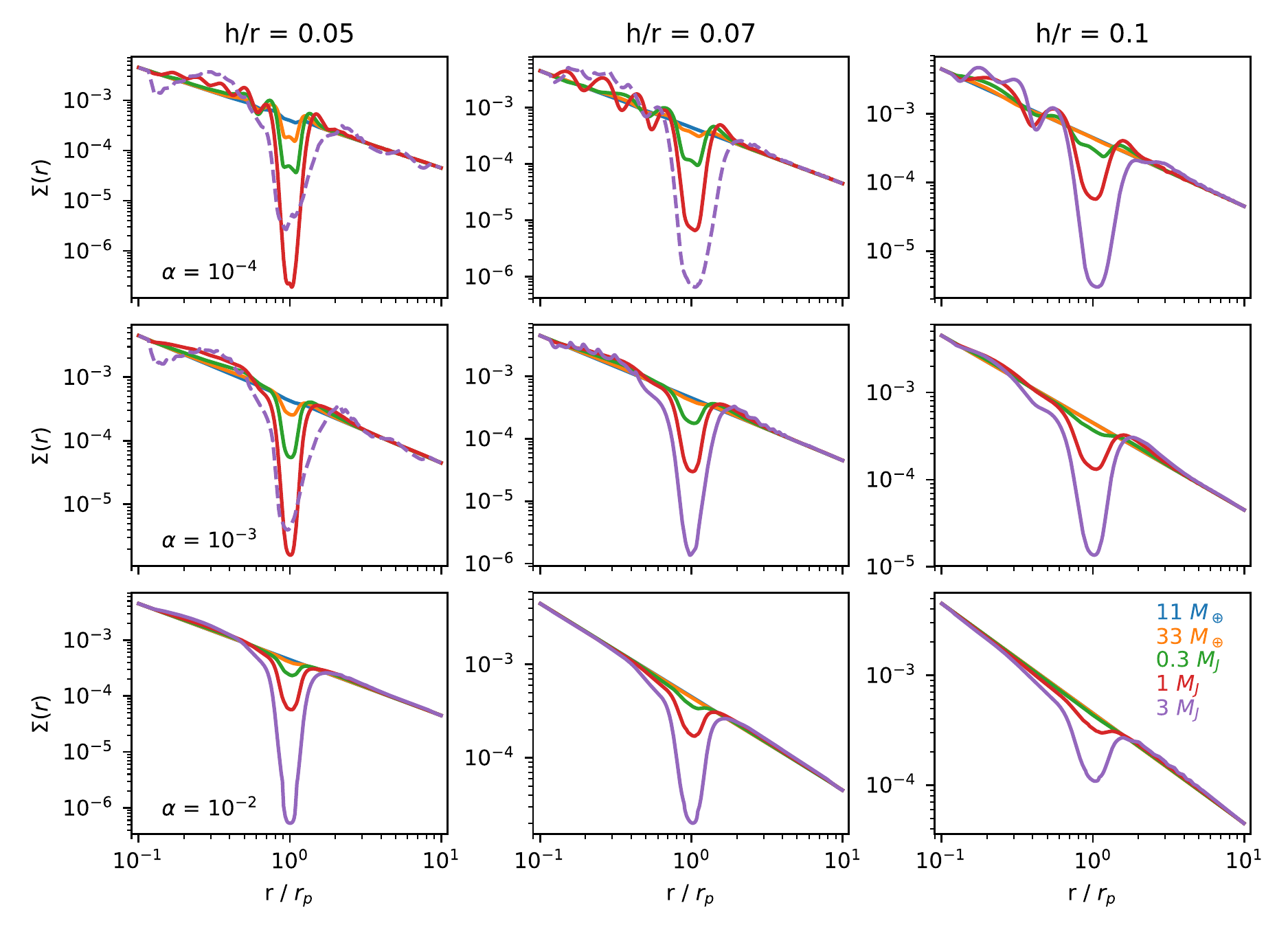}
\figcaption{The azimuthally-averaged gas surface density for models of $h/r$=0.05, 0.07 and 0.1 are shown from left to right. Disks with $\alpha=10^{-4}\,,10^{-3}\,,10^{-2}$ are shown from top to bottom. Blue, yellow, green, red and purple curves represent the gas surface density for planet mass $M_p$ = 11 $M_\oplus$, 33 $M_\oplus$, 0.3 $M_J$, 1 $M_J$ and 3 $M_J$ respectively, if $M_*$ = $M_\odot$. 
The dashed curves show the cases with visible asymmetry at the gap edge in Figure \ref{fig:2Dgas}. 
\label{fig:gasprofile}}
\end{figure*}

2) Then, with a given $\Sigma_{g,0}$, we use the assumed particle size distributions (DSD1 and DSD2) in the initial condition to calculate the mass weight for each super-particle in the simulation. Note that during the simulation, the resulting dust size distribution at each radius is different from the initial dust size distribution since particles drift in the disk. As mentioned above, we  divide the dust component in the disk into two parts: (a) the small dust particles ($s$ $<$ $s_{min}^{code}$) represented by the gas component in the simulation and (b) large dust particles ($s$ $\geq$ $s_{min}^{code}$) represented by the super-particles in the simulation. We calculate the initial mass fractions of the dust contributed by part (a) and (b). The mass fraction of small particles (part a) with respect to the total dust mass is
\begin{equation}
    f_{sd} = \int_{min\{s_{min}, s_{min}^{code}\}}^{min\{s_{max},s_{min}^{code}\}} s^{3+p} ds \bigg/ \int_{s_{min}}^{s_{max}} s^{3+p} ds\,,
\end{equation}
and the mass fraction of large particles using dust super-particles (part b) is
\begin{equation}
    f_{ld} = 1 - f_{sd}\,.
\end{equation} 

We want to explore two dust size distributions $n(s)$ $\propto$ $s^{-3.5}$ and $s^{-2.5}$, given the minimum and maximum dust size $s_{min}$ and $s_{max}$. However, the super-particles in our setup have a different distribution. The number of super-particles $N(s)$ follows a uniform distribution in the $log(s)$ space, $\int N(s) ds$ $\propto$ $dlog(s)$. Thus for dust in part (b), if $f_{ld}$ $>$ 0, we give each particle (having size $s$) a mass weight to scale them into the desired distribution:

\begin{equation}
w_i(s) = \frac{M_{tot}}{N_{part}} \dfrac{s^{3+p}/{\int_{max\{s_{min},s_{min}^{code}\}}^{s_{max}}s^{3+p}ds}}{s^{-1}/{\int_{s_{min}^{code}}^{s_{max}^{code}}s^{-1}ds}}\,,
\end{equation} 
where $M_{tot}$ is the total dust mass in the disk and $N_{part}$ is the total number of super-particles in the simulation.

3) Next, we assign the opacity for each particle to derive the total optical depth. DSHARP opacities are produced by \citet{birnstiel18}, which contains a table of absorption and scattering opacities for a  given wavelength and grain size,  $\kappa(\lambda, s)$. For part (b) dust component, we assign each particle a DSHARP absorption opacity $\kappa_{abs,i}(s_i)$ at 1.27 mm based on the particle's size, where $s_i$ is the $s$ value in the table that is the closest to this particle size. If the particle size is smaller than the minimum size in the opacity table, we take the opacity for the minimum sized particle in the table, namely using a constant extrapolation, since the opacity is already independent of the particle size at the lower size end of the opacity table. We bin all super-particles in each numerical grid cell to derive the total optical depth through the disk for particles in part (b):
\begin{equation}
    \tau_{ld} = f_{ld} \frac{\sum_i w_i(s)\kappa_{abs,i}(s)}{A_{cell}}\,,
\end{equation}
where the sum is adding all particles in the cell, and $A_{cell}$ is the surface area of the grid cell. The optical depth contributed by part (a) is simply
\begin{equation}
    \tau_{sd} = f_{sd}\kappa_{ma}\Sigma_g / 100
\end{equation}
where
\begin{equation}
    \kappa_{ma} = \int_{min\{s_{min}, s_{min}^{code}\}}^{min\{s_{max},s_{min}^{code}\}} \kappa_{abs}(s) s^{3+p} ds 
\end{equation}
is the mass-averaged opacity of the small dust within the range of dust sizes in part (a). The final optical depth for each grid cell at ($r,\theta$) is the sum of both components,
\begin{equation}
    \tau (r,\theta) = \tau_{sd} (r,\theta) + \tau_{ld} (r,\theta)\,.
\end{equation}
Note that we do not consider dust and gas within one Hill radius $r_H$ around the planet for our analysis since our simulations are not able to resolve the circumplanetary region. Thus, we impose the optical depth there to be the minimum optical depth within the annulus ($r_0$ - $r_H$) $<$ r $<$ ($r_0$ + $r_H$).

4) Then, we calculate the brightness temperature or intensity for each grid cell as
\begin{equation}
    T_b(r,\theta) = T_d(r)( 1-e^{-\tau(r,\theta)})\,,
    \label{eq:normTb}
\end{equation}
and we assume that the midplane dust temperature follows the assumed disk temperature. Thus, 
\begin{equation}
    T_d(r)=T_d(r_0)\left(\frac{r}{r_0}\right)^{-0.5}\,.
\end{equation}
 Because we seek to derive a scale-free intensity for different systems, the Rayleigh-Jeans approximation is made here. For the young solar system
and the HR 8799 calculations in \S \ref{sec:syandhr}, and the detailed modeling of AS 209 and Elias 24 in \S\ref{sec:as209} and \S\ref{sec:elias24}, we use the full Planck function at $\nu$ = 240 GHz to derive more accurate intensities.

The normalized brightness temperature ($T_b(r,\theta)/T_d(r_0)$) is adequate for the gap width and depth calculation in \S \ref{sec:gapring}. But for individual sources, we would like to calculate the absolute brightness temperature. Then, we need to multiply the normalized brightness temperature
by the disk temperature at $r_0$ ($T_d(r_0)$). We estimate $T_d(r_0)$ using
\begin{equation}
    T_d(r_0) = \Big(\frac{\phi L_*}{8\pi r_0^2 \sigma_{SB}}\Big)^{1/4}
    \label{eq:disktemp}
\end{equation}
where $L_{*}$ is the stellar luminosity and $\phi$ is a constant of 0.02 coming from an estimate from Figure 3 in \cite{dalessio01}. This disk mid-plane temperature is the same as Equation 5 in \cite{dullemond18b}, and more details can be found there. We calculate $T_{d}(r)$ for each DSHARP source using the stellar properties ($L_*$)  listed in \citet{andrews18b}. Knowing $T_{d}(r)$, we can simply derive  $h/r$ at the gap position using $h/r$ = $c_s$/$v_\phi$ (the $M_*$ that is used to calculate $v_{\phi}$ is also given in \citet{andrews18b}.).  

5) Finally, we convolve these intensity maps with two different Gaussian beams. The beam size is $\sigma=0.06 r_p$ and $\sigma=0.025 r_p$ respectively. For a protoplanetary disk 140 pc away, this is equivalent to FWHM (Full Width Half Maximum) beam size of 0.1" and 0.043" if $r_p=100$ au, or 0.05" and 0.021" if $r_p=50$ au.

\section{Simulation Results \label{sec:results}}

\subsection{Gas}
We will first present results for the gas component in the simulations, including gaseous gap profiles (\S \ref{sec:gasdensity}) and the sub/super Keplerian gas motion at the gap edges (\S \ref{sec:gaskinematics}). 

\subsubsection{Density\label{sec:gasdensity}}
Figure \ref{fig:2Dgas} shows
the two-dimensional gas density maps for all the simulations at 1000 planetary orbits. The left, middle, and right panel blocks show simulations with $h/r$=0.05, 0.07, and 0.1. Within each panel block, $\alpha=10^{-4}$, $10^{-3}$, and $10^{-2}$ cases are shown from left to right. Some large-scale azimuthal structures are evident in the figure. First, low $\alpha$ disks exhibit noticeable horseshoe material within the gap. Since the planet is at ($x$=1, $y$=0) and orbiting around the star in the counterclockwise direction, most horseshoe material is trapped behind the planet (around the L5 point). This is consistent with the shape of the horseshoe streamlines around a non-migrating planet in a viscous disk \citep{masset2002b}. Second, the gap edge becomes more eccentric and off-centered for smaller $h/r$, smaller $\alpha$ and larger planet mass cases (especially for $M_p\geq$ 3 $M_{J}$).
Such an eccentric gap edge for the $M_p\geq$ 3 $M_{J}$ planet is consistent with previous studies \citep{lubow1991a,lubow1991b,kley06,Teyssandier2017}. Third, large-scale vortices can be seen at the gap edges
for some of the $\alpha=10^{-4}$ cases. Although they are not very apparent in the gas surface density maps, they can trap dust particles azimuthally, causing a large azimuthal contrast in the dust continuum images (as shown in \S \ref{sec:dustthermal}). 

The azimuthally averaged gas surface density profiles for all the models are shown in Figure \ref{fig:gasprofile}. Several noticeable trends in this figure are:

1) When the planet mass increases, the gap depth normally increases. However, when the gap is very eccentric (e.g. \texttt{h5am4p5}, \texttt{h5am3p5}), the azimuthally
averaged gas surface density at the gap is actually higher than the cases with lower mass planets. This is because azimuthal averaging over an elliptical gap smears out the gap density profile.  

2) With the same planet mass, gaps in $h/r=0.1$ cases are shallower but wider than the $h/r=0.05$ cases. This is consistent with previous studies \citep{fung14,kanagawa_depth,kanagawa_width}. 

3) For a given planet mass and $h/r$, the gaps are shallower and smoother with increasing $\alpha$. 
With $\alpha=0.01$ and $10^{-3}$, the gap edge is smooth, and there is only a single gap at $r/r_p \sim 1$.
With $\alpha=10^{-4}$, there are clearly two shoulders at two edges of the gap, and the material in the horseshoe region still remains in some cases. 
Especially, for low mass planets in  $\alpha=10^{-4}$ disks, the gap
at $r/r_p\sim 1$ appears to split into two adjacent gaps. This is consistent with non-linear wave steepening theory \citep{goodman2001, muto2010, dong2011, duffell2012, zhu2013} which suggests that the waves launched by a low mass planet in an inviscid disk need to propagate for some distance to shock and open gaps, leaving the horseshoe region untouched.  

5) For $\alpha=10^{-4}$ cases, we see secondary gaps at $r/r_p \sim 0.6$
in $h/r$=0.05 disks, $r/r_p \sim 0.5$ in $h/r=0.07$ disks, and $r/r_p \sim 0.4$ in $h/r=0.1$ disks. For some cases, we  can even see tertiary gaps at smaller radii. 
These are consistent with simulations by \cite{bae2017,dong2017} and these gaps are due to the formation of shocks from the secondary and tertiary spirals \citep{bae2018a, bae2018b}.

\subsubsection{Kinematics Across the Gap \label{sec:gaskinematics}}

\begin{figure}[t!]
\includegraphics[width=\linewidth]{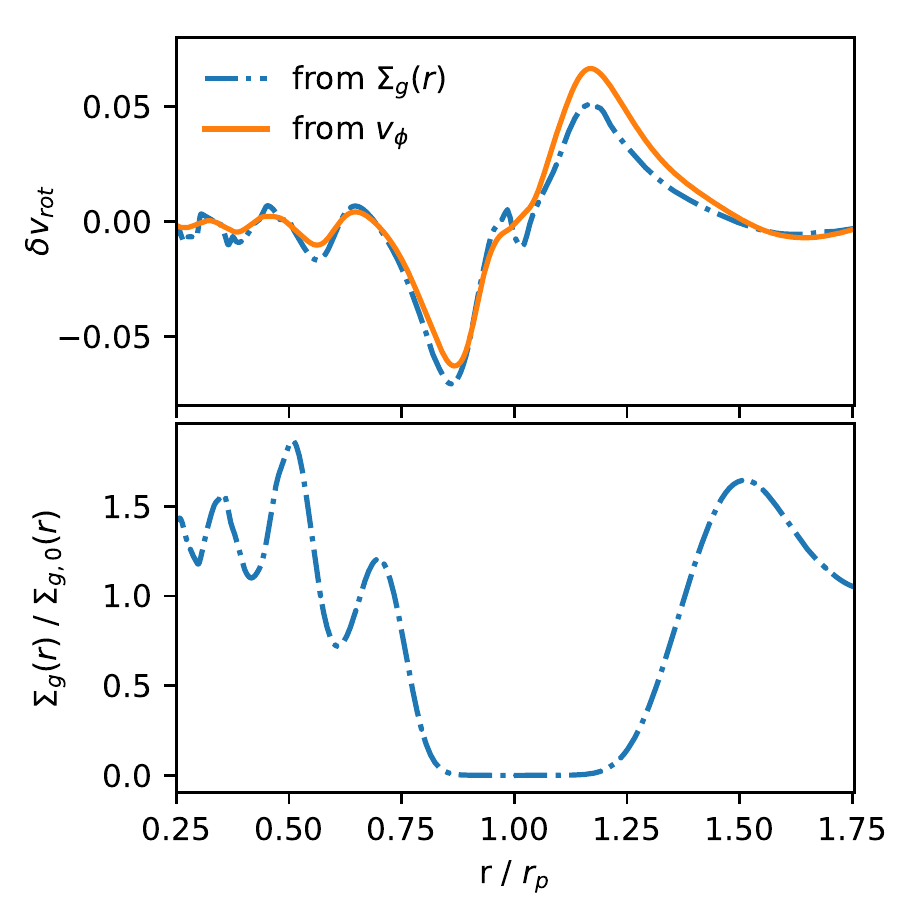}
\figcaption{The deviation from the Keplerian velocity $\delta v_{rot}$ (the upper panel) and the normalized disk surface density (the disk density over the initial disk density, the lower panel) across the gap for 
model \texttt{h5am4p4}. In the upper panel, the directly measured $\delta v_{rot}$ is plotted as the orange curve, while the $\delta v_{rot}$ derived from the radial force balance is plotted as the blue curve. 
\label{fig:subkep}}
\end{figure}

\begin{figure*}[t!]
\includegraphics[trim=0cm 0cm 0cm 0cm, width=1.\textwidth]{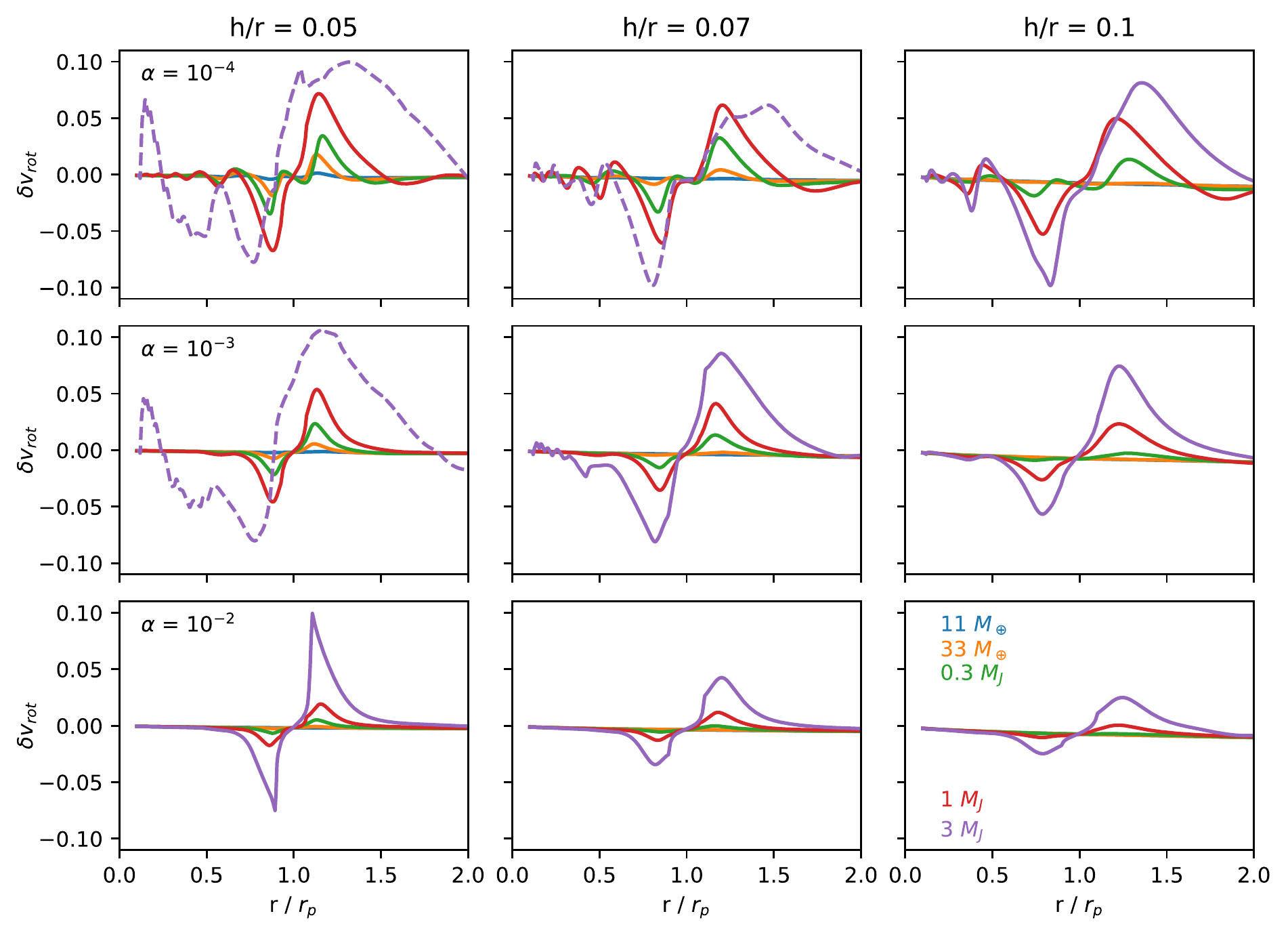}
\figcaption{The deviation from the Keplerian velocity for all runs, where $\delta v_{rot}$ = ($v_{\phi}$ - $v_K$) / $v_K$.
\label{fig:subkep_all} The layout is the same as Figure \ref{fig:gasprofile}.}
\end{figure*}

Recent works by \cite{teague2018} and \cite{pinte2018} have shown that, using molecular lines, ALMA can detect the velocity deviation from  Keplerian rotation in protoplanetary disks. Such deviations are caused by the radial pressure gradient at the gaseous gap edges,
\begin{equation}
    \frac{v_{\phi}^2}{r}=\frac{v_{K}^2}{r}+\frac{1}{\rho_{gas}}\frac{\partial P}{\partial r}\,.\label{eq:vphi}
\end{equation}
In our 2-D simulations, $v_{K}$ is simply $\sqrt{GM_*/r}$ and $P$ is $\Sigma c_s^2$. 
Equation \ref{eq:vphi} suggests that the deviation from the Keplerian motion is  
\begin{equation*}
    \frac{\Delta v_{\phi}}{v_{K}}\sim \frac{r}{2\rho_{gas}v_{K}^2} \frac{\partial P}{\partial r}\,,
\end{equation*}
 where $\Delta v_{\phi}=v_{\phi}-v_{K}$. In a smooth disk where $\partial P/\partial r\sim P/r$, this deviation is very small, on the order of $(h/r)^2$ or $\leq$1$\%$ in a typical protoplanetary disk. But if the gaseous disk has a sharp pressure transition (e.g. at gap edges), the deviation from the Keplerian rotation can be significantly larger. In Figure \ref{fig:subkep}, we plot the azimuthally averaged $\delta v_{rot}\equiv(v_{\phi}-v_{K})/v_{K}$ and $\Sigma$ in run \texttt{h5am4p4}. The directly measured $\delta v_{rot}$ is plotted as the orange curve in the upper panel, while the calculated $\delta v_{rot}$ using the disk surface density profile (presented in the lower panel) and Equation \ref{eq:vphi} is plotted as the blue curve in the upper panel. We can see that Equation \ref{eq:vphi} reproduces the measured azimuthal velocity very well, confirming that the sub/super Keplerian motion is due to the radial pressure gradient. 
 
\begin{figure}[t]
\includegraphics[width=\linewidth]{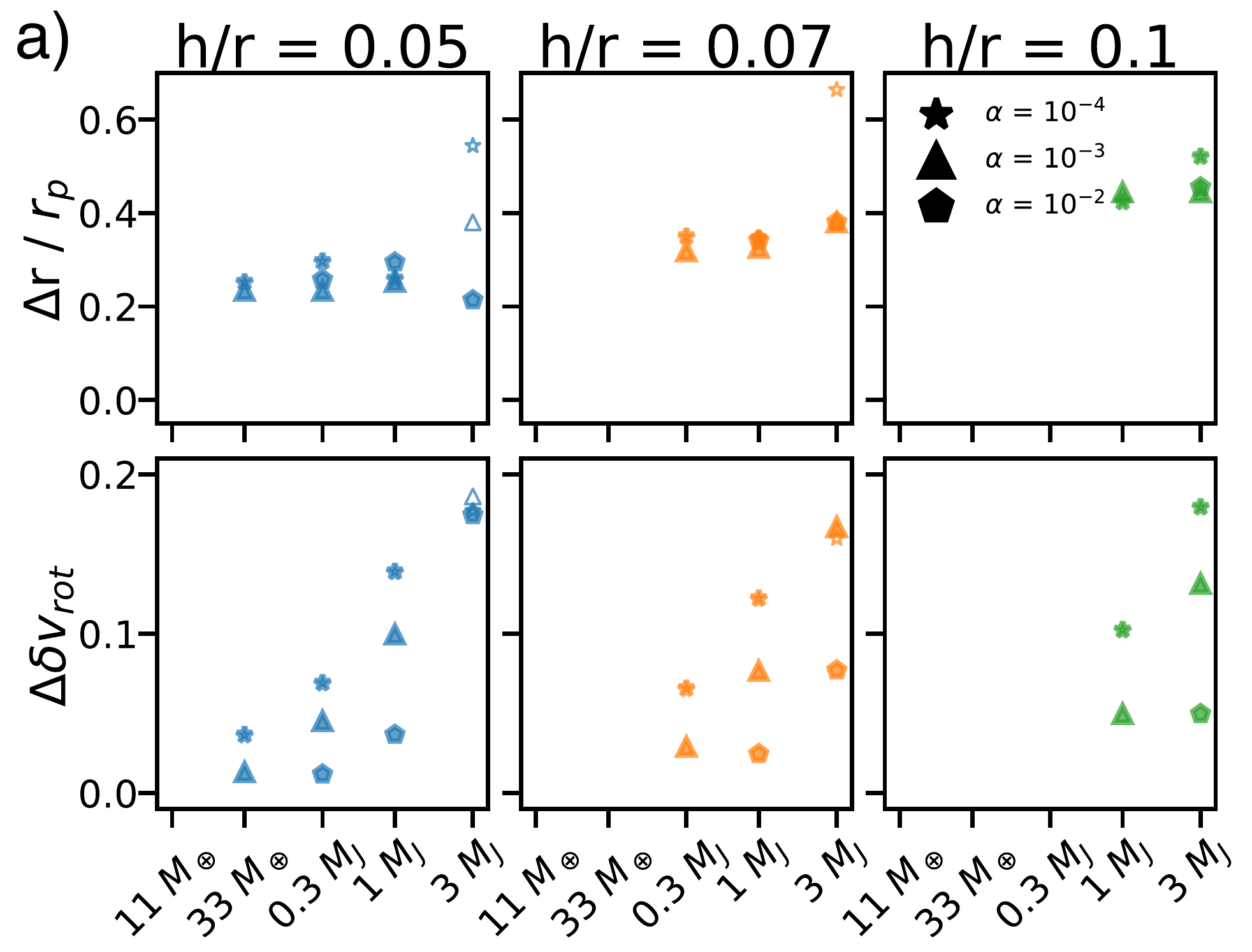}
\includegraphics[width=\linewidth]{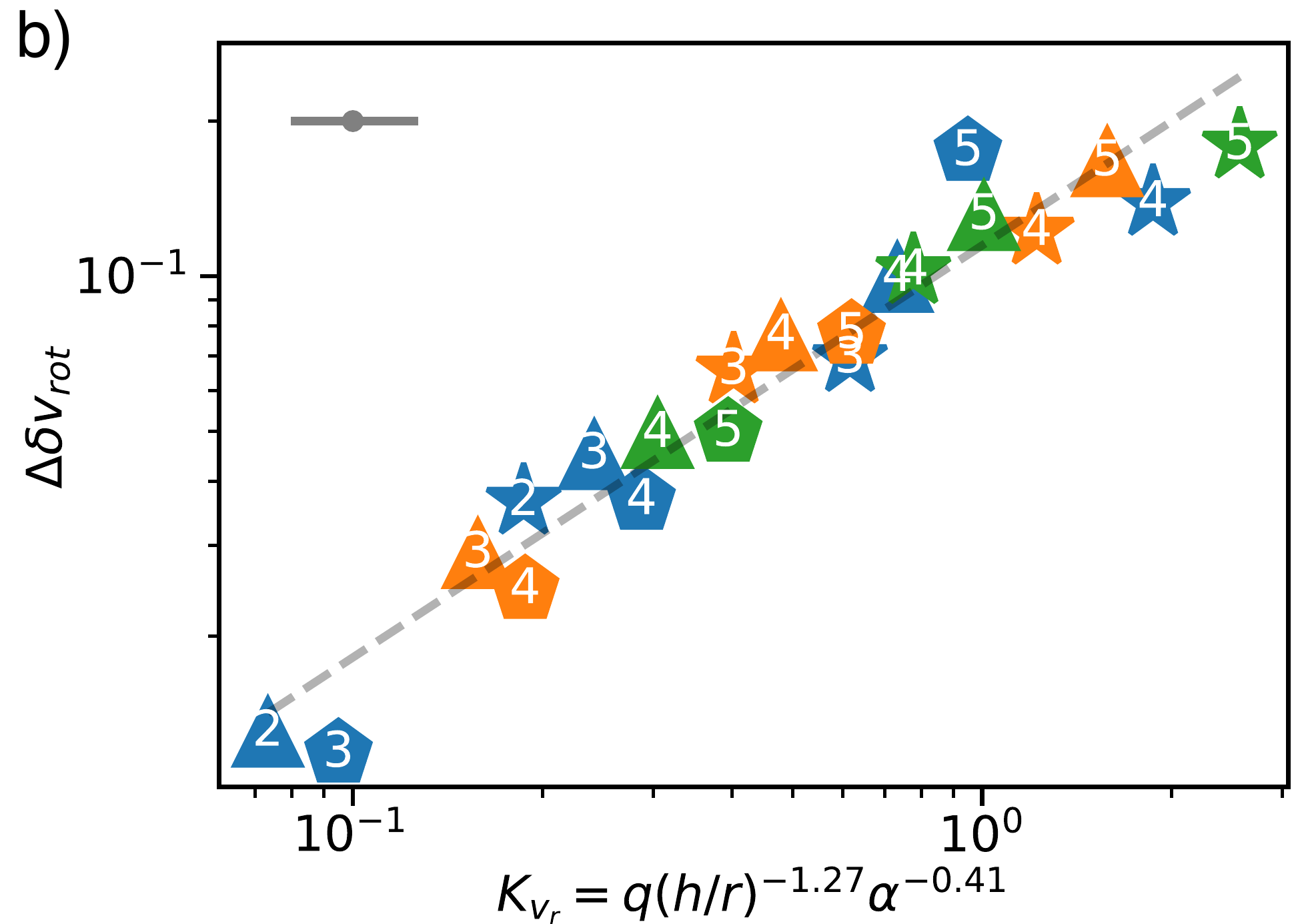}
\caption{Panel a: upper panels show the radial distance between the positions of $\delta v_{rot}$ maximum and minimum peaks ($\Delta r$). 
Bottom panels show the difference between $\delta v_{rot}$ at its maximum and minimum  values ($\Delta\delta v_{rot}$). The star, triangle and pentagon markers represent models with $\alpha$ = $10^{-4}$, $10^{-3}$ and $10^{-2}$, respectively. The unfilled markers are eccentric cases the same as in Figure \ref{fig:gasprofile} and \ref{fig:subkep_all} shown in dashed lines. Panel b: the fitting formula (Equation \ref{eq:Kvr}) with all measured $\Delta\delta v_{rot}$ in panel A. The numbers inside the symbols represent cases with different planet masses in ascending order (e.g., "1" stands for 11 $M_\oplus$). The error-bar is shown at the upper-left corner. \label{fig:deltadrdv}}
\end{figure}

 Figure \ref{fig:subkep_all} shows $\delta v_{rot}$ for all our cases. As expected, when the gap is deeper due to either smaller $\alpha$, smaller $h/r$, or a more massive planet, the amplitude of $\delta v_{rot}$ is larger. However, when the gap becomes very eccentric and off centered (e.g. \texttt{h5am4p5}, \texttt{h5am3p5}), the azimuthally averaged $\delta v_{rot}$ shows a much wider outer bump, indicating an eccentric outer disk. We label these cases as dashed curves in Figure \ref{fig:subkep_all} and unfilled markers in panel a of Figure \ref{fig:deltadrdv}. 
 Another interesting feature shown in Figure \ref{fig:subkep_all}  is that the presence of the gap edge vortices in $\alpha=10^{-4}$ cases does not affect the azimuthally averaged $\delta v_{rot}$ very much. They look similar to the larger $\alpha$ cases without vortices. We interpret this as: if the vortex is strong with fast rotation, it has a smaller aspect ratio so that it is physically small \citep{lyra13} and contributes little to the azimuthally averaged gas velocity profile; and if the vortex is weak, although it has a wider azimuthal extent  its rotation is small compared with the background shear so again it contributes little to the global velocity profile. 

The radial distance and amplitude of the sub/super Keplerian peaks are plotted in  panel a of Figure \ref{fig:deltadrdv}. $\Delta\delta v_{rot}$ is the difference between the maximum $\delta v_{rot}$ (at $r>r_p$) and the minimum  $\delta v_{rot}$ (at $r<r_p$) from Figure \ref{fig:subkep_all}.  Note that these velocity peaks are not peaks (or rings) at mm intensity images that will be presented in \S \ref{sec:dustthermal}. We first notice that the distance between these peaks in $\Delta r/r$ is roughly 4.4 times $h/r$, which is not sensitive to either the planet mass or $\alpha$ (upper panel a). Thus, we can use the distance of these sub/super-Keplerian peaks to roughly estimate the disk temperature. 
On the other hand, the amplitude of the sub/super Keplerian peaks depends on all of these parameters (lower panel a). With increasing planet mass, the amplitude increases until the gap edge becomes eccentric. For the same mass planet in the same $h/r$ disk, the amplitude decreases with increasing $\alpha$. For the same mass planet in the same $\alpha$ disk, the amplitude decreases with increasing $h/r$. 
 
Thus, using gas kinematics,  we can first use the distance between the peaks to estimate $h/r$, and then we can use the amplitude together with the estimated $h/r$ and assumed $\alpha$ value to derive the planet mass. 

Following \cite{kanagawa_depth,kanagawa_width}, we seek simple power laws to fit various observable quantities throughout the paper so that the fittings can be easily used by the community. Here, we try to find the best fit for $\Delta\delta v_{rot}$. We define a $K_{v_r}$ parameter that is proportional to $q$ and has power law dependence on $h/r$ and $\alpha$,
\begin{equation}
    K_{v_r} = q(h/r)^{ph}\alpha^{pa}\,.\label{eq:Kdef}
\end{equation}
We try to find the best fitting parameters $ph$ and $pa$. If $ph$=0 or $pa$=0, it means that the fitting does not depend on the disk $h/r$ or $\alpha$, respectively. 
First, we assign values to $ph$ and $pa$, and we can make the log $\Delta\delta v_{rot}$ - log $K_{v_r}$ plot for all the data points. Then, we do a linear-regression fitting for these data points using
\begin{equation}
    \Delta\delta v_{rot} = AK_{v_r}^{B} \,.
\end{equation}
The coefficients in the fitting ($A$ and $B$) are thus determined. The sum of the square difference of the vertical distance between the data points and the fitting is $\sigma$. Finally, we vary $ph$ and $pa$ and follow  the same fitting procedure until the minimum $\sigma$ is achieved. The resulting $ph$ and $pa$ are the best degeneracy parameters, and $A$ and $B$ are the best fitting parameters. For $\Delta\delta v_{rot}$, the fitting formula is:
\begin{equation}
    K_{v_r} = q(h/r)^{-1.27}\alpha^{-0.41}\nonumber
\end{equation}
with
\begin{equation}
    \Delta\delta v_{rot} = 0.11 K_{v_r}^{0.80} \label{eq:Kvr}\,.
\end{equation}
Thus, the sub/super Keplerian motion is most sensitive to $h/r$, followed by $q$ and $\alpha$. 
The fitting formula is shown in panel b of Figure \ref{fig:deltadrdv} together with all measured $\Delta\delta v_{rot}$. 
The uncertainty in $K_{v_r}$ is estimated by measuring the horizontal offset between each data point and the fitting line. From the distribution of the offset, the left side error is estimated by the 15.9 percentile of the distribution and the right side error is 84.1 percentile of the distribution. The uncertainty in $log_{10}(K_{v_r})$ is $_{-0.099}^{+0.103}$, which is about a factor of 1.25 of $K_{v_r}$.

\subsection{Dust Thermal Emission \label{sec:dustthermal}}
After exploring the gaseous gaps, we study the gaps in mm dust continuum maps
in \S \ref{sec:features}. We detail our method to fit the gap width and depth in \S \ref{sec:gapring}. 

\subsubsection{Axisymmetric and Non-axisymmetric Features \label{sec:features}}
As discussed in \S \ref{sec:scaling}, we have 45 simulations with different $h/r$, $\alpha$, and $M_p$.
For each simulation, we generate seven continuum maps for seven $\Sigma_{g,0}$ with the DSD1 dust size distribution and five continuum maps for five $\Sigma_{g,0}$ with the DSD2 dust size distribution. Thus, we produce 45$\times$12 mm maps.

\begin{figure*}[t!]
\includegraphics[width=\linewidth]{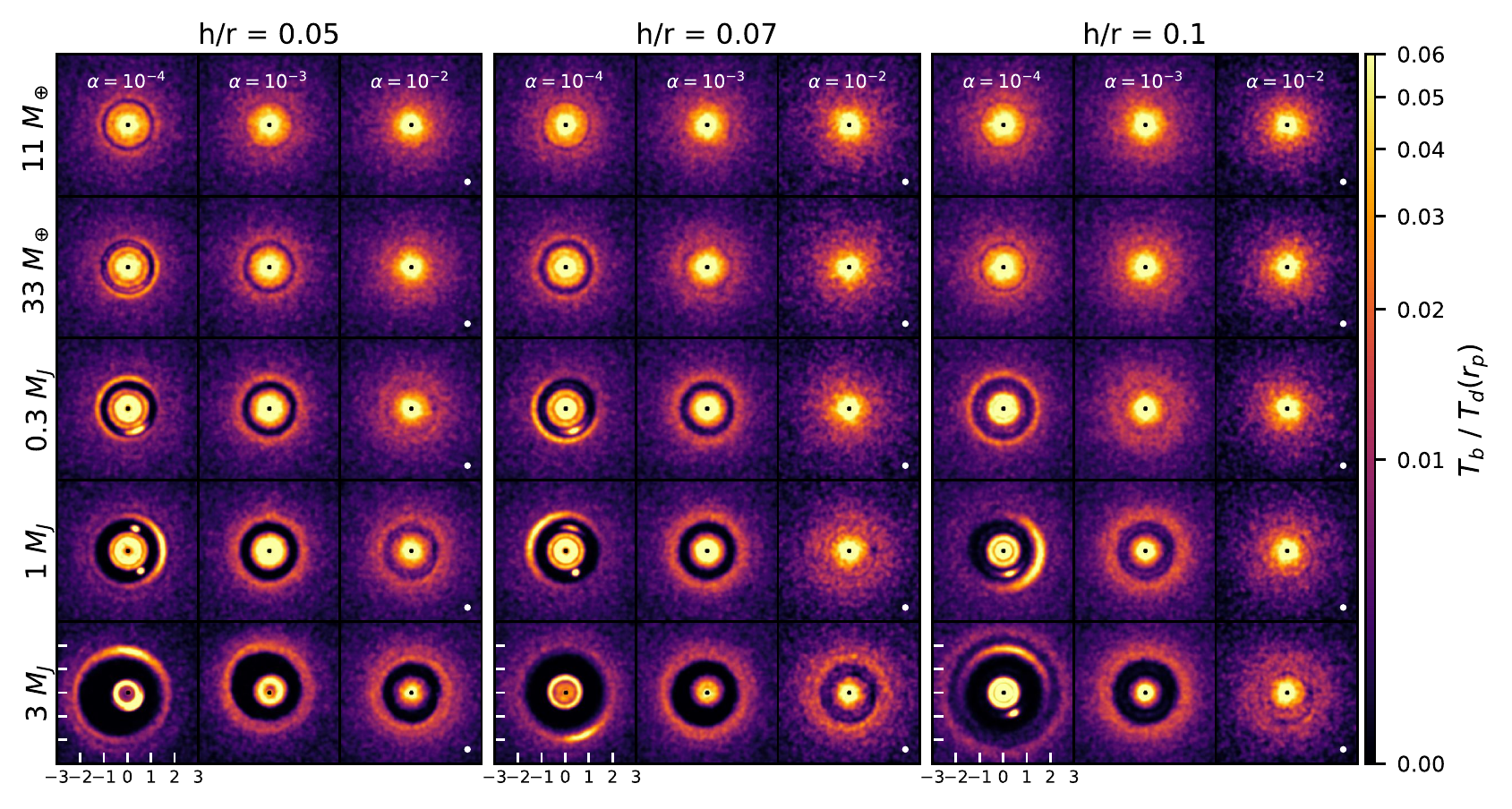}
\figcaption{The dust continuum emission maps for cases with $h/r$=0.05 (left panels), $h/r$=0.07 (middle panels) and $h/r$=0.1 (right panels) at 1.27 mm. The initial gas surface density
at the planet position $\Sigma_{g,0}$ is 3 $\mathrm{g\, cm^{-2}}$. The initial dust size distribution is assumed to follow $n(s)\propto s^{-3.5}$ with the maximum grain size of 0.1 mm (DSD1). The layout is the same as Figure \ref{fig:2Dgas}. The images are convolved with a Gaussian kernel with $\sigma$ of 0.06 $r_p$ (or FWHM of 0.14 $r_p$), which is shown in the bottom right of the panels. 
\label{fig:slope3p5max100}}
\end{figure*}

\begin{figure*}[t!]
\includegraphics[width=\linewidth]{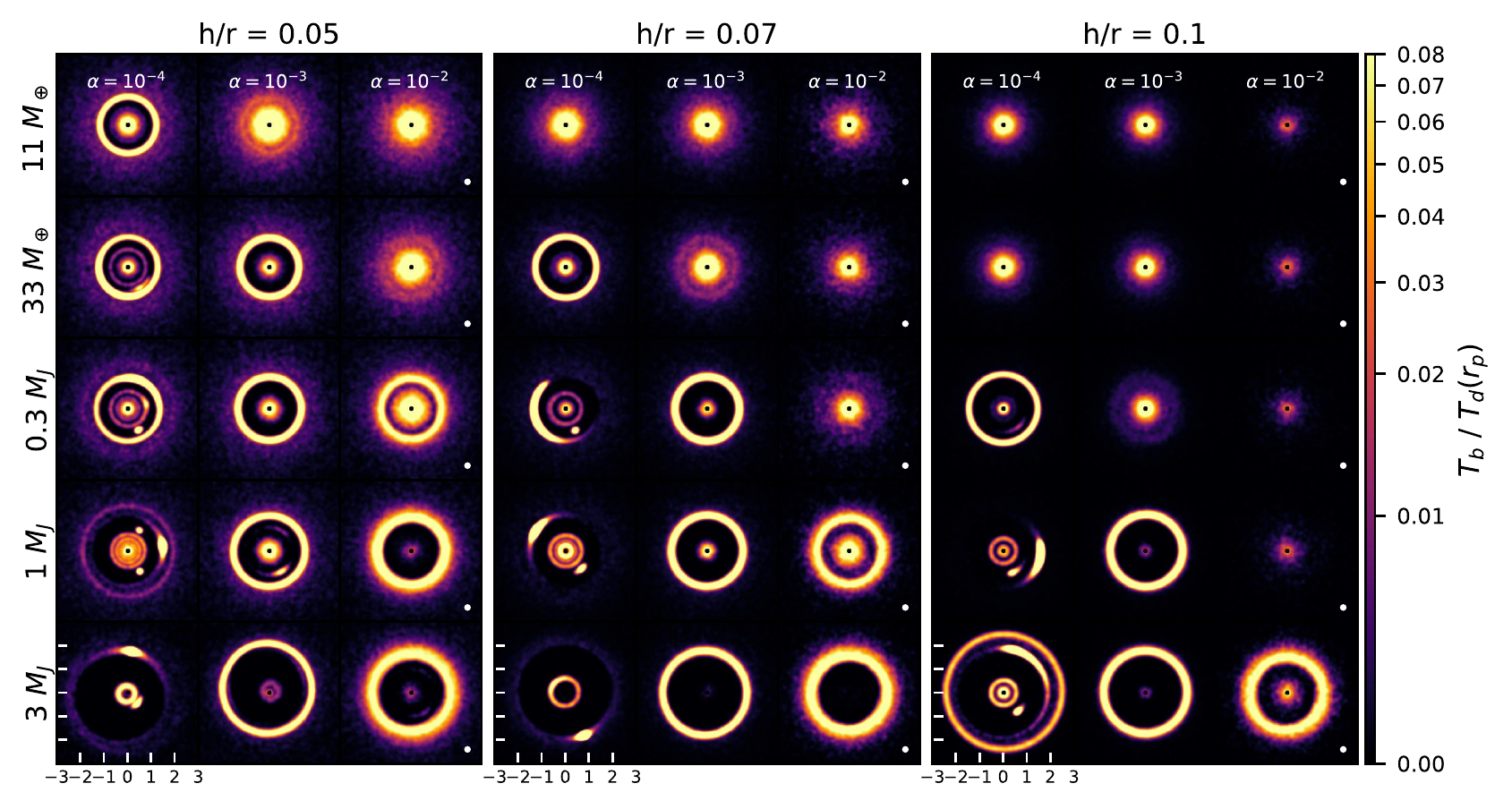}
\figcaption{Similar to Figure \ref{fig:slope3p5max100}, except that
the initial dust size distribution is assumed to follow $n(s)\propto s^{-2.5}$ with the maximum grain size of 1 cm (DSD2). 
\label{fig:slope2max1000}}
\end{figure*}

The mm intensity maps for a $\Sigma_{g,0}=3\, \mathrm{g\, cm^{-2}}$ disk with DSD1 and DSD2 dust size distributions are presented in Figures \ref{fig:slope3p5max100} and \ref{fig:slope2max1000}, respectively. We want to emphasize that, if the opacity is a constant with the maximum dust size (which roughly stands when  the maximum dust size, $s_{max}$, is not significantly larger than the  wavelength of observation), there is a degeneracy in the relative intensity maps between different $\Sigma_{g}$ and $s_{max}$ because only the Stokes number matters for the gas dynamics.  
For example, the shapes of intensity maps for the $\Sigma_{g,0}=3\, \mathrm{g\, cm^{-2}}$ and $s_{max}=0.1\, \mathrm{mm}$ cases are very similar to the $\Sigma_{g,0}=300\, \mathrm{g\, cm^{-2}}$ and $s_{max}=1 \,\mathrm{cm}$ cases, since they have the same Stokes number. Thus, Figure \ref{fig:slope3p5max100} should be regarded as the dust well-coupled limit, while Figure \ref{fig:slope2max1000} should be regarded as the dust fast-drifting limit.

Regarding the gaps and rings, there are several noticeable trends: 

1) By comparing these two figures, we can see that
the rings are more pronounced when particles with larger Stokes numbers are present in the disk. 
For the well-coupled case (Figure \ref{fig:slope3p5max100}), 
the gap edge is smoothly
connecting to the outer disk and the outer disk is extended. However, for the fast-drift particle cases (Figure \ref{fig:slope2max1000}), there is a clear dichotomy: either the disk does not show the gap or the gap edge becomes a narrow ring. This is because the gap edge acts as a dust trap so that a small gaseous feature can cause significant pileup for fast-drifting particles. 

2) The marginal gap opening cases are in panels that are along the diagonal line in Figures \ref{fig:slope3p5max100} and \ref{fig:slope2max1000}, which are similar to the trend for the gaseous gaps in Figure \ref{fig:2Dgas}.   

3) The narrow gap edge of the the fast-drifting particle cases (Figure \ref{fig:slope2max1000}) becomes wider with a higher $\alpha$ due to turbulent diffusion. Thus, if we know the particles' Stokes number at the gap edge, we can use the thickness of the ring to constrain the disk
turbulence, as shown in \cite{dullemond18b}. 

Besides axisymmetric structures, there are also several non-axisymmetric features to notice:

\begin{figure}[t]
\includegraphics[width=\linewidth]{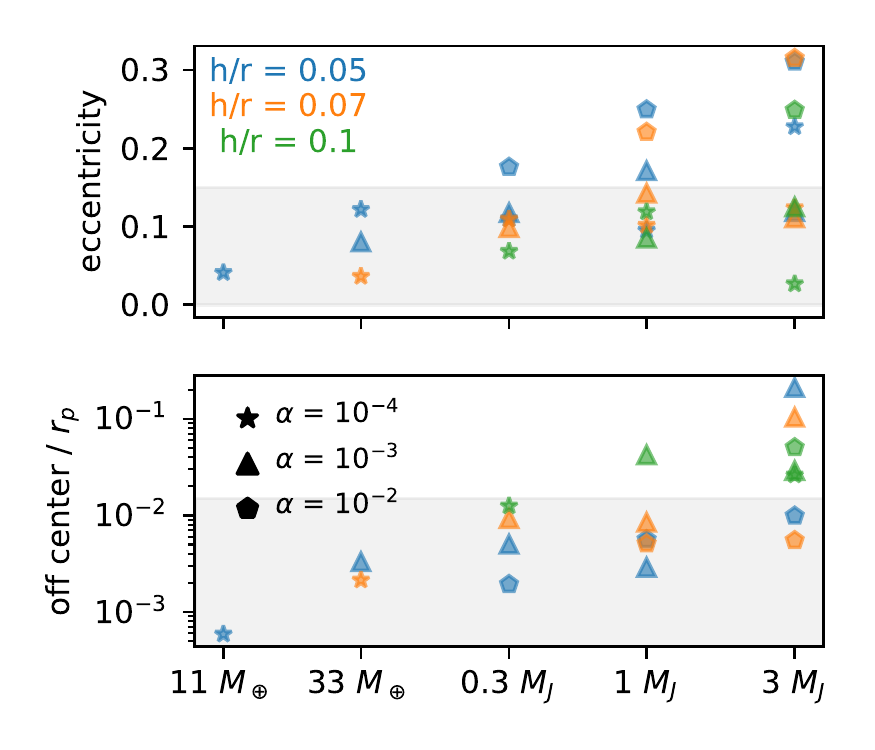}
\caption{Eccentricity (upper panels) and distance between the ellipse center and the central star (lower panels) for intensity images from $\Sigma_{g,0}$ = 3 $\mathrm{g\,cm^{-2}}$ (Figure \ref{fig:slope3p5max100} and \ref{fig:slope2max1000}). \label{eq:eccentricity}} 
\end{figure}

1) The gaps in the lower left panels (\texttt{h5am4p5}, \texttt{h5am3p5}) are clearly eccentric
and off-centered. We may be able to use the ellipticity of the gap edges to infer the planet properties. Thus, for every mm intensity map,
we find the local maximum in each azimuthal angle and use linear fitting method to measure the gap eccentricity and the distance between the center of the  ellipse and the star. We find that, even in mm images generated from disks having dramatically different Stokes numbers, the gap eccentricity and off-centered distance are quite similar.  However, the lower planet mass cases for the DSD1 have mild dust trapped rings thus having lower SNR, while the higher mass cases for the DSD2 have strong asymmetry, thus leading to half of the rings with the low SNR. Thus, we combine the fitting results for both DSD1 and DSD2 at $\Sigma_{g,0}$ = 3 g cm$^{-2}$, and pick up the smaller values for eccentricity and the off-centered distance (Figure \ref{eq:eccentricity}). 
We also test several cases with the ring-fitting method described in \S 3.1 in \cite{huang18b} (a MCMC fitting of the offset $\Delta x$, $\Delta y$, the semi-major axis, the aspect ratio and the position angle) and find that the derived eccentricity and the distance from the central star are very similar to those derived here. Clearly, both eccentricity and off-centered distance increase with the planet mass, which is consistent with gas only simulations in \cite{kley06,ataiee2013,Teyssandier2017,ragusa2018}. These quantities do not quite depend on $h/r$ and $\alpha$ except a weak trend that gaps in larger $\alpha$ disks have higher eccentricities. Unfortunately, due to the limited number of super-particles in the simulations, the Poisson noise in the intensity maps  prevents us from measuring the eccentricity very accurately. The adopted Gaussian convolution kernel to reduce the Poisson noise has a $\sigma_c$ of  0.06 $r_p$. If the major-axis and the minor-axis have an error of $\sigma_c$/2, the uncertainty of the eccentricity is $\Delta  e=\big(1- (1-0.03/2)^2\big)^{\frac{1}{2}}=0.17$. Thus, any measured eccentricity smaller than 0.15 is consistent with zero eccentricity. For the same reason, any off-centered distance smaller than half of the pixel size (0.015) is consistent with zero. We mark these uncertainties as the light grey area in Figure \ref{eq:eccentricity}. On the other hand, if the eccentricity and the off-centered distance is above these limits, our results suggest that the eccentric gap edge may be a signature of a massive planet in disks. Eccentric and off-centered gap edges have been measured in HL Tau \citep{ALMA2015} and HD 163296 \citep{isella16}, which may suggest that these gaps are induced by planets.

\begin{figure*}[t!]
\includegraphics[width=\linewidth]{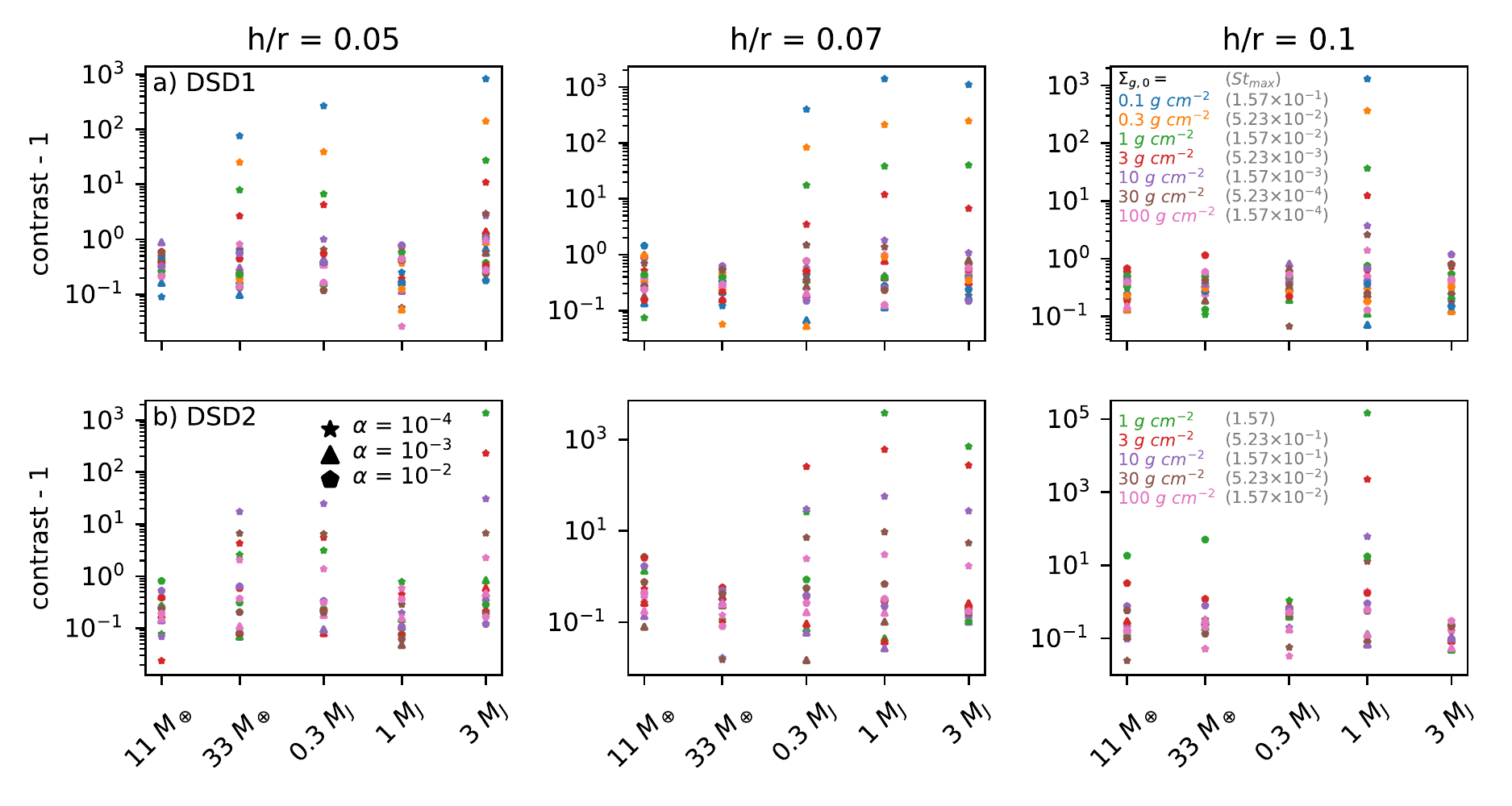}
\figcaption{The contrast at the outer gap edge for every model. The upper panels use DSD1 dust size distribution while the lower panels use DSD2 dust size distribution. Contrast is the intensity of the brightest part of the ring over the intensity at $\Delta\theta$ = 180$^\circ$ opposite location on the ring.
\label{fig:contrast}}
\end{figure*}

2) For the lowest viscosity cases ($\alpha=10^{-4}$), particle concentration within vortices can be seen at the gap edge. Even a 33 M$_{\earth}$ planet
can induce particle-concentrating vortices. Interestingly, the vortex sometimes is inside the gap edge, e.g.  $h/r=0.05$, $M_p=1 M_J$ case and $h/r=0.1$, $M_p=3 M_J$ case. This is probably because large particles are trapped at the gap edges, while small particles move in and get trapped into the vortex. 
For the majority of cases, the vortices that cause significant asymmetry in mm intensity maps are at the gap edge where $dP/dr=0$. To characterize such large-scale asymmetries, Figure \ref{fig:contrast} shows the contrast at the gap edge, which is the ratio between the intensity of the brightest part of the ring over the intensity  180 degree opposites on the previously fitted ellipse. The figure shows that the case with a smaller gas surface density tends to show a higher contrast. 
We note that
the contrast is very large in some cases. A 33 $M_{\earth}$ planet can lead to a factor of 100 contrast at the gap edge for a $h/r=0.05$ disk with $St=0.16$ particles.
Thus, a low mass planet may also explain some of the extreme asymmetric systems: e.g. IRS 48  \citep{vandermarel13} and HD 142527 \citep{casassus13}.

3) The dust concentration at L5 or both L4/L5 is seen in some $\alpha=10^{-4}$ cases, consistent with previous simulations \citep{lyra2009}. These features are more apparent than
those in the gas (Figure \ref{fig:2Dgas}). As pointed out by \cite{Ricci2018}, such features may be observable.  On the other hand, we want to emphasize that the dust concentration at Lagrangian points is not in a steady state, and the amount of dust at those points decreases with time. Thus, in this paper, we will not use these feature to constrain the planet properties.

\subsubsection{Fitting Gaps/Rings \label{sec:gapring}}

\begin{figure*}[t!]
\includegraphics[trim=0cm 0cm 0cm 0cm, width=1.\textwidth]{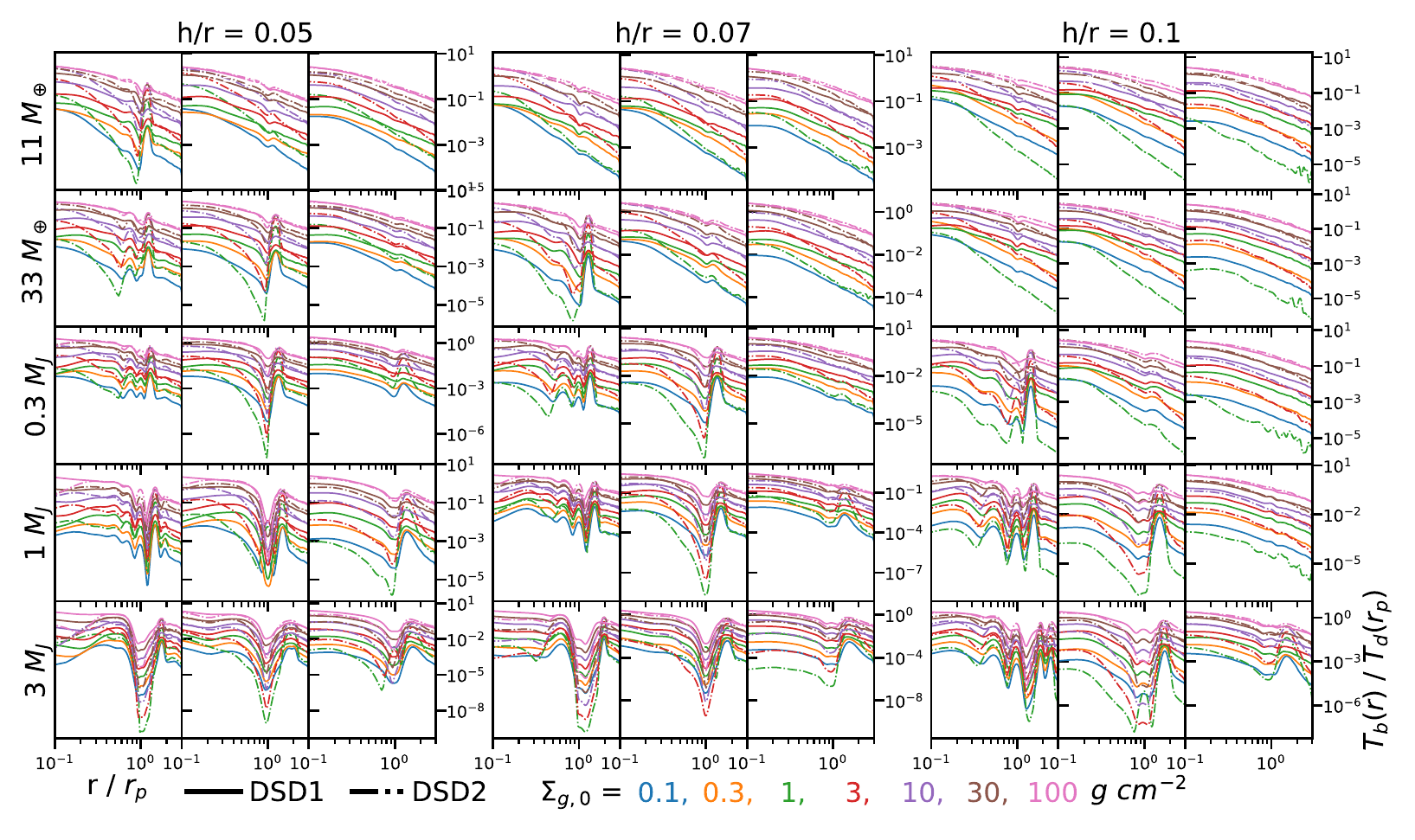}
\figcaption{The `normalized' radial intensity profile for cases with $h/r$=0.05 (left panels) $h/r$=0.07 (middle panels), and $h/r$=0.1 (right panels). From left to right in each panel block, $\alpha=10^{-4}\,,10^{-3}\,,10^{-2}$ in disks. From top to bottom, the planet mass increases (the layout is similar to Figure \ref{fig:2Dgas}, \ref{fig:slope3p5max100} and \ref{fig:slope2max1000}.).
The solid curves are calculated with the DSD1 dust size distribution, while the dot-dashed curves are calculated with the DSD2 dust size distribution. The seven colors of lines denote different initial gas surface densities ($\Sigma_{g,0}$). The profiles are smoothed with a Gaussian kernel with  $\sigma=$0.06 $r_p$. 
\label{fig:Tb1d}}
\end{figure*}

To derive the relationship between the gap profiles and the planet mass, we azimuthally average the mm intensity maps as shown in Figure \ref{fig:Tb1d}. 
The solid curves are for models with
$s_{max}=0.1$ mm (DSD1), while the dashed curves are for models with $s_{max}=1$ cm (DSD2). 

We try to find the relationship between the planet mass and the gap properties (such as the gap width $\Delta$ and depth $\delta$), using the dust intensity profiles in Figure \ref{fig:Tb1d}. Previous works such as \citealt{kanagawa_width, kanagawa_depth, DongFung} studied the relationship between the planet mass and the gaseous gap width and depth. However, mm observations are probing dust with sizes up to mm/cm and these dust can drift in the gaseous disk. Thus, studying only the gaseous gap profiles is not sufficient for explaining mm observations and carrying out a similar study but directly for dust continuum maps is needed.  We seek to first find a relationship between disk and planet properties ($\alpha$, $h/r$, and $M_p$) using the fitting of the azimuthally averaged gas surface density profile, and characterize those three parameters using a single parameter $K$ (for the depth-$K$ relation) or $K'$ (for the width-$K$' relation). Then, we fit the azimutally averaged dust intensity profile for our grid of models and find their depth-$K$ and width-$K'$ relations. Overall, our fitting follows \cite{kanagawa_width} and \cite{kanagawa_depth} but extend those relationships to dust particles with different sizes. 

\begin{figure}[t]
\includegraphics[width=\linewidth]{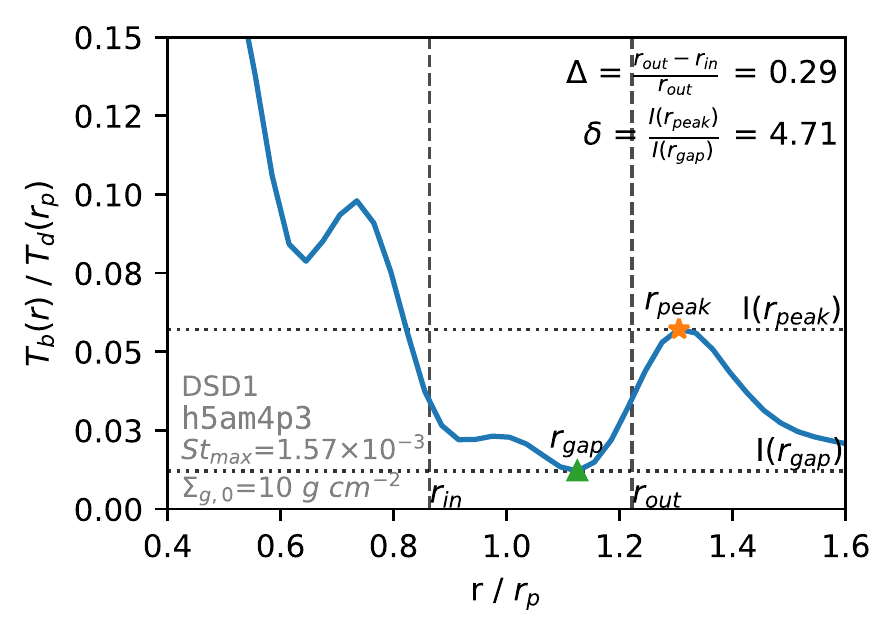}
\caption{\label{fig:gap_def} An example of our definition of the gap depth ($\delta$) and width ($\Delta$). $r_{peak}$ (marked by a star) and $r_{gap}$ (marked by a triangle) are first found and are used to calculate $I_{edge}$, which is the average between $I(r_{peak})$ and $I(r_{gap})$. $r_{out}$ and $r_{in}$ are positions where the intensity equals $I_{edge}$. The gap width ($\Delta$) is $(r_{out}$-$r_{in})/r_{out}$. The depth ($\delta$) is $I(r_{peak}) / I(r_{gap})$. (This example is taken from model \texttt{h5am4p3} with $\Sigma_{g,0}$ = 10 $\mathrm{g\,cm^{-2}}$ and DSD1.) } 
\end{figure}

The detailed steps are the following:

(1) We measure the gap depth ($\delta$) for both gas surface density profiles (Figure \ref{fig:gasprofile}) and  mm intensity profiles (Figure \ref{fig:Tb1d}). From the outer disk to the inner disk, we first find the outer peak (the first local maximum, which corresponds to where dust piles up due to the dust trapping) and mark this point as $r_{peak}$, and then find the bottom of the gap (local minimum) inside $r_{peak}$ and mark it as $r_{gap}$. $r_{gap}$ is not necessarily $r_p$. As demonstrated in Figure \ref{fig:gap_def}, the gap can have the deepest point further out than $r_p$. This is because some gaps have significant horseshoe material in between. In some extreme cases with very shallow gaps, only the outer portion of the gap that is outside the horseshoe region is visible (e.g. the top middle panel in Figure \ref{fig:Tb1d}). We define the gap depth $\delta$ as
\begin{equation}
    \delta_\Sigma = \Sigma(r_{peak}) / \Sigma(r_{gap})\,,
\end{equation}
for the gas surface density profiles, and
\begin{equation}
    \delta_I = I_{mm}(r_{peak}) / I_{mm}(r_{gap})\,,
\end{equation}
for the dust mm intensity profiles.

(2) Measuring the gap width ($\Delta$) for these profiles. To calculate the width, we first define the edge quantities as the average between the peak and gap surface densities (for gas) or the mm intensities (for intensity maps):
\begin{equation}
    \Sigma_{edge} = \frac{\Sigma(r_{peak}) + \Sigma(r_{gap})}{2}\,,
\end{equation}
and
\begin{equation}
    I_{edge} = \frac{I_{mm}(r_{peak}) + I_{mm}(r_{gap})}{2} \,.
\end{equation}

Then, we find one edge $r_{in}$ at the inner disk and the other $r_{out}$ at the outer disk, where $\Sigma$($r_{in}$) = $\Sigma$($r_{out}$) = $\Sigma_{edge}$ for the gas surface density or $I(r_{in})$ = $I(r_{out})$ = $I_{edge}$ for the dust intensity (Figure \ref{fig:gap_def}). Thus, we define the gap width $\Delta$ for either the gas surface density or the dust intensity as
\begin{equation}
    \Delta = (r_{out} - r_{in}) / r_{out}\,.
\end{equation}

\begin{figure*}[t!]
\includegraphics[width=\linewidth]{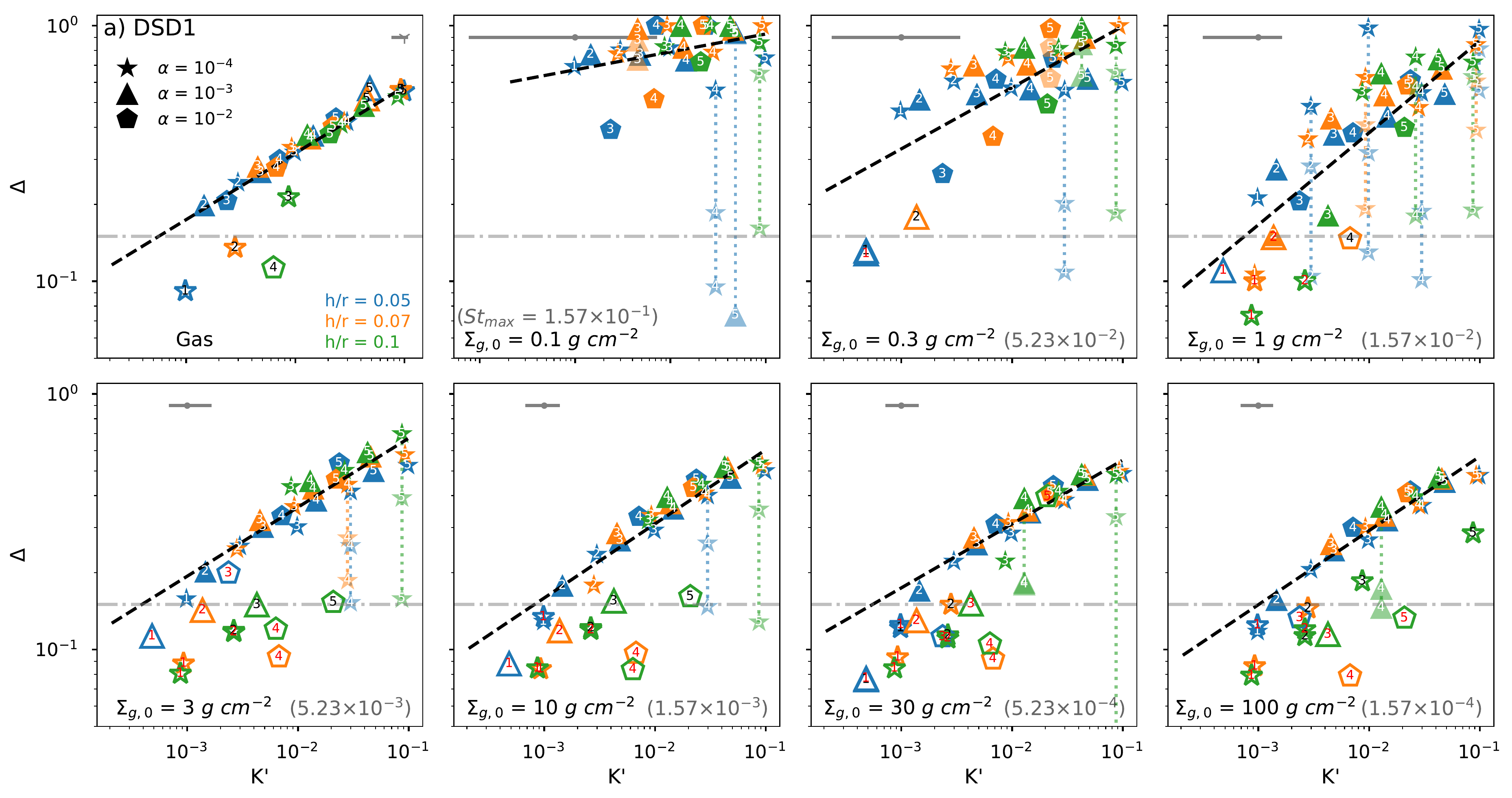}\\
\includegraphics[width=\linewidth]{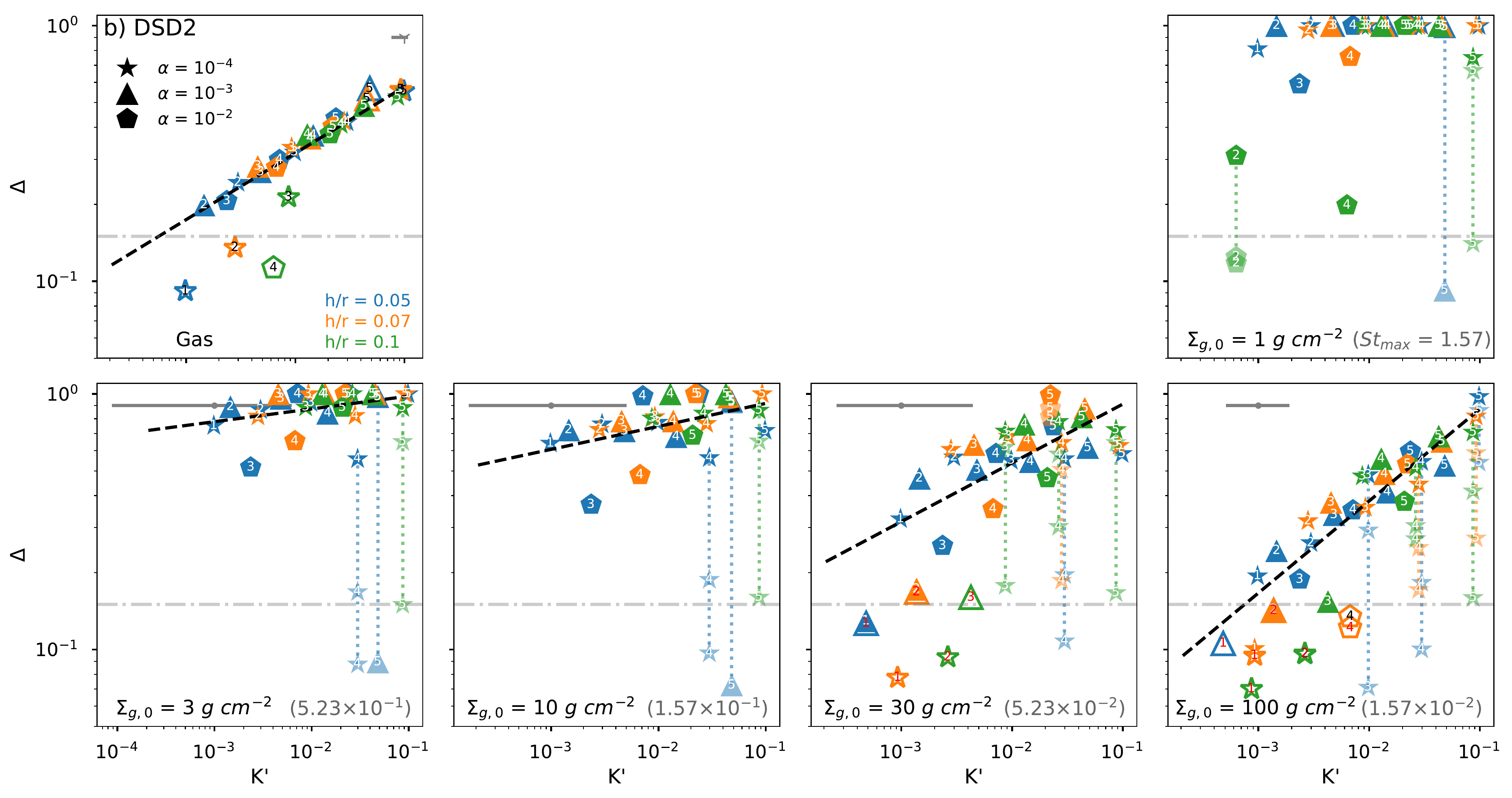}
\figcaption{Fitting of gap widths $\Delta$ vs. $K'$ for different models with dust size distribution $\{s_{max},\ p\}$ = $\{0.1\ mm, -3.5\}$ (panel a) and $\{s_{max},\ p\}$ = $\{1\ cm, -2.5\}$ (panel b). The first panel is the fitting of the gas surface density, which is used to calibrate the index above $h/r$ and $\alpha$. The best fit is $K'=q(h/r)^{-0.18}\alpha^{-0.31}$. The stars, triangles, and pentagons represent models of $\alpha$ = $10^{-4}$, $10^{-3}$, and $10^{-2}$, respectively. Models for $h/r$ = 0.05, 0.07 and 0.1 are in blue, orange, and green respectively. The label 1, 2, 3, 4 and 5 within symbols represent the planet mass from 10 $M_{\earth}$ to 3 $M_J$ increasingly. The rest of panels are fits of gaps in dust intensity profiles. From left to right and top to bottom, they are models scaled to the initial gas density $\Sigma_{g,0}$ = $0.1\, \mathrm{g\,cm^{-2}}$, $0.3 \, \mathrm{g\,cm^{-2}}$, $1 \, \mathrm{g\,cm^{-2}}$, $3 \, \mathrm{g\,cm^{-2}}$, $10 \, \mathrm{g\,cm^{-2}}$, $30 \, \mathrm{g\,cm^{-2}}$, $100 \, \mathrm{g\,cm^{-2}}$.  The best fits using Equation \ref{eq:k_width} are plotted as the dashed lines and the constants $A$ and $B$ are shown in Table \ref{table:relationwidth}.
We  neglect outliers (shown in unfilled markers) when fitting the line. The outliers either have very shallow gaps, or have double gaps (horseshoe in between), thus have widths smaller than their counterparts. For cases which clearly show that the major gap is split into two by the horseshoe region, the widths of the two individual gaps around the horseshoe are also presented and they are connected to the main gap width with the vertical dotted line. The open symbols with red numbers in them are derived from images which are convolved with a smaller beam of $\sigma=0.025 r_p$. The grey errorbar on top of each plot shows the uncertainty of the fitting.
\label{fig:gap_width_fit}}
\end{figure*}

 Figure \ref{fig:gap_width_fit} shows $\Delta$ for all $\Sigma_{g,0}$ cases with DSD1 (panel a) and DSD2 (panel b) dust distributions.
If there is some horseshoe material around $r$=$r_p$ separating the main gap into two gaps, the horizontal $\Sigma_{edge}$ or $I_{edge}$ line will cross through the horseshoe and we treat two individual gaps as a single one (i.e., the $r_{in}$ is taken to be the $r_{in}$ of the inner gap and $r_{out}$ is taken to be the $r_{out}$ of the outer gap), but the individual gaps on either side of the horseshoe region are also plotted in Figure \ref{fig:gap_width_fit}  as fainter makers and they are connected to the main gap width using dotted lines. 

Note that our definition of gap width is more convenient to use than that in \cite{kanagawa_width}, because the width here is normalized by $r_{out}$ instead of $r_p$ as in \cite{kanagawa_width}.
In actual observations, we do not have the knowledge of the planet position $r_p$ within the gap. Another difference between our defined gap width and the one used in \cite{kanagawa_width} is that we use $(\Sigma(r_{peak}) + \Sigma(r_{gap}))/2$ to define $\Sigma_{edge}$ while \cite{kanagawa_width} use $\Sigma_0/2$ to define the gap edge. Our definition enables us to study shallow gaps that are shallower than $\Sigma_0/2$.

\begin{deluxetable*}{ccccccccc}
\tabletypesize{\scriptsize}
\tablecaption{The relation between the gap width $\Delta$ and $K'$ \label{table:relationwidth}}
\tablehead{
\colhead{Parameters} &
\colhead{$\Delta_g$} & \colhead{$\Delta_{d,0p1}$} & \colhead{$\Delta_{d,0p3}$} & 
\colhead{$\Delta_{d,1}$} & \colhead{$\Delta_{d,3}$} &
\colhead{$\Delta_{d,10}$} & 
\colhead{$\Delta_{d,30}$} & \colhead{$\Delta_{d,100}$}
}
\colnumbers
\startdata
$St_{max}$ (DSD1) & -- & 1.57 $\times$ $10^{-1}$ & 5.23 $\times$ $10^{-2}$ & 1.57 $\times$ $10^{-2}$ 
& 5.23 $\times$ $10^{-3}$ & 1.57 $\times$ $10^{-3}$ & 5.23 $\times$ $10^{-4}$ & 1.57 $\times$ $10^{-4}$ \\ 
$A$ $\{p=-3.5, s_{max}=0.1 mm\}$ & 1.05 & 1.09 & 1.73 & 2.00 & 1.25 & 1.18 & 0.98 & 1.11 \\
$B$ $\{p=-3.5, s_{max}=0.1 mm\}$ & 0.26 & 0.07 & 0.24 & 0.36 & 0.27 & 0.29 & 0.25 & 0.29 \\
Uncertainty in $log_{10}(K')$  & $^{+0.03}_{-0.12}$  & $^{+0.86}_{-1.12}$ & $^{+0.53}_{-0.63}$ & $^{+0.21}_{-0.50}$ & $^{+0.22}_{-0.16}$ & $^{+0.14}_{-0.17}$ & $^{+ 0.16}_{-0.14}$ & $^{+0.13}_{-0.16}$ \\
\hline
$St_{max}$ (DSD2) & -- & -- & -- & 1.57
& 5.23 $\times$ $10^{-1}$ & 1.57 $\times$ $10^{-1}$ & 5.23 $\times$ $10^{-2}$ & 1.57 $\times$ $10^{-2}$ \\ 
$A$ $\{p=-2.5, s_{max}=1 cm\}$ & -- & -- & -- & -- & 1.10 & 1.13 & 1.55 & 2.00 \\
$B$ $\{p=-2.5, s_{max}=1 cm\}$ & -- & -- & -- & -- & 0.05 & 0.09 & 0.23 & 0.36 \\ 
Uncertainty in $log_{10}(K')$  & --  & -- & -- & -- & $^{+0.80}_{-1.06}$ & $^{+0.70}_{-0.77}$ & $^{+0.65}_{-0.59}$ & $^{+0.28}_{-0.29}$ \\
\enddata
\tablecomments{$\Delta$ = $AK'^{B}$, where $A$, $B$ are fitting parameters here. $K'=q(h/r)^{-0.18}\alpha^{-0.31}$.}
\end{deluxetable*}

(3) Fitting the width ($\Delta$)-$K'$ relation.
We first use the width $\Delta$ measured from the gas surface density profiles to find the optimal degeneracy parameter
$K'$ following the same procedure as in Equation \ref{eq:Kdef}.
Similarly, a least squares fitting was done to minimize the sum of the square difference of the vertical distance between the points and the linear-regression line log($\Delta$) vs. log(K'). 
With this procedure, we derive that the optimal $K'$ is 
\begin{equation}
    \frac{K'}{0.014} = \frac{q}{0.001}\Big(\frac{h/r}{0.07}\Big)^{-0.18}\Big(\frac{\alpha}{10^{-3}}\Big)^{-0.31}\,.
    \label{eq:k_width}
\end{equation}
With this definition of $K'$, the best fitting relationships ( $\Delta=AK'^{B}$ ) are found for each initial gas densities with two dust size distributions DSD1 and DSD2. 
The resulting $A$ and $B$ for these fits are listed in Table \ref{table:relationwidth}. Note that our definition of $K'$ is equivalent to the square root of $K'$ defined in \cite{kanagawa_width}. Compared with the fitting formula for the gas surface density in \cite{kanagawa_width}, our $K'$ is less sensitive to $h/r$ and the gaseous gap width is less sensitive to $q$. We confirm that this is largely due to our different definition of the gap width (compared with their definition, our normalized gap width is smaller for wide gaps and larger for shallow gaps that are normally narrow.). 

Figure \ref{fig:gap_width_fit} shows the fits for all the cases with DSD1 (panel a) and DSD2  (panel b) dust size distributions. We can see that uncertainties of these fittings become large when $\Delta\lesssim 0.15$. Thus, our fitting procedure  does not involve widths that are smaller than 0.15. For these narrow gaps whose widths are smaller than 0.15 (labeled as the open symbols with back numbers in them), their gap profiles start to be affected by the smoothing kernel with $\sigma=0.06 r_p$. Thus in Figure \ref{fig:gap_width_fit}, we also plot the widths measured from the profiles that are convolved with a $\sigma$ = 0.025 $r_p$ kernel. These widths are plotted as open symbols with red numbers in them. 


\begin{figure*}[t!]
\includegraphics[width=\linewidth]{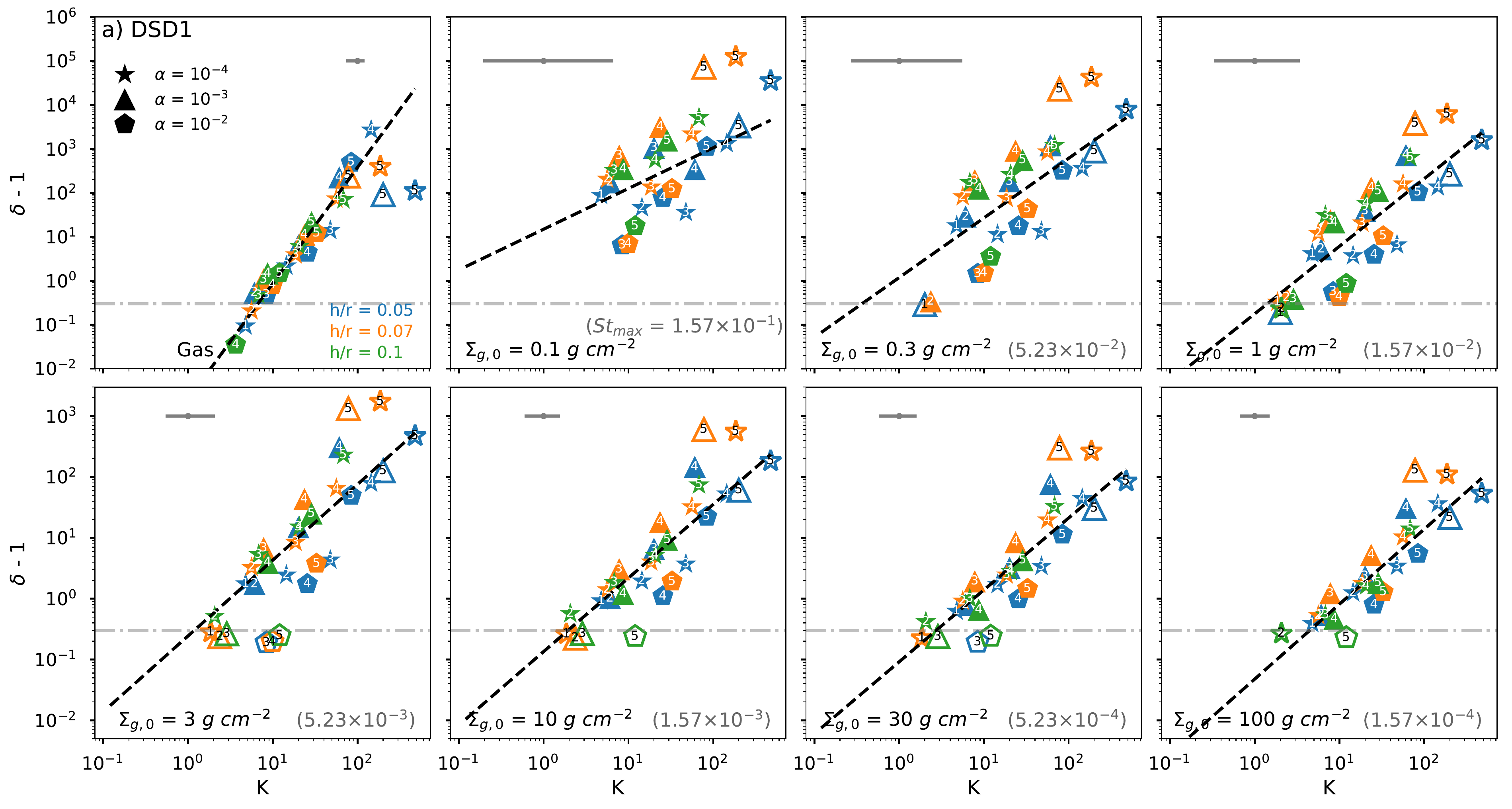}\\
\includegraphics[width=\linewidth]{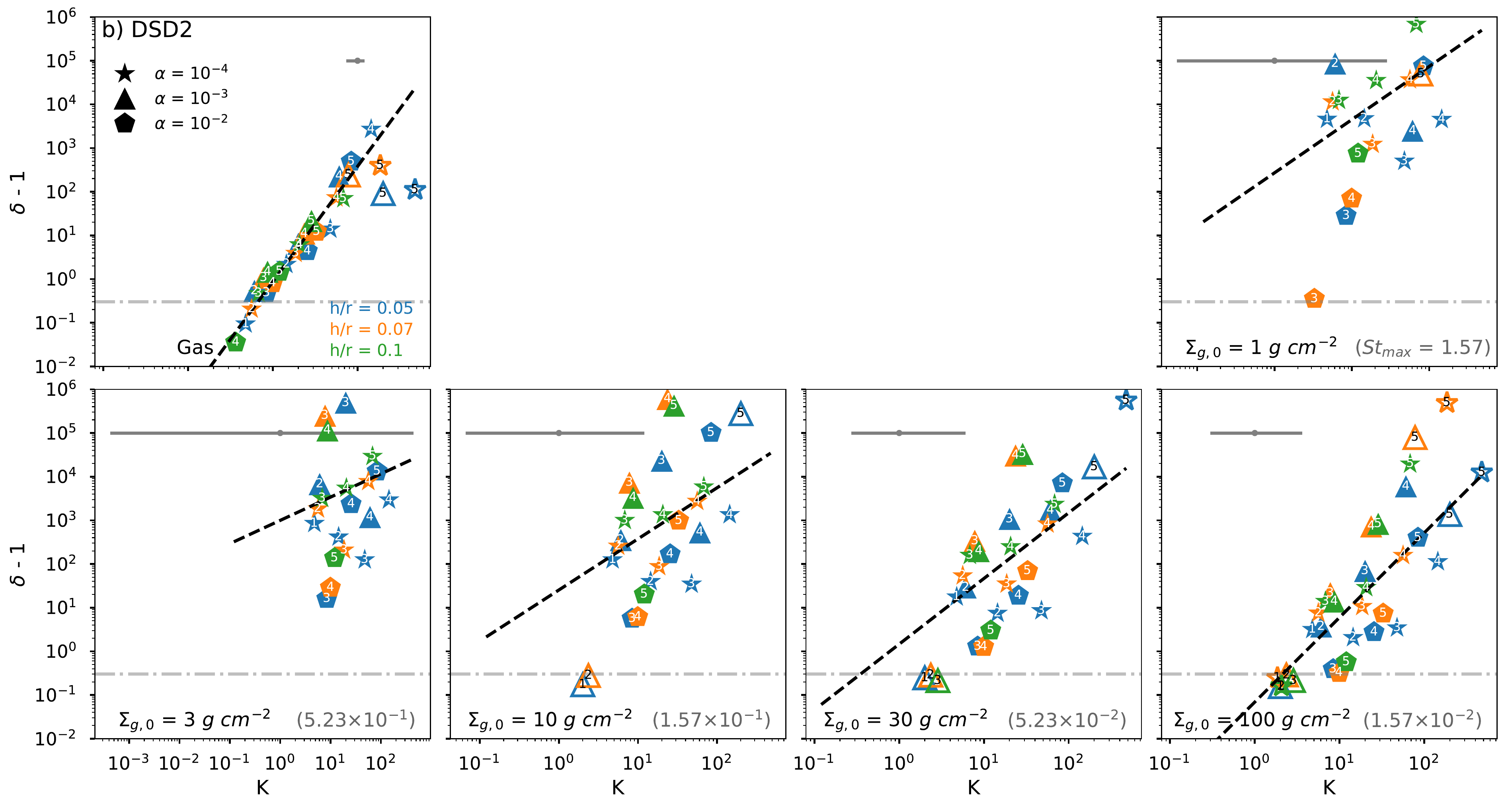}
\figcaption{ Similar to Figure \ref{fig:gap_width_fit} but for fits of the gap depths minus one ($\delta$ - 1) vs. $K$. The panel a) adopts the dust size distribution of DSD1 $\{s_{max},\ p\}$ = $\{0.1\ mm, -3.5\}$ while the panel b) adopts DSD2 $\{s_{max},\ p\}$ = $\{1\ cm, -2.5\}$. The best-fit parameters are listed in Table \ref{table:relationdepth}. 
\label{fig:gap_depth_fit}}
\end{figure*}

\begin{deluxetable*}{ccccccccc}
\tabletypesize{\scriptsize}
\tablecaption{The relation between $\delta$ -1  and $K$, where $\delta$ is the gap depth. \label{table:relationdepth}}
\tablehead{
\colhead{Parameters} &
\colhead{$\delta_g$ - 1} & \colhead{$\delta_{d,0p1}$ - 1} & \colhead{$\delta_{d,0p3}$ - 1} & 
\colhead{$\delta_{d,1}$ - 1} & \colhead{$\delta_{d,3}$ - 1} &
\colhead{$\delta_{d,10}$ - 1} & 
\colhead{$\delta_{d,30}$ - 1} & \colhead{$\delta_{d,100}$ - 1}
}
\colnumbers
\startdata
$St_{max}$ (DSD1) & -- & 1.57 $\times$ $10^{-1}$ & 5.23 $\times$ $10^{-2}$ & 1.57 $\times$ $10^{-2}$ 
& 5.23 $\times$ $10^{-3}$ & 1.57 $\times$ $10^{-3}$ & 5.23 $\times$ $10^{-4}$ & 1.57 $\times$ $10^{-4}$ \\ 
C $\{$3.5, 0.1 mm$\}$ & 0.002 & 14.9 & 1.18 & 0.178 & 0.244 & 0.135 & 0.0917 & 0.0478  \\ 
D $\{$3.5, 0.1 mm$\}$ & 2.64  & 0.926 & 1.36 & 1.54 & 1.25 & 1.21 & 1.18 & 1.23 \\
Uncertainty in $log_{10}(K)$  & $^{+0.08}_{-0.13}$  & $^{+0.82}_{-0.71}$ & $^{+0.74}_{-0.57}$ & $^{+0.53}_{-0.48}$ & $^{+0.32}_{-0.26}$ & $^{+0.19}_{-0.23}$ & $^{+1.20}_{-0.24}$ & $^{+0.18}_{-0.18}$ \\
\hline
$St_{max}$ (DSD2) & -- & -- & -- & 1.57
& 5.23 $\times$ $10^{-1}$ & 1.57 $\times$ $10^{-1}$ & 5.23 $\times$ $10^{-2}$ & 1.57 $\times$ $10^{-2}$ \\ 
C $\{$2.5, 1 cm$\}$ & -- & -- & -- & 271 & 998 & 25.5 & 1.46 & 0.069 \\ 
D $\{$2.5, 1 cm$\}$ & -- & -- & -- & 1.22 & 0.533 & 1.17 & 1.50 & 1.94 \\
Uncertainty in $log_{10}(K)$  & --  & -- & -- & $^{+1.45}_{-1.26}$ & $^{+2.64}_{-3.38}$ & $^{+1.08}_{-1.18}$ & $^{+0.78}_{-0.56}$ & $^{+0.56}_{-0.52}$ \\
\enddata
\tablecomments{$\delta$-1 = $CK^{D}$, where $C$ and $D$ are fitting parameters here. $K = q(h/r)^{-2.81}\alpha^{-0.38}$.}
\end{deluxetable*}


(4) Fitting the depth ($\delta$)-$K$ relation. 
We adopt the same procedure to fit the depth-$K$ as the width-$K'$ aforementioned. 
Since no-gap is equivalent to $\delta$=1, we try to find the optimal degeneracy parameter $K$ by
a least squares fitting for log($\delta$ - 1) vs. log($K$), 
\begin{equation}
    \delta - 1 = CK^D\,,
    \label{eq:depth_cd}
\end{equation}
for various $K$.
The optimal $K$ is fitted to be
\begin{equation}
    \frac{K}{24} = \frac{q}{0.001}\Big(\frac{h/r}{0.07}\Big)^{-2.81}\Big(\frac{\alpha}{10^{-3}}\Big)^{-0.38}\,.
    \label{eq:k_depth}
\end{equation}

After $K$ is fixed, we use Equation \ref{eq:depth_cd} to fit the relationship between $\delta$ - 1 and $K$ for the dust intensity profiles from different $\Sigma_{g,0}$ with DSD1 and DSD2. $C$ and $D$ are found using linear regression. The resulting $C$ and $D$ in different $\Sigma_{g,0}$ cases with either DSD1 or DSD2 are listed in Table \ref{table:relationdepth}. Figure \ref{fig:gap_depth_fit} show $\delta$ - 1 for all $\Sigma_{g,0}$ cases with DSD1 and DSD2. The best fits are also plotted for each panel. Note that open symbols are not involved in the fitting since these gaps are eccentric and their depths do not follow the trend for other gaps. Clearly, with the Stokes number increasing, the fitting becomes worse. This is expected since particles with larger Stokes numbers drift faster and  the gap profile becomes more irregular.

(5) The uncertainty of the fittings.
We apply the same measure to calculate the uncertainty of  the gap width/depth fitting as that of $\Delta\delta v_{rot}$-$K_{v_r}$ relation mentioned in \S\ref{sec:gaskinematics}. That is, we measure the horizontal offset (in $log_{10}(K')$ or $log_{10}(K)$) between each point and the fitting line at each sets of dust configurations and also the gas surface density. From the distribution of the offset, the left side error is estimated by the 15.9 percentile of the distribution and the right side error is 84.1 percentile of the distribution. These uncertainties are summarized in Table \ref{table:relationwidth} and \ref{table:relationdepth} and marked in grey color at the top of each panel in Figure \ref{fig:gap_width_fit} and \ref{fig:gap_depth_fit}. For widths that are larger than 0.15, the uncertainties for the fittings are less than a factor of two for $K'$ (or $q$) when $St\lesssim5\times10^{-3}$ and around a factor of three for $K'$ (or $q$) when $5\times10^{-3}<St\lesssim5\times10^{-2}$. When $St\gtrsim10^{-1}$, particles drift to the central star quickly and most of the gaps only have a single ring left at the outer disk so that $\Delta\sim$1 and the uncertainties for $K'$ at a given $\Delta$ is very large. For these cases, we cannot use the gap width to estimate the planet mass.

\begin{figure}[t]
\includegraphics[width=\linewidth]{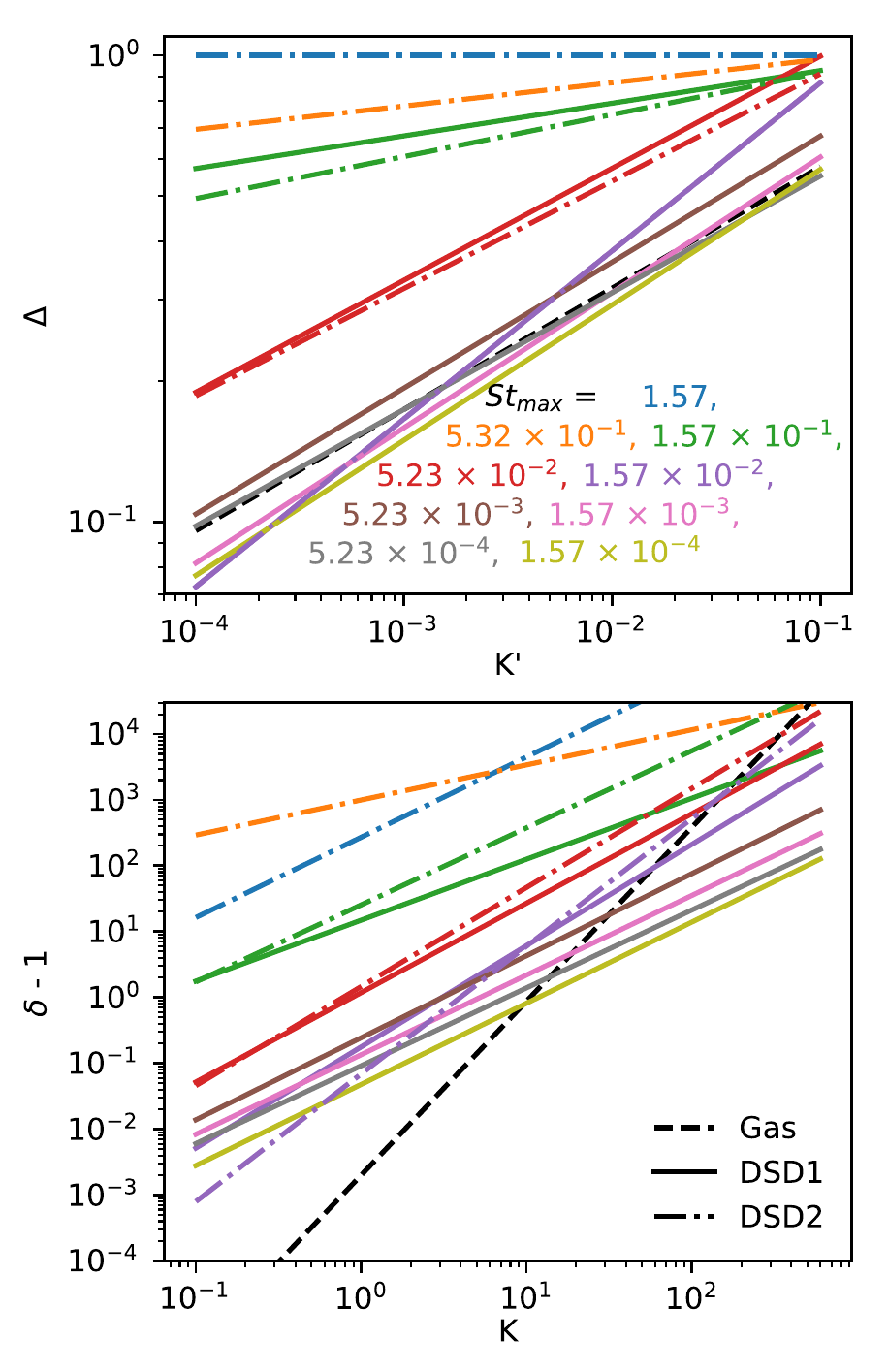}
\caption{\label{fig:alllines} Upper panel: $\Delta$-$K'$. Lower panel: ($\delta$-1) - $K$. The fits for the gas surface density are shown as the black dashed lines. The fits for the dust continuum intensity are shown as the solid lines for DSD1 ($\{s_{max},\ p\}$ = $\{0.1\ mm, -3.5\}$), and the dashed-dotted lines for DSD2 ($\{1\ cm, -2.5\}$). Maximum Stokes numbers ($St_{max}$) under $\Sigma_{g,0}$ (DSD1, DSD2) are 1.57 (--, 1 $\mathrm{g\,cm^{-2}}$), 5.32 $\times$ $10^{-1}$ (--, 3 $\mathrm{g\,cm^{-2}}$), 1.57 $\times$ $10^{-1}$ (10 $\mathrm{g\,cm^{-2}}$, 0.1 $\mathrm{g\,cm^{-2}}$), 5.23 $\times$ $10^{-2}$ (30 $\mathrm{g\,cm^{-2}}$, 0.3 $\mathrm{g\,cm^{-2}}$), 1.57 $\times$ $10^{-2}$ (100 $\mathrm{g\,cm^{-2}}$, 1 $\mathrm{g\,cm^{-2}}$), 5.23 $\times$ $10^{-3}$ (3 $\mathrm{g\,cm^{-2}}$), 1.57 $\times$ $10^{-3}$ (10 $\mathrm{g\,cm^{-2}}$, --), 5.23 $\times$ $10^{-4}$ (30 $\mathrm{g\,cm^{-2}}$, --), 1.57 $\times$ $10^{-4}$ (100 $\mathrm{g\,cm^{-2}}$, --).} 
\end{figure}

Finally, we summarize all the fits for the width and depth in Figure \ref{fig:alllines}. 
In the Appendix, we provide gap depth $\delta$ and width $\Delta$ of our whole grid of models.
In spite of the dramatically different dust size distributions between DSD1 and DSD2, the fits for DSD1 are quite close to fits for DSD2  as long as the Stokes number for the maximum-size particles is the same (e.g. red solid and dot-dashed lines). This is reasonable since only the Stokes number matters for the dust dynamics, and DSD1 have a similar opacity as DSD2. For 1 mm observations, the opacity is roughly a constant when $s_{max}\lesssim$ 1 cm (the opacity is slightly higher when $s_{max}\sim$ 1mm, see \citealt{birnstiel18}). Thus, different disks with different surface densities ($\Sigma_{g,0}$) and different dust size distributions have the same intensity profiles as long as their Stokes numbers for maximum-size particles (where most of the dust mass is) are the same and $s_{max}\lesssim$ 1 cm. Thus,
our derived relationships can be used in other disks with different surface densities and dust size distributions as long as the Stokes number of the maximum-size particles is in our simulated range (1.57$\times$10$^{-4}$ to 1.57). For disks with Stokes number smaller than 1.57$\times$10$^{-4}$, their gap profiles should be similar to the disks with $St$=1.57$\times$10$^{-4}$ since dust is well coupled to the gas.

\subsubsection{Secondary Gaps/Rings \label{sec:secondarygaps}}


\begin{figure*}[t!]
\includegraphics[width=\linewidth]{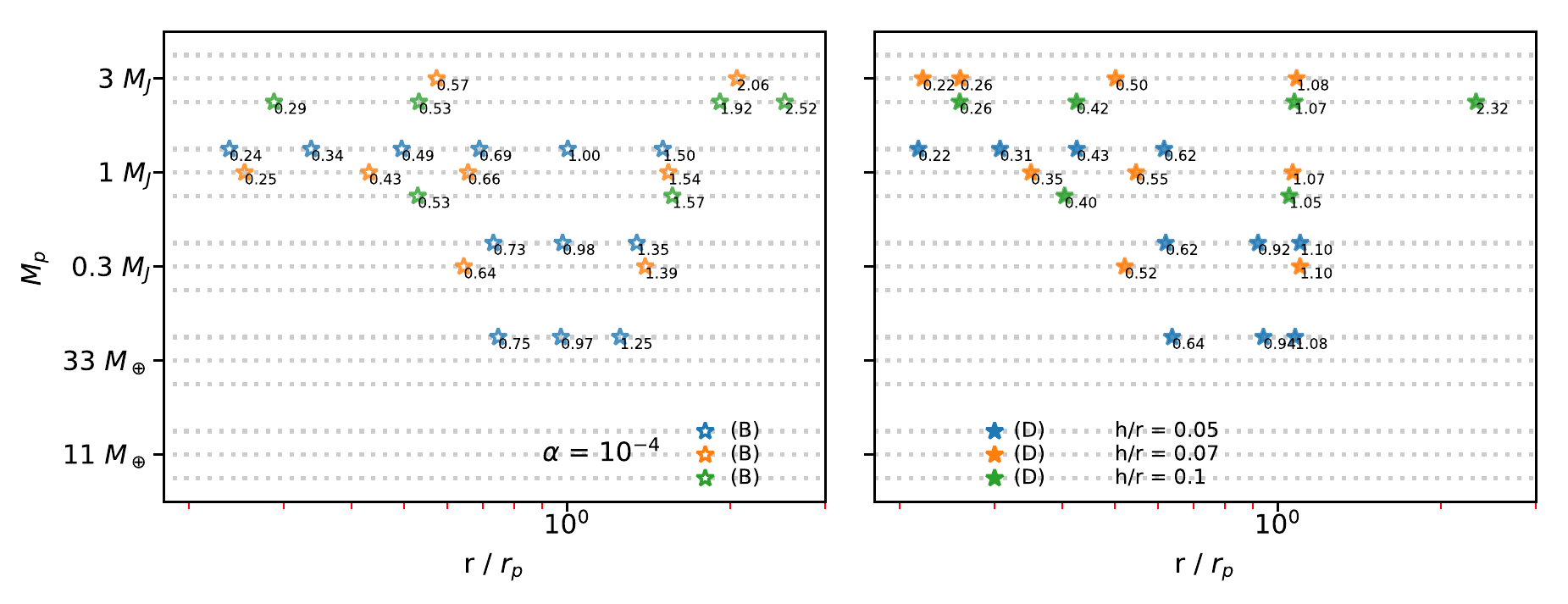}\figcaption{Position of gaseous rings (left panels, B: Bright ring) and gaps (right panels, D: Dark annulus) for simulations having $\alpha$ = $10^{-4}$. Note that, in the right panel, two cases with
$h/r$=0.05 have two minima around $r=r_p$ because the horseshoe region splits the primary gap into two smaller gaps.
\label{fig:gap_position}}
\end{figure*}

\begin{figure}[t!]
\includegraphics[width=\linewidth]{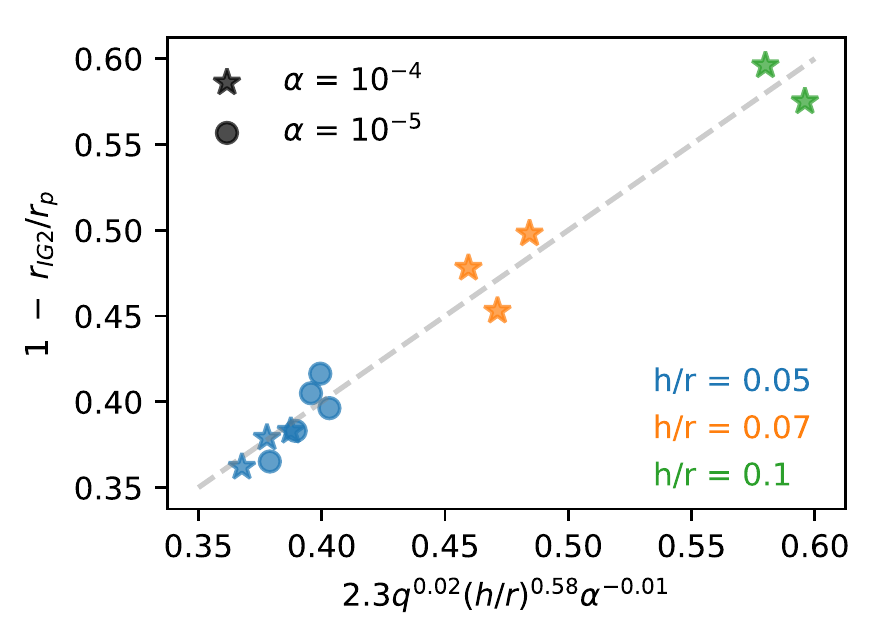}\figcaption{The fit of the position of secondary gaps as a function of $q$, $h/r$ and $\alpha$.
\label{fig:gap_position_fit}}
\end{figure}

Previous simulations have shown that a planet can introduce many gaps/rings in disks having very low viscosities \citep{Zhu14,dong2017,bae2017}.
These gaps can be grouped into two categories: 1) two gaps adjacent to the planet that are separated by the horseshoe material (e.g. two troughs at 0.9 $r_p$ and 1.1 $r_p$ in Figure \ref{fig:gap_def}, also mentioned in \S \ref{sec:gasdensity}), and
2) secondary shallower gaps much further away into the inner and outer disks (e.g. the gap at 0.6 $r_p$ in Figure \ref{fig:gap_def}).
The two gaps in the first category form because: a) the spiral waves, especially excited by low mass planets, 
need to propagate in the radial direction for some distance to steepen into spiral shocks and induce gaps \citep{goodman2001}, b) the horseshoe material
has a slow relative motion with respect to the spiral shocks thus this material takes a long time to be depleted. Eventually, these two gaps
may merge into one single main gap, which is studied in \S \ref{sec:gapring}. The gaps in the second category are induced by additional spiral
arms from wave interference \citep{bae2018a}. Instead of disappearing, these gaps will become deeper with time in inviscid disks. Thus, they are useful 
to constrain the planet and disk properties \citep{bae2018b}. 

We  label the positions of all these additional gaps and rings in Figure \ref{fig:gap_position}. We find that the positions of these rings and gaps in dust intensity radial profiles are similar to those in gas surface density profiles. Thus, we plot the positions based on the gas density profiles. It turns out that only disks
with $\alpha\leq 10^{-4}$ can form noticeable multiple gaps. Thus, if we find a system with multiple gaps induced by a single planet (e.g. AS 209 in the next section), the disk viscosity has to be small. From Figure \ref{fig:gap_position}, we can see that distance between the secondary gap and the main gap mainly depends on the disk scale height ($h$).

For the secondary gap at $\sim 0.5-0.7$, following our fitting procedure before, we find that the position of the secondary gap ($r_{IG2}$) and $r_p$ is best fitted with
\begin{equation}
    1 - \frac{r_{IG2}}{r_p} = 2.3\ q^{0.02} (h/r)^{0.58} \alpha^{-0.01}\,.
    \label{eq:secondary_gap_fit}
\end{equation}
This clearly shows that the position of the secondary gap  is almost solely determined by the disk scale height. Thus,
if the secondary gap is present, we can use its position to estimate the disk scale height ($h/r$). The fitting is given in Figure \ref{fig:gap_position_fit}. The $\alpha=10^{-5}$ cases are the AS 209 cases which will be discussed in the next section. We caution that the fitting has some scatter. Within each $h/r$ group in Figure \ref{fig:gap_position_fit}, the $r_{IG2}/r_p$ depends on the planet mass. But this dependence seems to be different for different $h/r$ groups, so that the fitting using all $h/r$ suggests a weak dependence on the planet mass.
We also note that our fit is different from the recent fit by \cite{Dong2018b} which has a $q^{-0.2}(h/r)^{1.3}$ dependence (note that their planet mass is normalized by the thermal mass).
The difference may be due to: 1) The disks in \cite{Dong2018b} are thinner, where their main set of simulations uses $h/r$=0.03, 2) \cite{Dong2018b} fit the gap positions at different times for different simulations while we fit the gap positions at the same time in the simulations.

\section{Planet Properties \label{sec:flowchart}}
\begin{figure}[t!]
\includegraphics[width=\linewidth]{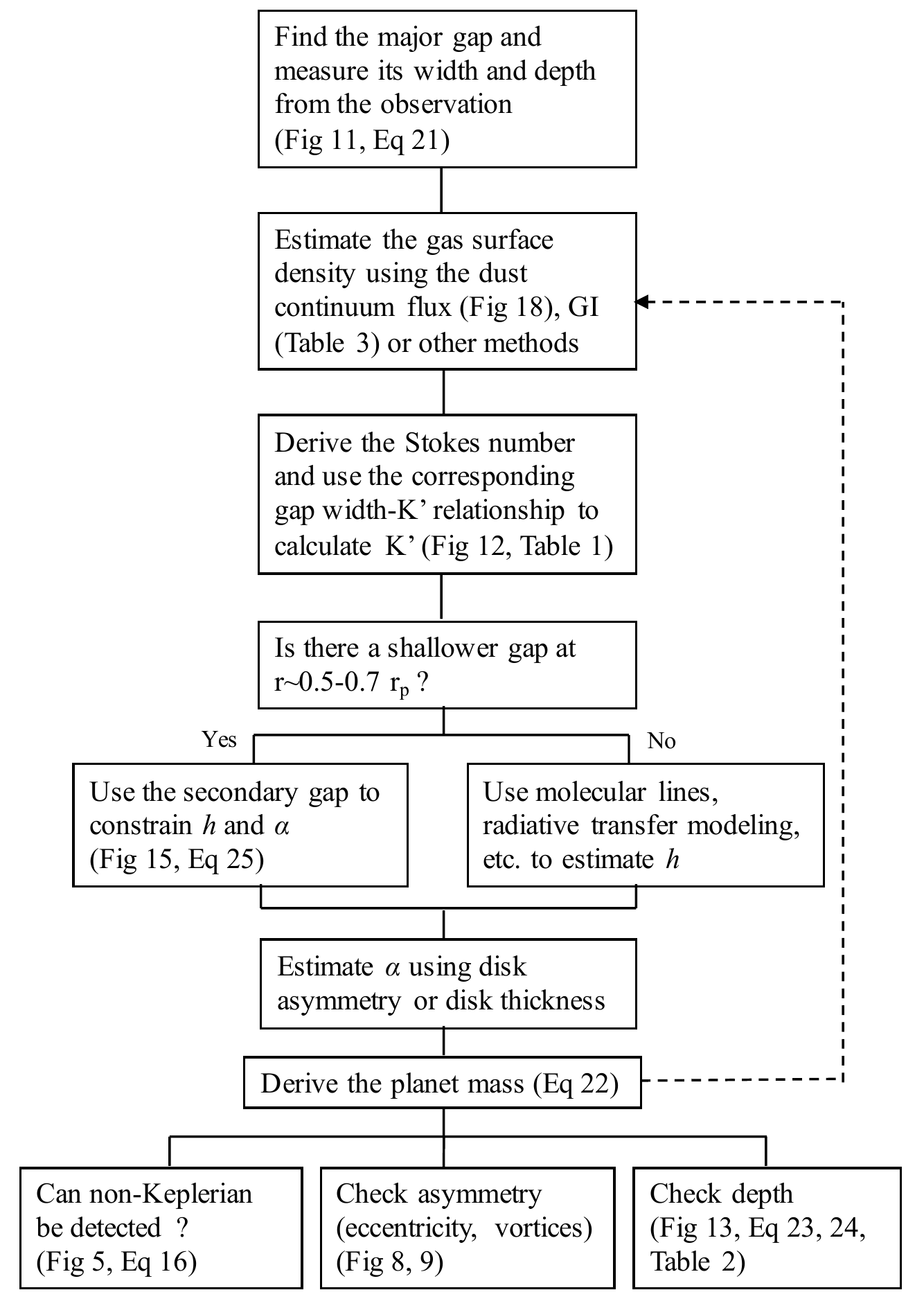}
\figcaption{The flow chart to derive the planet mass. 
\label{fig:flowchart}}
\end{figure}

With all the relationships derived in previous sections regarding the planet mass and gap profiles, we can now put them together to constrain the mass of potential planets in the DSHARP disks. We use the measured radial intensity profiles from Figure 2 in \citet{huang18b}. These profiles are derived by deprojecting the observed images to the face-on view and then averaging the intensity in the azimuthal direction. Details regarding generating the radial intensity profiles are given in \citet{huang18b}. By using these intensity profiles, we can derive the planet mass following
the flowchart given in Figure \ref{fig:flowchart}.

First, for each source, we plot the observed radial intensity profile and 
identify gaps that have $\Delta$ $\geq$ 0.15.  As shown in Figure \ref{fig:gap_width_fit}, $\Delta$ $\lesssim$ 0.15 have large scatter and are sensitive to the size of the convolution beam. By examining the surface density profiles in detail, we find that such narrow gaps are also very shallow and they are actually the outer one of the double gaps around the horseshoe region. Since these gaps are very shallow, the inner one does not cause enough disk surface density change to be identified as a gap. 
Thus, for narrow gaps with $\Delta$ $\lesssim$ 0.15, we do not use the fitting formula to derive the planet mass. Instead, we try to directly match the gap $\Delta$ with data points in Figure \ref{fig:gap_width_fit} by eye to get a rough planet mass estimate. For these narrow gaps, the size of the convolution beam matters. Thus, if the gap is at 10s of au, we use the widths derived in images with the $\sigma=0.06 r_p$ beam, and if the gap is at $\sim$100 au we use the widths derived in images with the $\sigma=0.025 r_p$ beam.

\begin{figure*}[t!]
\includegraphics[width=\linewidth]{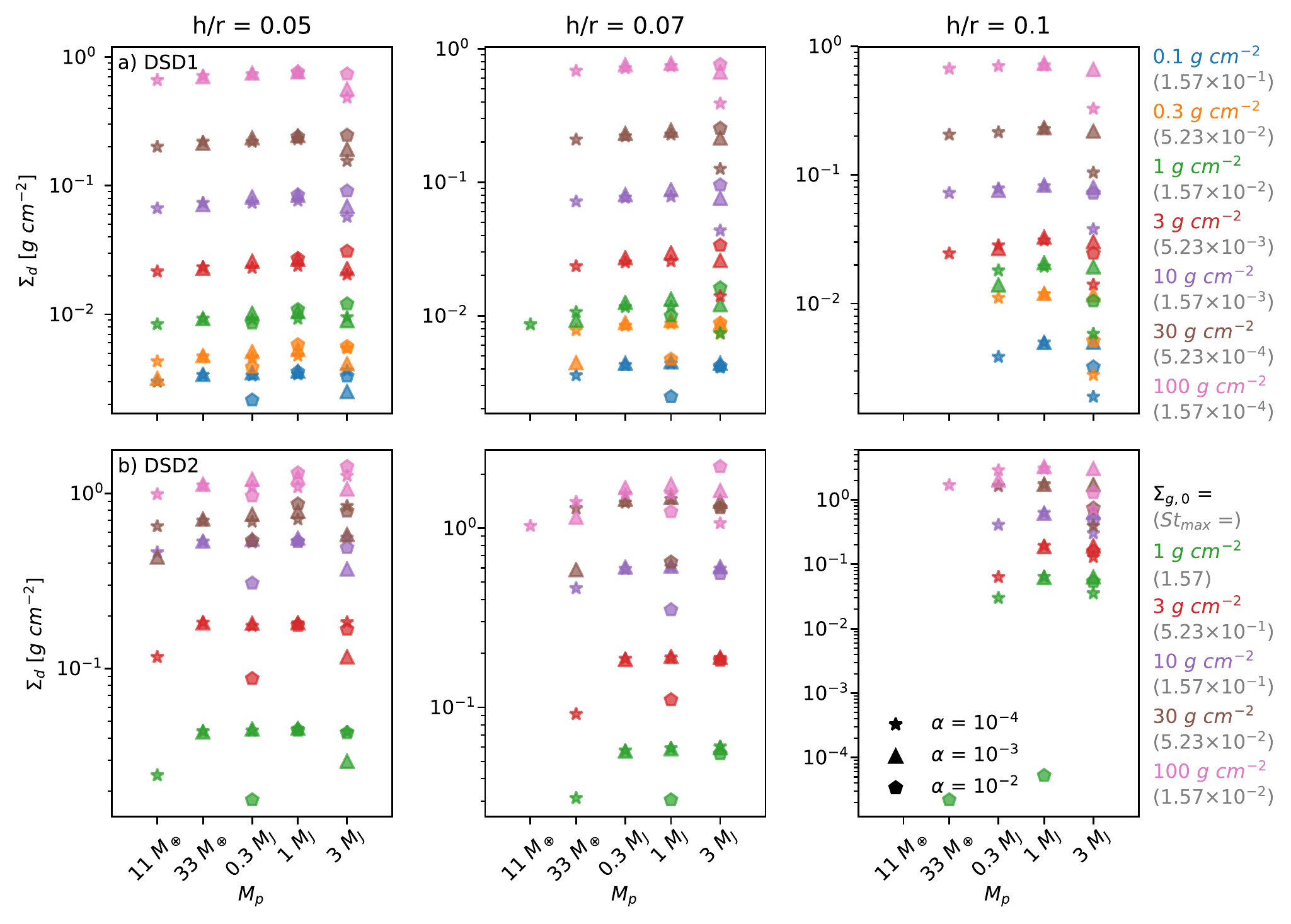}\figcaption{The averaged dust surface density at the outer disk, integrated from 1.1 $r_p$ to 2 $r_p$, for all the models with DSD1 (upper panels) and DSD2 (lower panels).
\label{fig:gasdustratio}}
\end{figure*}

Second, we estimate the gas surface density, using the observed mm flux at the outer disk and/or some other constraints. 
We integrate the observed intensity from  1.1 $r_{gap}$
to 2 $r_{gap}$ where $r_{gap}$ is the gap center. 
Using $T_{d}$ derived by Equation \ref{eq:disktemp} and the dust opacity of 0.43 $\mathrm{cm^{2}\,g^{-1}}$ (\S \ref{sec:scaling}),  we calculate the averaged dust surface density ($\Sigma_d$) from 1.1 to 2 $r_{gap}$. We have done the same exercise for all our simulations, and
Figure \ref{fig:gasdustratio} shows the relationship between $\Sigma_{g,0}$ and the averaged $\Sigma_d$ at the outer disk
for the simulations. Figure \ref{fig:gasdustratio} indicates that, with a smaller gas surface density or larger particles (higher Stokes numbers), the ratio between $\Sigma_d$ and $\Sigma_{g,0}$ increases because particles with larger Stokes numbers are more easily trapped at the gap edges. 
We can then use Figure \ref{fig:gasdustratio} to 
estimate $\Sigma_{g,0}$ based on 
the derived $\Sigma_{d}$ from the observation, the estimated $h/r$, and the assumed $\alpha$ and planet mass. After we  derive the planet mass, we will go back to this step to see if the derived planet mass is consistent with our assumed mass. Otherwise, we iterate these processes again with the new assumed planet mass. On the other hand, this estimate is prone to large errors. If we have more ways to estimate the gas surface density, such as using molecular tracers or constraints from the gravitational instability, we should adopt these constraints.

Third, with known $\Sigma_{g,0}$ and the assumed dust size distribution, we can calculate 
$St_{max}$ and use the $\Delta$-$K'$ relationship (\S \ref{sec:gapring} and Table \ref{table:relationwidth}) to derive the $K'$ parameter. Given the sensitivity limits of ALMA, we decide not to use the gap depth ($\delta$) to estimate the $K$ parameter. For example, two gaps with different depths, one being a factor of 10$^5$ deep and the other being a factor of 10$^3$ deep,
can look similar if the S/N of the observation is 100. 

Next, we need to constrain the disk scale height and the disk $\alpha$ parameter to break the degeneracy of
$K'$ in order to derive $q$. For each major gap, if there is a shallower gap at $r/r_p \sim$0.5-0.7, the shallower gap may be the secondary gap induced by the planet. The distance between the secondary gap and $r_p$ is very sensitive to $h$ (\S \ref{sec:secondarygaps} and Equation \ref{eq:secondary_gap_fit}). Thus, the presence of the secondary gap at the right radii not only makes the planet gap-opening scenario more plausible but also gives constraints on the disk scale height. If there is no secondary gap, we may need to use radiative transfer calculations or Equation \ref{eq:disktemp} to estimate the disk temperature. The existence of the secondary gap also implies that the disk viscosity parameter $\alpha\lesssim10^{-4}$. Without the presence of the secondary gap, the $\alpha$ parameter can then be constrained by the symmetry of the disk structures. If the rings/gaps are highly axisymmetric, $\alpha$ is likely to be larger than $10^{-4}$.

Finally, we can use Equation \ref{eq:k_width} to calculate $q$ and thus the planet mass. 
With $M_p$ derived, we can go back to Step 2 to estimate a more accurate gas surface density. We can also do a consistency check with the derived $M_p$. For example, we can check if the sub/super-Keplerian motion at the gap edge could be detected (\S \ref{sec:gaskinematics}, Equation \ref{eq:Kvr}), if the planet should produce large-scale asymmetries (e.g. eccentricity, vortices \S \ref{sec:features}, Figure \ref{eq:eccentricity}), and if the gap depth
is consistent with observations (Table \ref{table:relationdepth}). 

\begin{deluxetable*}{lcccccccclllll}
\tabletypesize{\scriptsize}
\rotate
\tablecaption{Inferred Planet Mass from 19 Gaps\label{table:Mp}}
\tablehead{
\colhead{Name} & \colhead{$M_{*}$} &
\colhead{$r_{gap}$} & \colhead{width} & \colhead{$\Sigma_{dust}$} & 
\colhead{$h/r$} & \colhead{$\Sigma_{g,0}$} & \colhead{$\Sigma_{GI}$} & \colhead{$\Sigma_{g,0}^{used}$} &
\colhead{$St_{max}^{used}$} &
\colhead{$M_{p,am4}$} & 
\colhead{$M_{p,am3}$} & \colhead{$M_{p,am2}$} & \colhead{Uncertainty}
\\
\colhead{} & \colhead{($M_{\odot}$)} &
\colhead{(au)} & \colhead{($\Delta$)} & \colhead{($\mathrm{g\, cm^{-2}}$)} & 
\colhead{} & \colhead{($\mathrm{g\, cm^{-2}}$)} & \colhead{($\mathrm{g\, cm^{-2}}$)} & \colhead{($\mathrm{g\, cm^{-2}}$)} & \colhead{($5.23 \times 10^{-4}$)} &
\colhead{($M_{Jup}$)} & 
\colhead{($M_{Jup}$)} & \colhead{($M_{Jup}$)} & \colhead{($log_{10}(M_p)$)}
}
\colnumbers
\startdata
AS 209    & 0.83  & 9    & 0.42  & 1.23        & 0.04 & $>$100, 100, 100 &  1278.4 & 100, 100, 100 & 0.33, 3, 30 & 1.00, 0.81, 0.37   & 2.05, 1.66, 0.76   & 4.18, 3.38, 1.56 & $^{+0.13}_{-0.16}$, $^{+0.14}_{-0.17}$, $^{+0.28}_{-0.29}$\\
AS 209    & 0.83  & 99   & 0.31  & 0.17        & 0.08 & 30, 10, 3  &  19.2\tablenotemark{$\dagger$}  & 10, 10, --  & 3, 30, -- & 0.32, 0.18, --   & 0.65, 0.37, --   & 1.32, 0.75, -- &  $^{+0.14}_{-0.17}$, $^{+0.21}_{-0.50}$,  -- \\
Elias 24  & 0.78 & 57   & 0.32  & 0.52        & 0.09 & 100, 30, 10  & 58.6\tablenotemark{$\dagger$} & 30, 30, -- & 1, 10, --  & 0.41, 0.19 --  & 0.84, 0.40, --  & 1.72, 0.81, -- &  $^{+ 0.16}_{-0.14}$, $^{+0.22}_{-0.16}$,  --\\
Elias 27  & 0.49 & 69   & 0.18  & 0.48        & 0.09 & 100, 30, 10   & 25.6\tablenotemark{$\dagger$} & 10, 10, -- & 3, 30, --  & 0.03, 0.02, --  & 0.06, 0.05, --  & 0.12, 0.10, -- & $^{+ 0.16}_{-0.14}$, $^{+0.21}_{-0.50}$, -- \\
GW Lup\tablenotemark{*}    & 0.46 & 74   & 0.15  & 0.13 & 0.08 & 10, 3, 3  & 19.8 & 10, --, -- & 3, --, -- & 0.01, --, --   & 0.03, --, --  & 0.06, --, --  & $^{+0.14}_{-0.17}$, --, --\\
HD 142666 & 1.58 & 16   & 0.20   & 1.63        & 0.05 & $>$100, 100, 100  & 814.0 & 100, 100, 100 & 0.33, 3, 30 & 0.15, 0.12, 0.09   & 0.30, 0.25, 0.19  & 0.62, 0.50, 0.38 & $^{+0.13}_{-0.16}$, $^{+0.14}_{-0.17}$, $^{+0.28}_{-0.29}$\\
HD 143006 & 1.78 & 22   & 0.62  & 0.20         & 0.04 & 30, 10, 3  &  442.7 & 30, 10, --  & 1, 30, -- & 9.75, 2.35,  --  & 19.91, 4.80, -- & 40.64, 9.81, -- & $^{+ 0.16}_{-0.14}$, $^{+0.21}_{-0.50}$, --  \\
HD 143006 & 1.78 & 51   & 0.22  & 0.14        & 0.05 & 30, 10, 3   &  101.6 & 30, 10, --  & 1, 30, -- & 0.16, 0.14 --  & 0.33, 0.28, --  & 0.67, 0.57, -- & $^{+ 0.16}_{-0.14}$, $^{+0.21}_{-0.50}$, -- \\
HD 163296 & 2.04 & 10   & 0.24  & 1.43        & 0.04 & $>$100, 100, 100  &  2273.0 & 100, 100, 100 & 0.33, 3, 30 & 0.35, 0.28, 0.19   & 0.71, 0.58, 0.39   & 1.46, 1.18, 0.79 & $^{+0.13}_{-0.16}$, $^{+0.14}_{-0.17}$, $^{+0.28}_{-0.29}$ \\
HD 163296 & 2.04 & 48   & 0.34  & 0.41        & 0.06 & 30, 10, 10  &  146.0 & 30, 10, -- & 1, 30, -- & 1.07, 0.54, --   & 2.18, 1.10, --   & 4.45, 2.24, -- &   $^{+ 0.16}_{-0.14}$, $^{+0.21}_{-0.50}$, --\\
HD 163296 & 2.04 & 86   & 0.17  & 0.15        & 0.07 & 30, 10, 3   &  52.6 & 30, 10, --  & 1, 30, -- & 0.07, 0.08, --  & 0.14, 0.16, --  & 0.29, 0.34, -- &  $^{+ 0.16}_{-0.14}$, $^{+0.21}_{-0.50}$, --  \\ 
SR 4      & 0.68 & 11   & 0.45  & 1.56        & 0.05 & $>$100, 100, 100 &  792.8 & 100, 100, 100 & 0.33, 3, 30 & 1.06, 0.86, 0.38  & 2.16, 1.75, 0.77  & 4.41, 3.57, 1.57 &  $^{+0.13}_{-0.16}$, $^{+0.14}_{-0.17}$, $^{+0.28}_{-0.29}$ \\
\hline
DoAr 25\tablenotemark{*}   & 0.95 & 98   & 0.15  & 0.48        & 0.07 & 100, 30, 10  &  20.0\tablenotemark{$\dagger$} & 10, 10, -- & 3, 30, -- &   (-- , 0.10, --)   & (0.10, --, --)  &(-- , 0.95, --) &  --, --, --\\
DoAr 25   & 0.95 & 125  & 0.08  & 0.14        & 0.07 & 30, 10, 3  &  13.1\tablenotemark{$\dagger$} & 10, --, -- & 3, --, -- &  (0.03, --, --)  & -- , --, -- &  -- , --, --  & --, --, --\\
Elias 20  & 0.48 & 25   & 0.13  & 0.80        & 0.08 & 100, 30, 30 &   171.9 & 100, 30, 30  & 0.33, 10, 100  &  --, --, --  &  (0.05, 0.05, 0.05)  & -- , --, --   & --, --, --\\
IM Lup    & 0.89 & 117   & 0.13  & 0.20        & 0.09 & 30, 10, 3   &  16.0\tablenotemark{$\dagger$} & 10, --, -- & 3, --, -- &  (0.09 , --, --)   &  (0.09, --, --)  & --, -- , --& --, --, --\\
RU Lup    & 0.63 & 29   & 0.14  & 1.13        & 0.07 & $>$100, 100, 100 &  144.1 & 100, 100, 100 & 0.33, 3, 30 & (0.07, --, --)  &  (--, 0.07, 0.07)  & -- , -- & --, --, --  \\
Sz 114    & 0.17 & 39    & 0.12  & 0.22        & 0.10 & 30, 10, 3  &    35.3  & 30, 10, -- & 1, 30, -- & (0.02 , 0.02, --)   &  --, --, --  & --, -- , -- & --, --, -- \\
Sz 129    & 0.83 & 41    & 0.08  & 0.47        & 0.06 & 100, 30, 10  &  77.7\tablenotemark{$\dagger$} & 30, 30, -- & 1, 10, -- &  (--, 0.03 , --)   &  (0.03, --, --)  & --, -- , --& --, --, -- \\
\enddata
\tablecomments{(1) Name of the object (2) Stellar mass in $M_\odot$ \citep{andrews18b} (3) Position of the gap in au (4) The width calculated using the same method in \S \ref{sec:gapring} (5) The averaged dust surface density from 1.1 $r_p$ to 2.0 $r_p$ using the observed profiles in Figure 6 of \citet{huang18b} and $\kappa$ = 0.43 $\mathrm{g\, cm^{-2}}$. Here we assume $r_p$ = $r_{gap}$.  (6) The aspect ratio at the position of the inferred planet using Equation \ref{eq:disktemp}; the mass and luminosity of the stars are taken from \citet{andrews18b}. (7) The closest gas density $\Sigma_{g,0}$ found from Figure \ref{fig:gasdustratio} for DSD1, "1 mm" and DSD2 (the following columns which have three entries separated by comma are all in this order.). (8) The maximum gas surface density calculated from the gravitational instability constraint $\Sigma_{GI} = C_s\Omega_K/(\pi G)$ (with Toomre $Q=1$). The difference between these values and those in \citet{dullemond18b} Table 3 is due to that Dullemond et al. calculated $\Sigma_{GI}$ using $Q=2$ and at the position of the ring instead of the gap. (9) The initial gas surface density $\Sigma_{g,0}$ constrained by $\Sigma_{GI}$, otherwise it is the same as (7).  (10) The $St_{max}$ (in unit of $5.23\times 10^{-4}$) used (constrained by the gravitational instability) to find the planet mass. (11) Planet mass assuming $\alpha$=$10^{-4}$, estimated from DSD1, "1 mm" and DSD2. (12) Similar to (11) but assuming $\alpha$=$10^{-3}$ (13) Similar to (11) but assuming $\alpha$=$10^{-2}$. The 12 inferred planets above the horizontal line are estimated from the fits, while the 7 below are estimated by directly comparing the individual models with the observations (See Figure \ref{fig:flowchart} for the flow chart). (14) The uncertainty of the estimated planet masses given the $\alpha$ and $h/r$.}
\tablenotetext{*}{The gap of the GW Lup at 74 au has width $\Delta$ $>$ 0.15, while the gap of the DoAr 25 at 98 au has $\Delta$ $<$ 0.15 before rounding.}
\tablenotetext{\dagger}{$\Sigma_{GI}$ is used to constrain the initial gas density $\Sigma_{g,0}$, thus the Stokes number. Rows with $\dagger$ have at least one of the DSD1 or "1 mm" model exceed the gravitational instability limit, thus lower available $\Sigma_{g,0}$ (i.e., higher Stokes number) are adopted (listed in column 9).}
\end{deluxetable*}

Following this procedure (Figure \ref{fig:flowchart}), we identify potential planets in the DSHARP disks (as summarized in Table \ref{table:Mp}) using the intensity profiles from \citet{huang18b}. All the gaps  with $\Delta$ $\geq$ 0.15 in the DSHARP sample have been carefully measured for their widths and then we use the fitting formula to estimate the planet mass based on their widths. These are shown in the upper part of Table \ref{table:Mp}. Since each fitting line with a Stokes number comes with an uncertainty in $K'$ (See \S\ref{sec:gapring}, and Table \ref{table:relationwidth}), the uncertainties of the planet mass with the given $\alpha$ and $h/r$ are also included in the table. For shallow gaps with $\Delta$ $\leq$ 0.15, our fitting formulae fail to fit the gap widths from the simulations and the gap width is also sensitive to the convolution beam size (Figure \ref{fig:gap_width_fit}). Thus,
we only choose those that look similar to shallow gaps in our grid of numerical simulations and compare them directly with simulations. Thus, only a subset of the shallow gaps in DSHARP sample have been fitted. They are shown in the lower part of Table \ref{table:Mp}. Since we compare these shallow gaps with the simulations by eye, proper error estimate can not be provided. Thus, they are considered not robust and complete, and will not be included in the statistical study later. This also means that our statistical study may miss low mass planets. In the next section, we will comment on each case in detail. 

Table \ref{table:Mp} gives the gap positions, measured gap widths, outer disk dust surface densities and estimated $h/r$. Using the dust-to-gas mass ratio (Figure \ref{fig:gasdustratio}) in simulations with different dust size distributions (DSD1 and DSD2), the gas surface densities are also provided. If the gas surface density is above the gravitational instability (GI) limit with $Q=1$, we use the GI limit as the gas surface density. Then with $St_{max}$ calculated for DSD1 and DSD2, we derive $K'$ for DSD1 and DSD2 using $\Delta-K'$ relationships. To break the degeneracy in $K'$ to derive $q$, we need to know the disk viscosity. Thus, for either DSD1 or DSD2, we provide three possible planet masses with the disk $\alpha$=10$^{-2}$, 10$^{-3}$, and 10$^{-4}$. These three masses are labeled as $M_{p,am2}$, $M_{p,am3}$, and $M_{p,am4}$, which are listed in Table \ref{table:Mp}. The inferred planet mass is roughly twice as high if $\alpha$ is 10 times larger. This is because $K'= q (h/r)^{-0.18}\alpha^{-0.31}$, so that $q\propto \alpha^{0.31}$ with a given $K'$ and $h/r$. As shown in Table \ref{table:Mp}, many gaps (especially having low $\Sigma_{g,0}$) cannot be fit using DSD2 dust size distribution. This is because the Stokes number for dust in DSD2 is very large, so that particles in the inner disk quickly drift to the central star forming a cavity with a single ring at the gap edge. This is consistent with the conclusion in \cite{dullemond18b} that large particles (cm-sized) are not preferred in the DSHARP disks. 

As can be seen from Equation \ref{eq:stokes} and Table \ref{table:Mp}, the Stokes number estimated from DSD1 and DSD2 can differ by three orders of magnitude. DSD1 with $s_{max}$ = 0.1 mm and DSD2 with $s_{max}$ = 1 cm can be seen as two extreme cases. Dust with $s_{max}<$ 0.1 mm should have similar profiles as DSD1 since 0.1 mm particles already couple with the gas well in the sample. Dust with $s_{max}$ = 1 cm already drifts very fast and we can hardly find a mass solution for most of our disks. To cover a more comprehensive parameter space, we add a new set of planet masses estimated assuming $s_{max}$ = 1 mm ("1 mm" hereafter). The estimated initial gas density $\Sigma_{g,0}$ are used between the values of DSD1 and DSD2. Holding $\Sigma_{g,0}$ constant, $St_{max}$ for "1 mm" is 10 times larger than that of the DSD1 or 10 times smaller for DSD2. Thus, the Stokes number of the "1 mm" models are in between those two extremes. The gap width-K' relation of the "1 mm" models are taken from the corresponding $St_{max}$ fits in DSD1. The justification is that only the Stokes number matters regarding the gap width, as discussed at the end of \S\ref{sec:gapring} and demonstrated in Figure \ref{fig:alllines}. The estimated $\Sigma_{g,0}$, $St_{max}$, three planet masses given $\alpha = 10^{-4},  10^{-3},  10^{-2}$ and their uncertainties are all given in Table \ref{table:Mp} in the order of DSD1, "1 mm" and DSD2 (ascending $s_{max}$). 
Among the nine planet masses estimated for each source, we prefer $M_{p,am3}$ with DSD1 size distribution. The main reason that $\alpha=10^{-3}$ is preferred is that most rings of the DSHARP sample do not show significant asymmetry, indicating that $\alpha\gtrsim 10^{-3}$. On the other hand, if the gaps are shallow, low mass planets in $\alpha=10^{-4}$ disks can also produce axisymmetric gaps/rings. 

\subsection{Comments on Individual Sources \label{sec:individuals}}

\subsubsection{AS 209 \label{sec:as209}}

\begin{figure*}[t!]
\includegraphics[width=\linewidth]{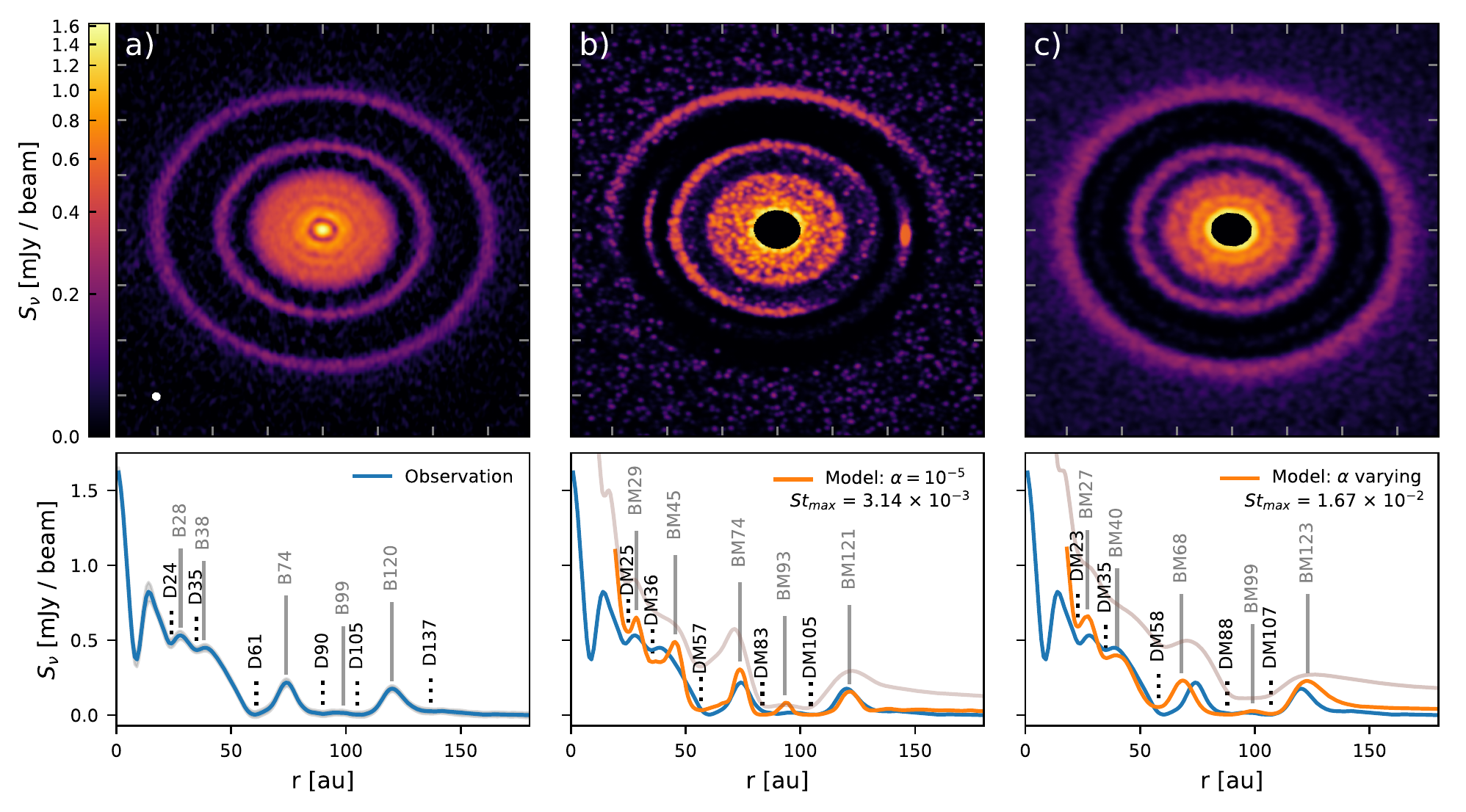}\figcaption{Top panels: a) The observation image of AS 209 (See \citealt{guzman18}, \citealt{huang18b}). The distance between two ticks on the axes is 40 au.  b) The synthetic image from the simulation with a single planet ($M_p/M_*$ = 0.1 $M_J/M_{\odot}$) at 99 au in a $\alpha$ = $10^{-5}$,  $\Sigma_{g,0}$ = 15 $\mathrm{g\,cm^{-2}}$,  $s_{max}$ = 0.3 mm and $p=-3.5$ disk at 2000 orbits ($\sim$ 2Myrs). c) The synthetic image from the simulation with a single planet ($M_p/M_*$ = 0.1 $M_J/M_{\odot}$) at 99 au in a varying $\alpha$,  $\Sigma_{g,0}$ = 6.4 $\mathrm{g\,cm^{-2}}$, n(s) $\propto$ $s^{-3.5}$, $s_{max}$ = 0.68 mm disk at 1350 orbits ($\sim$ 1.35 Myrs). Bottom panels: the azimuthally-averaged intensity profiles. Panel a) is the profile from the observation, and b) and c) are the profiles from the simulations above. The ``DM" and ``BM" stand for Dark annulus and Bright ring in the Model, respectively; the digits coming after mark the position in au. The gas density profiles of two models are overplotted on the bottom panels in grey color in arbitrary unit. \label{fig:as209}}
\end{figure*}

AS 209 is a system with many gaps.
\cite{fedele2018} found two gaps at 62 au and 103 au and
they proposed that a 0.7 $M_{Saturn}$ planet at $\sim$103 au can explain both gaps. \citet{huang18b} and \citet{guzman18} identified many gaps in this system including dark annuli at 9, 24, 35, 61, 90, 105 and 137 au. Following our procedure (Figure \ref{fig:flowchart}), we first derive the $K'$ parameter for the main gap at $\sim$100 au. The  narrow width of the gap  suggests that it is a sub-Jupiter mass planet. Then 
we find that the gap at r = 61 au is shallower than the main gap, and it is at  0.5-0.7 $r_p$. Thus, we treat it as a secondary gap induced by the planet.  The distance between the secondary and primary gaps suggests that $h/r\sim 0.05-0.06$ (Equation \ref{eq:secondary_gap_fit} and Figure \ref{fig:gap_position_fit}).
This $h/r$ is slightly smaller than the simple estimate with Equation \ref{eq:disktemp}, but the faint emission at the near-IR scattered light image \citep{avenhaus2018} may support that the disk is indeed thin (another possibility is that the disk is significantly less flared.). With this $h/r$ and $K'$, we derive that the 100 au planet has a mass of $q=3\times10^{-4}$ in a $\alpha=10^{-4}$ disk or $q=10^{-4}$ in a $\alpha=10^{-5}$ disk. Motivated by the smaller gaps at 24 and 35 au from the DSHARP data \citep{guzman18}, we carry out several additional simulations 
extending the range of $\alpha$ to $10^{-5}$. Since a smaller $\alpha$ is used, we double the numerical resolution for all simulations that are constructed for AS 209.
Surprisingly, the $q=10^{-4}$ planet in a $\alpha=10^{-5}$ and $h/r=0.05$ disk can explain all 5 gaps at 24, 35, 62, 90 and 105 au (Figure \ref{fig:as209}). Although we assume that there is another planet at 9 au to explain the 9 au gap, it is possible that the 9 au gap is also produced by the main planet at 99 au, considering that our simulation domain does not extend to 9 au. We want to emphasize that our simulation with one planet at 99 au not only matches the primary gap around 100 au, but also matches the position and amplitude of secondary (61 au), tertiary (35 au) and even the fourth (24 au) inner gaps. This makes AS 209 the most plausible case that there is indeed a planet within the 100 au gap. 

Although the above model reproduces the positions and intensities of gaps and rings very well, its synthetic image (the upper middle panel in Figure \ref{fig:as209})
shows a noticeable horseshoe region and some degree of asymmetry in the rings. Such asymmetry disappears when $\alpha\gtrsim 10^{-3}$. On the other hand, the presence of the tertiary and the forth inner gaps requires a small $\alpha$. Thus, we carry out a simulation with a radially varying $\alpha$ ($\alpha=3\times 10^{-4}(r/r_p)^2$). This model reproduces the 2-D intensity maps better, as shown in the right panels of Figure \ref{fig:as209} and also presented in \citet{guzman18}. Such a radially varying $\alpha$ disk has also been suggested to explain HD 163296 \citep{liu18}. If these models are correct, they suggest that $\alpha$ in protoplanetary disks is not a constant throughout, supporting the idea that different accretion mechanisms are operating at different disk regions \citep{turner14}.

\citet{dullemond18b} constrained that the $\alpha/St$ for the ring at 74 au has a range roughly between 0.03 and 0.7 from the limits of pressure bump width argument (See Table 3 therein). Such constraint is derived using the particle trapping model and does not depend on the origin of the ring. In our $\alpha = 10^{-5}$ model, $\alpha/St_{max} \approx 0.003$ and in our $\alpha$ varying model, $\alpha/St_{max} \approx 0.02$. The actual characteristic $St$ can be smaller, considering that the $St_{max}$ here is the maximum Stokes number at the position of the planet in the initial condition ($t_0$). Since for both models $n(s)\propto s^{-3.5}$, 50\% of the dust mass in $t_0$ at $r_p$ have $St \leq 0.25\ St_{max}$. Adopting these values, their $\alpha/St \approx$ 0.012 and 0.08, respectively. Thus, the $\alpha = 10^{-5}$ model is off the lower limit of $\alpha/St$ by a factor of 3, whereas the $\alpha$ varying model is safely above the lower limit. Considering that the turbulent diffusion with the small $\alpha$ ($\alpha=10^{-5}$) in our simulations may have not reached to a steady state, we conclude that these models are consistent with \citet{dullemond18b}.

\subsubsection{Elias 24 \label{sec:elias24}}
\begin{figure}[t]
\includegraphics[width=\linewidth]{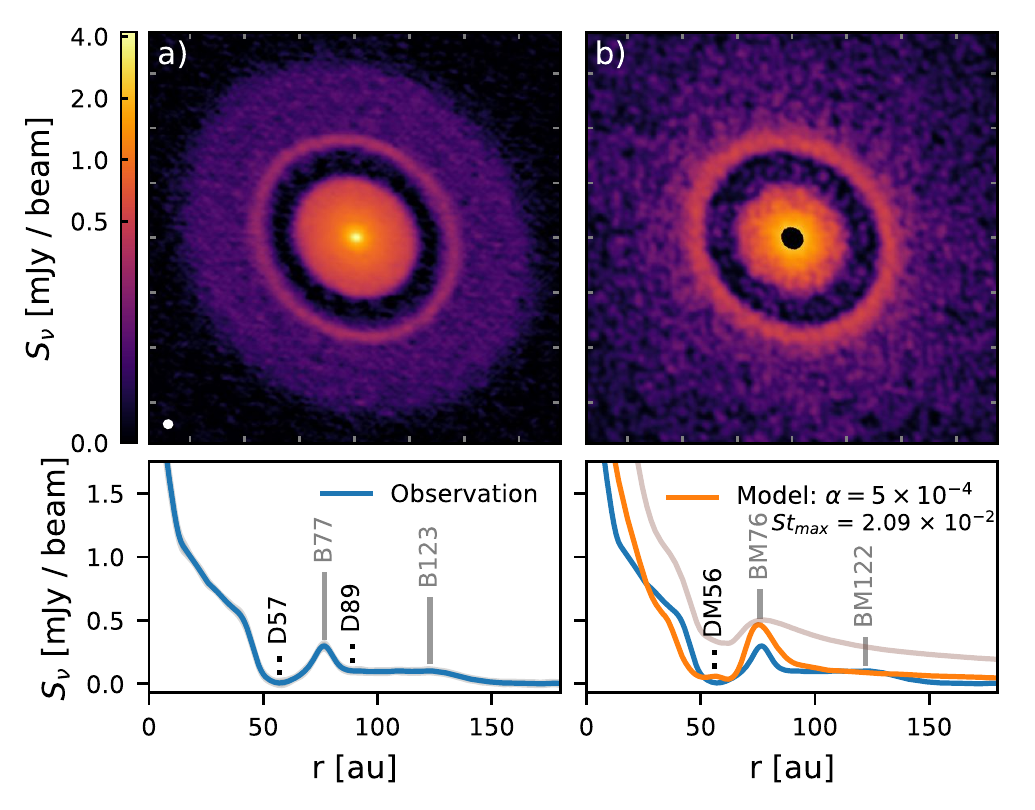}
\figcaption{\label{fig:elias24} The comparison between the observation and the simulation of Elias 24. Top panels: a) Observation images of the Elias 24 \citep{andrews18b} and b) our simulation with a single planet at 57 au. The model image is produced at 1000 planetary orbits, effectively 0.43 Myrs at 57 au. The distance between two ticks on the axes is 40 au. Lower panels: a) The radial profile of Elias 24 \citep{huang18b}, b) the radial profile of our simulation. The gas density profile in arbitrary unit is overplotted in grey color. The bright rings and dark annulus are marked the same way as in Figure \ref{fig:as209}.}
\end{figure}

Elias 24 \citep{cieza2017} is another system that looks very similar to our planet-disk interaction simulations. It has a deep gap at 57 au, a narrow ring at 77 au, and an extended outer disk \citep{huang18b}. The narrowness of the ring is suggestive of particle trapping at the gap edge. \citet{dipierro2018} estimated that there is a 0.7 $M_J$ mass planet at 57 au, while \citet{cieza2017} suggested that the mass of the 57 au planet is 1-8 $M_J$. Our estimate is roughly consistent with these previous estimates. The planet mass is $\sim$ 0.8 $M_J$ with $\alpha=10^{-3}$ and DSD1. On the other hand, the clear signature of dust pile-up at the outer gap edge may indicate that dust is larger than 0.1 mm as used in DSD1. If dust particles in Elias 24 are larger than 0.1 mm, the planet mass can be lower than our estimates. Based on our grid of simulations, we run an additional simulation with $\alpha=5\times 10^{-4}$, $h/r=0.07$ and $M_p = 0.16\ M_J$($q=0.2\ M_J/M_*$). We put the single planet at the 57 au gap and the result is shown in Figure \ref{fig:elias24}. The dust distribution is $n(s)\propto s^{-3.5}$, $s_{max}$= 2 mm, and initial gas surface density $\Sigma_{g,0} = 15$ $\mathrm{g\,cm^{-2}}$, hence $St_{max} = 2.09 \times 10^{-2}$. \citet{dullemond18b} estimated that the $\alpha/St$ is between 0.077 to 0.66 at the 77 au bright ring. Our estimated $\alpha/St_{max} = 2.39 \times 10^{-2}$ is roughly consistent with their lower limit considering that 50\% of the dust mass has $\alpha/St > 0.096$ under the dust size distribution $p=-3.5$.

\subsubsection{Elias 27}
The spiral arms detected in Elias 27 \citep{perez16} suggest that the disk may be undergoing gravitational instability or there is a massive companion at the outer disk \citep{meru17}. Besides the spirals, there is a shallow annular gap at 70 au \citep{huang18c}. If we follow our procedure to fit this gap, the planet mass is 0.06 $M_J$ using $\alpha=10^{-3}$ and DSD1. Such a low mass planet can not induce the large-scale spirals as observed \citep{zhu2015}. On the other hand, detecting this shallow gap means that if there are massive companions in the system within 200 au (e.g. with masses larger than 0.06 $M_J$), we should be able to see the induced gaps at the mm continuum images. The lack of deep gaps suggests that there are no massive companions in this disk within 200 au. The spirals must be induced by a massive companion outside 200 au or by some other mechanisms (e.g. GI). 

\subsubsection{GW Lup}
GW Lup has two narrow gaps at 74 and 103 au. The former gap is barely above $\Delta =$ 0.15 and the latter is extremely narrow with $\Delta\lesssim$ 0.15. We decide to only fit the 74 au gap since the 103 au gap is too shallow to fit with any of our models. 
To produce the 74 au gap, the planet mass must be very small ($\sim$ 0.03 $M_J$ or 10 $M_\earth$). If both 74 and 103 au gaps are part of a wide gap separated by the horseshoe region, the planet will be at $\sim$ 85 au with $M_{p,am3}$ = 0.36 $M_J$ or $M_{p,am4}$ = 0.18 $M_J$. The $K$ parameter (Equation \ref{eq:Kdef}) is thus $\sim$ 11 and the gaseous gap depth $\delta$ is $\sim$ 2, which is roughly consistent with the observations \citep{huang18b}. Thus, this more massive planet solution remains a possibility.

\subsubsection{HD 142666}
HD 142666 has several shallow dark annuli at 16, 36, and 55 au \citep{huang18b}. 
The outer two dark annuli (36 and 55 au) as identified in \citet{huang18b} have widths of 0.05 and 0.04 by our definition, less than the minimum width measured in our models. Thus, we do not fit those two gaps either. We only fit the 16 au gap, and it suggests that $M_{p,am3}$ is 0.3 $M_J$ with DSD1 and 0.2 $M_J$ with DSD2. 

\subsubsection{HD 143006}
HD 143006 has two wide gaps at r = 22 au and r = 51 au \citep{perez18}. The gap at r = 22 au has the widest relative width ($\Delta$) in all DSHARP disks, which also leads to the highest inferred planet mass with $M_{p,am4}$ = 10 $M_J$ and $M_{p,am3}$ = 20 $M_J$.  Both submm continuum observations \citep{perez18} and the near-IR scattered light observations \citep{Benisty2018} have suggested that the inner disk inside 10 au is misaligned with the outer disk. If such misalignment is caused by a planet on an inclined orbit, the planet mass needs to be larger than 2 $M_J$ in an $\alpha=10^{-3}$ disk (Zhu 2018), which is consistent with the high planet mass derived from fitting 
the gap profile here. 
With such a massive planet predicted, HD 143006 is a prime target to look for exoplanets with direct imaging techniques.

The outer gap at 51 au can be explained by a sub-Jovian planet in the disk. The 51 au gap also has an interesting arc feature  at the outer edge, which implies that the disk viscosity may be low ($\alpha\lesssim 10^{-4}$) and $M_{p,am4}$ are preferred in this system. 

Note that such high inferred planet-stellar mass ratio at 22 au exceeds the largest $q$ (3 $M_J/M_*$) in our grid of simulations. This brings more uncertainties to the estimated planet mass. Nevertheless, we believe that our extrapolation of Equation \ref{eq:k_width} to $q=0.01$ is justifiable since the dust is well coupled to the gas due to the small Stokes number under DSD1, and the previous study with a grid of much higher $q$ \citep{fung14} showed that the relation between gaseous gap properties and the planet mass can extend to $q$=0.01.

\subsubsection{HD 163296}
HD 163296 is another system with multiple gaps. 
The DSHARP observations (\citealt{huang18b, isella18b}) reveal 4 gaps at 10 au, 48 au, 86 au and 145 au. Based on the gap widths, we estimate that the planets at 10 au, 48 au, and 86 au have masses of 0.71, 2.18, 0.14 $M_J$ in an $\alpha=10^{-3}$ disk with DSD1 dust. If the disk $\alpha=10^{-4}$, the planet masses are 0.35, 1.07, 0.07 $M_J$ with DSD1 dust. 
Except the 10 au gap, the rest gaps have been revealed by previous ALMA observations \citep{isella16}.
\cite{isella16} estimated that the 48 au planet has a mass between 0.5 and 2 $M_J$ and the 86 au planet has a mass between 0.05 and 0.3 $M_J$, which are roughly consistent with our estimate. Our derived gas surface density ($\Sigma_{g,0}$) of 3-30 g cm$^{-2}$ at 48 au and 86 au is also consistent with $\sim$ 10 g cm$^{-2}$ derived in \cite{isella16}.
\cite{teague2018} studied the deviation from the Keplerian velocity profile as measured from CO line emission and inferred that the planet at 86 au has a mass around $M_J$, which is larger than our derived $M_{p,am2}$ by a factor of 3. However, the planet mass assuming $\alpha = 10^{-2}$ and 1 mm sized particles including $1\sigma$ error can reach to $\sim$ 0.6 $M_J$. Considering that the uncertainty is a factor of two in Teague et al. and also the uncertainties in our adopted gas density, dust size distribution and disk viscosity, these results are still consistent.
\cite{liu18} has adopted a disk with an increasing $\alpha$ from  10$^{-4}$ at 48 au to $10^{-2}$ at 86 au, and 
estimated that planets at 48 au and 86 au have masses of 0.46 and 0.46 $M_J$ (their same values were purely a coincidence). This is consistent with our estimate if we adopt the same $\alpha$ values.  

An asymmetric structure is discovered at the outer edge of the 48 au gap \citep{isella18b}, implying that the disk viscosity $\alpha\lesssim10^{-4}$. Thus, the M$_{p,am4}$ may be more representative for the 48 au gap.

\subsubsection{SR 4}
SR 4 has a wide single gap at 11 au. We estimate its mass $M_{p,am3}$ = 2.16 $M_J$ with DSD1 and 0.77 $M_J$ with DSD2. The gap is also quite deep, consistent with the presence of a Jovian mass planet. Thus, SR 4 may be an interesting source to follow up to study its gas kinematics or detect the potential planet with direct imaging observations. 

\subsubsection{DoAr 25, Elias 20, IM Lup, RU Lup, Sz 114 and Sz 129}

These six systems have shallow gaps with $\Delta<0.15$. Thus, we compare the observed gap widths directly with those derived in numerical simulations (Figure \ref{fig:gap_width_fit}). The inferred planet mass is less than 0.1 $M_J$ for all these gaps.  The smallest planet is 0.02 $M_J$ or 6.4 $M_{\earth}$. Note also that IM Lup features intricate spiral arms inside the gap fit at 117 au \citep{huang18c}.

On the other hand, DoAr 25, Elias 20, and RU Lup have adjacent double gaps, similar to GW Lup.
If we treat these double gaps as one main gap which is separated by the horseshoe material, we can derive the planet mass under this scenario. To explain both the 98 and 125 au gaps in DoAr 25 using a single planet, the planet is at 111 au with  $M_{p,am3}=0.73\, M_J$ or $M_{p,am4}=0.36\, M_J$.
To explain the 25 and 33 au gaps in Elias 20, the planet is at 29 au with $M_{p,am3}=0.57\, M_J$ or $M_{p,am4}=0.28\, M_J$.
To explain the 21 and 29 au gaps in RU Lup, the planet is at 24 au with $M_{p,am3}=1.18\, M_J$ or $M_{p,am4}=0.58\, M_J$. To make the gaps as shallow as possible, we assume DSD1 dust distribution here. Even so,
the corresponding gap depth $\delta$ is larger than 2 with these planet masses. By comparing with the intensity profiles
in \citet{huang18b}, DoAr 25 has gaps that could be deep enough, while the gaps in both Elias 20 and RU Lup are too shallow and this scenario seems unlikely. 

\subsection{Young Planet Population}

\begin{figure*}[t!]
\includegraphics[width=\linewidth]{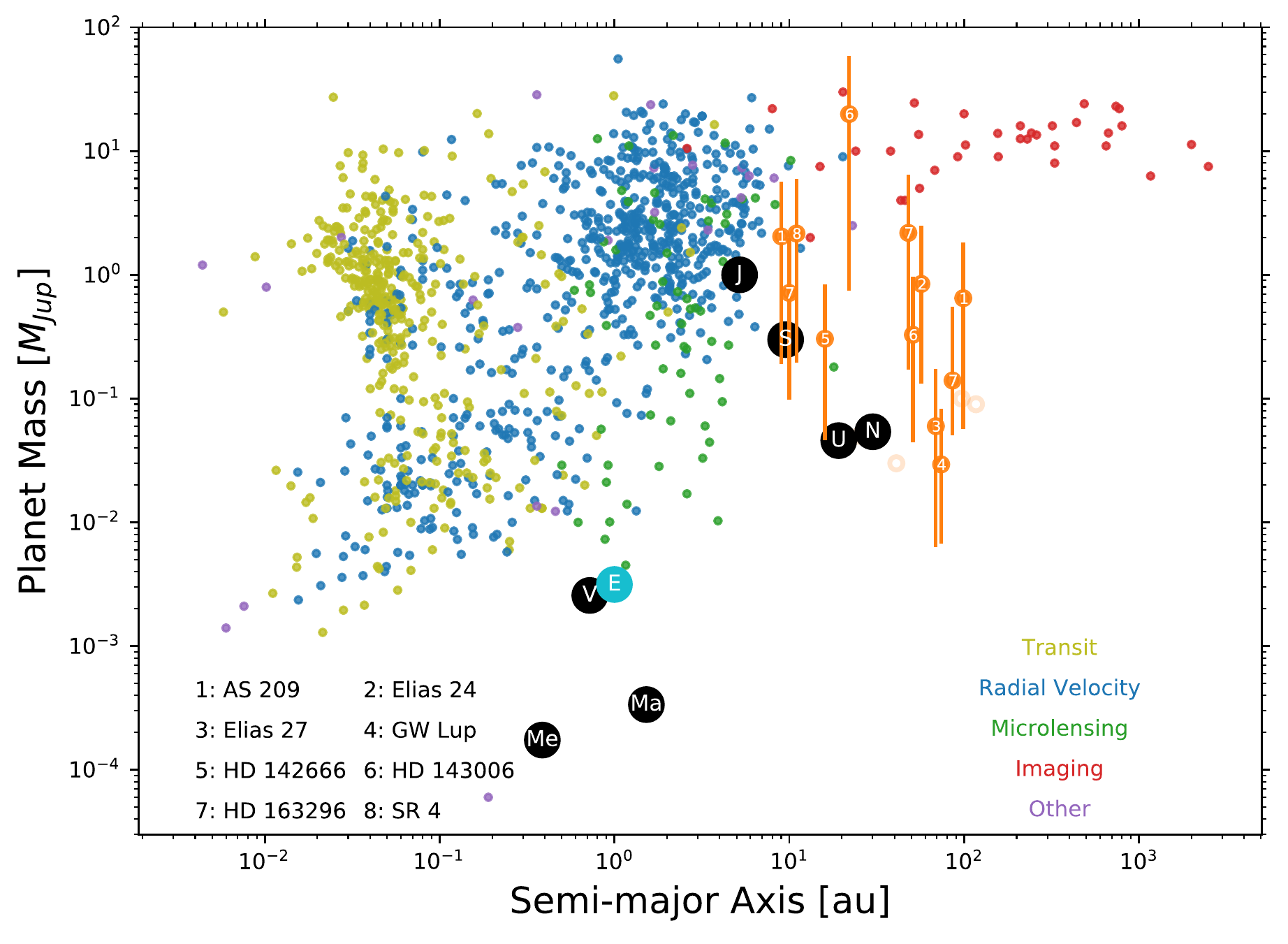}\figcaption{Planet mass vs. Planet semi-major axis.
\label{fig:moneyplot} Orange circles with errorbars are 12 inferred planets from 8 disks listed in Table \ref{table:Mp} using the mass $M_{p,am3}$, DSD1. The other inferred planet masses with the assumption of $\alpha$ = $10^{-2}$ and $10^{-4}$ (DSD1, "1 mm" or DSD2) are listed in Table \ref{table:Mp} as $M_{p,am2}$ and $M_{p,am4}$. We can see that ALMA is sensitive to planets which are not detectable using traditional methods. Young planetary systems may harbor Uranus and Neptune mass planets beyond 10 au similar to our Solar System. For reference, small dots with different colors are exoplanets confirmed as of August, 2018 (\url{https://exoplanetarchive.ipac.caltech.edu/}). Black circles with white labels are solar system planets, expect that the planet Earth is marked in light blue. Light orange open circles are planets inferred from shallow gaps (also $M_{p,am3}$, DSD1). They are not included in the statistics because we lack the knowledge of their uncertainties.}
\end{figure*}

Now, we can put these potential young planets in the exoplanet mass-semimajor axis diagram (Figure \ref{fig:moneyplot}). Considering most of these systems do not show asymmetric structures, we pick the planet mass that is derived using $\alpha=10^{-3}$ and DSD1. The mass errorbar is chosen as the minimum and maximum planet mass among all the nine masses that have constrained values in Table \ref{table:Mp} (columns 11 to 13), adding up the additional uncertainty due to the fitting from the column 14 of the table. Thus, this is a comprehensive estimate of the error covering different disk $\alpha$ (from 10$^{-4}$ to $10^{-2}$), particle sizes ($s_{max}$ from 0.1 mm to 1 cm), and the errors of the fitting. The planet masses that are from very narrow gaps in the lower part of Table \ref{table:Mp} (the ones with brackets)  are labeled with light circles, and we do not count them in the statistical study below since the narrowness of the gaps leads to large uncertainties in the mass estimate. \cite{Bae2018} has collected young planets from previous disk observations in the literature (most are Herbig Ae/Be stars). Here, we only consider the DSHARP sample \citep{andrews18b}. Although this sample is more homogeneous with similar observation requirements, it is still slightly biased towards bright disks and thus high accretion rate disks around more massive stars.

Since the DSHARP observations have resolutions of $\sim$ 3-5 au and most disks only extend to 200 au in the dust continuum images, the planet population we can probe lies between 5 and 200 au. The probed mass limit is around the Neptune mass in the outer disk and a little bit higher (a factor of $\sim$2) in the inner disk ($<$10 au, with a larger beam size). If there are planet-induced gaps in the disk, we should always detect them at almost all the viewing angles unless the disk is very edge on. Thus, the probability that we are missing gap-induced planets due to the observational bias is small. Under this circumstance, we can simply estimate the planet occurrence rate through dividing the number of planets by the total number of disks observed. Although DSHARP observes 20 disks, 2 are certainly in multiple star systems \citep{troncoso18}. Since we only focus on single star systems here, the total number of disks is 18. 

Since the gaps in protoplanetary disks may not be due to young planets, our derived planet occurrence rates should be considered as the upper limits. On the other hand, we may miss planets at the mass detection limit ($\sim$Neptune mass), as evidenced by that we do not include those planets that are fitted by eye and have no error estimates. Thus, the planet occurrence rates for Neptune mass planets may be higher than our estimates.

By comparing with exoplanets discovered with other methods, we find that:

First, we only have one planet that is more massive than 5 $M_J$. Thus, the occurrence rate for $>$5 $M_J$ planets beyond 5-10 au is 1/18 or 6\%. Wide-orbit giant planets are very rare. This is consistent with the direct imaging constraints that the occurrence rate for 5-20 $M_J$ planets at $>$5-10 au is 1-10\% \citep{meshkat2017, vigan2017, bowler2018}.

Second, using disk features, we may be probing a planet population that is not accessible by other planet searching techniques. These are Neptune to Jupiter mass planets beyond 10 au. Young planetary systems may harbor Uranus and Neptune mass planets beyond 10 au similar to our Solar System. The occurrence rate for 0.2 $M_J\lesssim M_p\lesssim$ 5 $M_J$ planets beyond 5-10 au is 8/18 or 44\%, and the occurrence rate for all the planets more massive than Neptune and less than 5 $M_J$ beyond 5-10 au is 10/18 or 56\%. These rates are comparable to the 31\% giant planet ($>$ 0.1 $M_J$) occurrence rates \citep{clanton2014} within $10^{4}$ days ($<$9 au for solar mass stars). 
If we consider that our derived planets spread from 5 au to 200 au, the occurrence rate per decade of semi-major axis is 27\% and 35\%, respectively. This rate is comparable to the occurrence rate (20\%) for giant planets ($>$ 0.1 $M_J$) with period between $10^3$ and $10^4$ days. Thus, giant planet distribution may be flat beyond several au to $\sim$ 100 au.

Finally, the planet's mass distribution is almost flat from Neptune to Jupiter mass. We have
$\sim$ 5 planets with 0.03 $M_J\lesssim M_p\lesssim$ 0.3 $M_J$, and 6 planets with
0.3 $M_J\lesssim M_p\lesssim$ 3 $M_J$. 

We bin the planet masses in decade in part due to the number of sources available and in part because of the uncertainties of the mass range for each planet (see Figure \ref{fig:moneyplot}). The uncertainties for most of the planet masses are around a factor of 10.
We want to emphasize that the derived planet mass has larger uncertainties due to the unknown disk $\alpha$ and dust size distribution. On the other hand, as long as all these disks have similar $\alpha$ values among each other, the derived planet mass will systematically shift up and down with the same fraction (e.g. decreasing the $\alpha$ value by a factor of 10 will decreasing the planet mass by a factor of two for all the planets).

\section{Discussion \label{sec:discussion}}

\subsection{Our Solar System and HR 8799 Analogs in Taurus \label{sec:syandhr}}
 Exoplanetary systems are very diverse with systems having multiple low-mass planets within 1 au (as probed by the $Kepler$ spacecraft) or systems having multiple giant planets beyond 10s of au (e.g. HR 8799). Our solar system has both terrestrial and giant planets. Are any of the DSHARP sources analogous to our Solar System when it was young? Is DSHARP capable of detecting young Solar System analog or HR 8799 analog?

To answer these questions, we embed planets in our Solar System and HR 8799 into a protoplanetary disk having a minimum mass solar nebulae surface density 
\begin{equation}
    \Sigma_g=1700 \left(\frac{r}{au}\right)^{-1.5} \mathrm{g\,cm^{-2}}\,.
\end{equation}
To maximize our chances to detect disk features, we use DSD2 dust size distribution ($s_{max}$=1 cm).
The initial dust-to-gas mass ratio is 1/100. We run simulations with both $\alpha=10^{-2}$ and $10^{-4}$ to explore the parameter space slightly. The mass of the HR 8799 central star is 1.47 M$_{\odot}$, and the four giant planets in HR 8799 are chosen as 7 $M_J$ at 14.5 au, 7 $M_J$ at 24 au,  7 $M_J$ at 38 au, and 5 $M_J$ at 68 au \cite{marios2010}. The inner and outer boundary of these simulations are 0.1 $r_0$ and 10 $r_0$, where $r_0$ = 10 au for two young solar system runs and $r_0$ = 20 au for two HR 8799 runs. The $\alpha = 10^{-4}$ run for the solar system has 1500 and 2048 grid points in the radial and $\theta$ direction, whereas the three other models have 750 and 1024 grids in the radial and $\theta$ direction. The Solar System simulation runs for $\sim$ 500 orbits  at 10 au (due to the higher resolution and computational cost) and the HR 8799 simulation runs for $\sim$ 1000 orbits at 20 au. The mm intensity images are calculated using the temperature structure from Equation \ref{eq:disktemp} with luminosities at 1 Myr found from \citet{dantona94} given current masses. Before making the ALMA synthetic images, the dust emission for the young solar system and HR 8799 runs are convolved with a 2-D Gaussian FWHM 1.4 au and 2.8 au, respectively.

\begin{figure*}[t!]
\includegraphics[width=\linewidth]{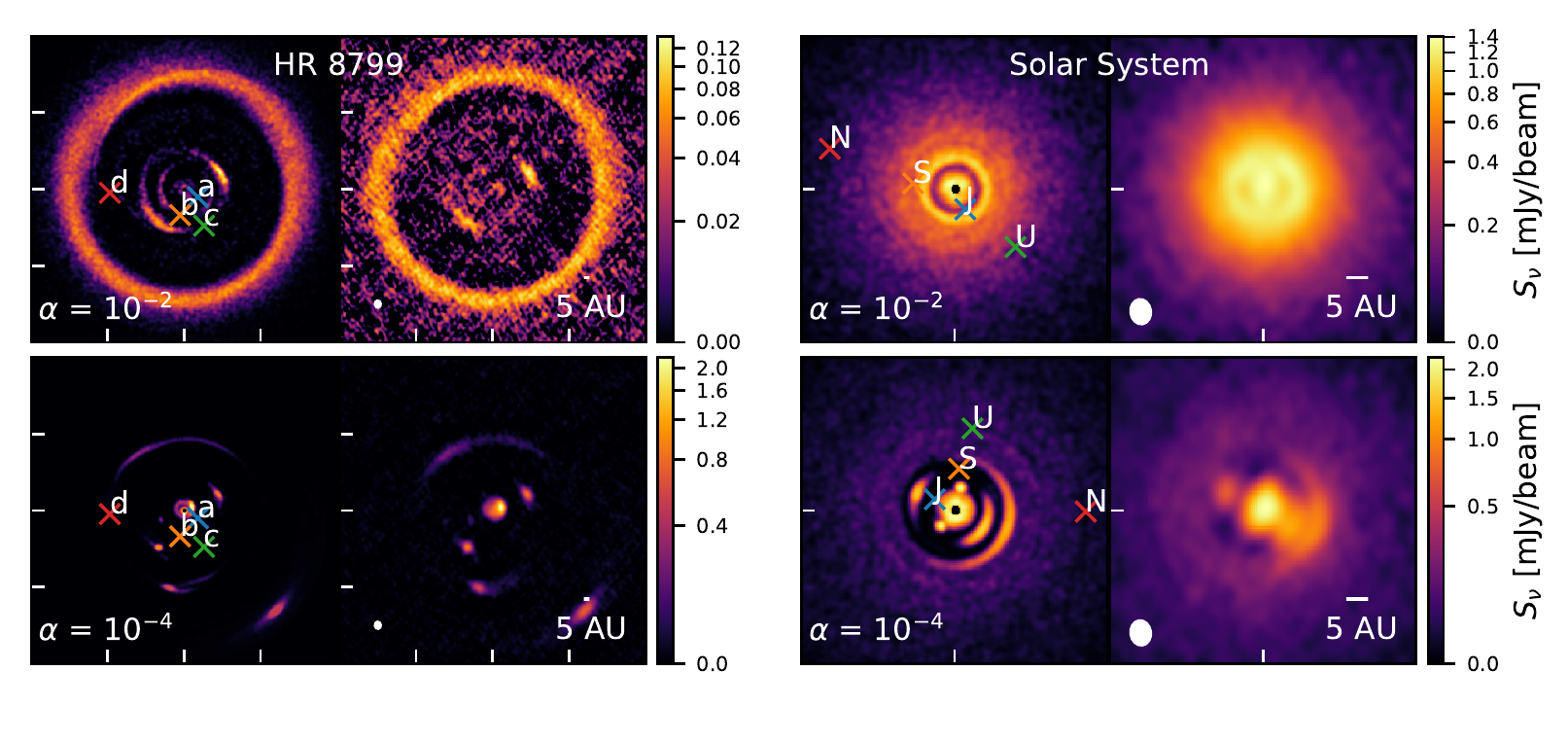}\figcaption{Simulation images (the left panel in each panel block) and synthetic observations (the right panel in each panel block, using the same configuration as the ALMA DSHARP observation) of HR 8799 and Solar System at a distance of 140 pc. The top panels adopt $\alpha$ = $10^{-2}$, while the bottom panels adopt $\alpha$ = $10^{-4}$.  The field of view for HR 8799 images are 2'' while that for Solar System is 0.5''. The distance between two ticks in HR 8799 is 0.5''. 
\label{fig:HRandSY}}
\end{figure*}

Then, we use the \texttt{CASA} \texttt{simobserve} task to generate synthetic observations with sensitivities and angular resolutions comparable to those of the DSHARP observations, which are shown in Figure \ref{fig:HRandSY}. The angular resolutions in FWHM are equivalent to $\sim$ 5 au in distance and are marked in the lower left corners in the figure. Each set of synthetic observations consist of 12 minutes of on-source integration time with the Cycle 5 C43-5 antenna configuration, 35 minutes on source in the C43-8 configuration, and 35 minutes on-source in the C43-9 configuration. A precipitable water vapor level of 1.0 mm is adopted throughout. The resulting synthetic visibilities are imaged in the same manner as the DSHARP sources, as described in \citet{andrews18b}. Clearly the DSHARP observational setup is capable of detecting both our Solar System analogs and HR 8799 analogs at a distance of 140 pc away. 

The four giant planets induce a wide gap in the HR 8799 analog. When the disk viscosity is high ($\alpha=10^{-2}$), the disk has an annular ring with an inner cavity, similar to transitional disks \citep{espaillat14}. When the disk viscosity is low ($\alpha=10^{-4}$), we see bright arcs. We also see bright sources at the inner disk, which are vortices at the gap edge between the adjacent pair of planets and the horseshoe region of the planets. In actual observations, we may misinterpret them as planets or circumplanetary disks. One way to distinguish these possibilities  is studying if the bright sources are spatially resolved \citep{zhu2018}.
Either the planet or circumplanetary disks should be smaller than the planet's Hill radius. If the structures within the gap are spatially resolved, it is likely that they are not from the planets or the circumplanetary disks. 

For the Solar System analog, when the disk viscosity is high ($\alpha=0.01$), we can only observe the gap induced by Jupiter. When the viscosity is low ($\alpha=10^{-4}$), the common gap induced by Jupiter and Saturn can be seen. Gap edge vortices and horseshoe regions can also be seen in this case. From the synthetic observations, we can barely see the disk features induced by Uranus and Neptune. Even by examining the radial intensity profiles, we can only see an extremely shallow dimple at the Neptune position. Thus, Uranus and Neptune in our Solar System analogs are not detectable with DSHARP. The reason we have Neptune mass planet candidates in Table \ref{table:Mp} and Figure \ref{fig:moneyplot} is because either the planet is further away or the central stellar mass is lower (so that $q$ is larger and gaps are deeper). 

\subsection{Caveats}
Although we seek to explain gaps with young planets, we want to point out that there are many other possible mechanisms to produce gaps and rings, such as ice lines \citep{zhang2015, okuzumi2016}, the dead zone transition \citep{pinilla2016}, MHD zonal flows \citep{flock15, ruge2016}, the secular gravitational instability \citep{takahashi2014}, disk winds \citep{bai2017,suriano2018} and so on. On the other hand, quantitative predictions from these mechanisms are desired for the future so that we can test various ideas and understand the nature of these gaps and rings.

Another major caveat in this work is that we fit the gap profiles at 1000 planetary orbits. The gap depth and width $do$ change with time \citep{rosotti2016}. To get a rigorous comparison between simulations and observations, we need to know when planets formed in the disk and how planets grew in time \citep{hammer2017}, which we have little knowledge about. We can only assume that the gap opening timescale is similar to the disk lifetime. Although 1000 orbits at $\sim$ 100 au is close to the disk lifetime, it is only 10\% of the disk lifetime for a planet at 20 au. A study similar to this work but also including the gap's change with time is needed in future. On the other hand, we can do some analytical estimates on the relationship between the gap width and time.
First, we do not expect that the gap profile can change dramatically over several thousand orbits if the disk has a large $\alpha$ (e.g. $\alpha>10^{-3}$) and small particles (e.g. $St<10^{-3}$). This is because, in these disks, the viscous timescale over the gap width is much shorter than 1000 planetary orbits and the gas disk has already reached the steady state. Small particles couple with the gas relatively well and their drift timescale is much longer than several thousand orbits. Dust turbulent diffusion with the large $\alpha$ can further smooth out dust features \citep{zhu12}. Second, for particles which are marginally coupled to the gas ($St\gtrsim10^{-2}$), they drift fast in the disk and we expect that the gap width will increase with time. As long as the gas profile is fixed (e.g. $\alpha\sim10^{-3}$), particles will drift twice further away from the planet over twice amount of time. On the other hand, particles with twice $St$ will drift twice further way from the planet over the same amount of time. Thus, we expect that the gap width is proportional to $St\times t$ for fast drifting particles. We have done a test for disks at different orbits and with fast drifting dusts having different Stokes numbers using the Elias 24 simulation above. We find that if the gas profile is about the same, the time $t$ and the Stokes number $St$ indeed play the same role in widening the gap: the gap width at $2t$ is similar to the gap width at $t$ from particles with $2St$. However, we have not explored the full parameter space, and the results may change with some other disk parameters. Especially, if $\alpha$ is small, 
the dramatic change in the gas profile with time will complicate the issue and break the degeneracy between $St$ and $t$. A detailed study requires adding the time dimension in the parameter space and is beyond the scope of this paper. 


Dust evolution and feedback to the gas is ignored in our study so that we can scale the simulations. In reality, particles are trapped at the gap edges which will promote its growth. When a significant amount of dust is trapped at the gap edge, the dust-to-gas feedback 
can affect the gap depth and width (Yang \& Zhu in prep.) or even trigger streaming instability \citep{youdin2005}. A proper study with all these effects considered is difficult for 2-D numerical simulations. But it can be incorporated into 1-D dust evolutionary models. 

We want to emphasize that, as shown in \S 4, it is straightforward to derive the planet mass assuming other dust size distributions besides DSD1 and DSD2. As shown in Figure \ref{fig:alllines}, only the maximum Stokes number affects the gap profiles. Thus, we can calculate the Stokes number for any given dust size distribution, and then use the fits to derive the planet mass.

\section{Conclusion \label{sec:conclusion}}

DSHARP provides a homogeneous sample
of young protoplanetary disks showing a variety of substructures, e.g. rings, gaps, spirals, and small scale asymmetry \citep{andrews18b}. If these substructures are induced by forming young planets, they are revealing a hidden young planet
population which has not been probed by direct planet searching techniques. 

To explore the potential planet population that is responsible to observed features in the DSHARP disks, we carry out two-dimensional hydrodynamical simulations including dust particles to study the relationships between the gap properties and the planet mass. We systematically study
a grid of 45 gas models (as in \S \ref{sec:gridofmodels}), with three values of $\alpha$ ($10^{-4}$, $10^{-3}$, $10^{-2}$), three values of
$h/r$ (0.05, 0.07, 0.10), and five values of planet mass (from 10$M_{\earth}$ to 3$M_{J}$).
For each model, we scale the dust distribution in the simulation to disks with different surface densities and different dust size distributions. Two different dust size distributions motivated by (sub-)mm polarization measurements (DSD1: $s_{max}$=0.1 mm, $p$=-3.5) and (sub-mm) dust thermal continuum observations (DSD2: $s_{max}$=1 cm, $p$=-2.5) are considered. 
Overall, for each model, we generate 12 millimeter images including 7 images using the DSD1 dust size distribution and 5 images using the DSD2 dust size distribution. 

\begin{itemize}
    \item First, we study the gas structure in these 45 simulations. 
    Overall, the gap becomes deeper with higher $q$, smaller $h/r$, and lower $\alpha$.  But when $q\gtrsim 3 M_J$ in a low $\alpha$ disk, the gap edge becomes eccentric and the gap depth starts to decrease. These are all consistent with previous studies.
    \item We study the sub/super-Keplerian motion at the gap edges. We confirm that the deviation from the Keplerian motion is due to the gas radial pressure gradient. The distance between the sub/super-Keplerian motion peaks is roughly 4.4 times $h$, with a weak dependence on $\alpha$ and $q$. The amplitude of the sub/super-Keplerian motion peaks is fitted with Equation \ref{eq:Kvr}, which shows a strong dependence on $h/r$.
    \item Then, we  study the mm intensity maps for all our simulations. The gap edge becomes more eccentric and off-centered with the increasing planet mass. The eccentricity and off-centered distance are provided (Figure \ref{eq:eccentricity}). Large eccentricity and off-centered distance may be indications of planets in disks. 
    \item Particle trapping in gap edge vortices and the horseshoe region are apparent in mm intensity maps for disks with $\alpha=10^{-4}$, leading to large-scale asymmetries in the images. For some parameters, even a $33 M_{\earth}$ planet can lead to a factor of 100 contrast between different azimuthal parts of the disk. In some cases, the vortex shows up at smaller radii than the gap edge (similar to the arc structure in HD 163296). 
    \item We derive several empirical relationships between the width/depth of the gaps in mm intensity maps and the planet/disk properties. All the fits for the width and depth are given in Table \ref{table:relationwidth} and \ref{table:relationdepth} and shown in Figure \ref{fig:alllines}. We  show that different disks with different surface densities and different dust size distributions have the same gap shape as long as their Stokes numbers for the maximum-size particles (where most of the dust mass is) are the same. Thus, our derived relationships can be used in other disks with different surface densities and dust size distributions.
    \item A single planet can open multiple gaps. The position of the secondary gap is fitted with Equation \ref{eq:secondary_gap_fit}. We find that the position of the secondary gap is almost solely determined by the disk scale height. Thus, if the secondary gap is present, we can use its position to estimate the disk scale height ($h/r$).
    \item With all these relationships, we lay out the procedure to constrain the planet mass using gap properties (the flowchart is presented in Figure \ref{fig:flowchart}).
    \item Applying these steps, we  identify potential planets in the DSHARP disks. We provide planet masses that are derived using three different values of $\alpha$ and three dust size distributions.
    \item We comment on the potential planets in each disk. Particularly, for AS 209, we point out that our simulation matches not only the primary gap, but also  the  position  and  amplitude  of the secondary (61  au),  tertiary  (35  au)  and  even  the  fourth  (24  au) inner gaps.  This makes AS 209 the most plausible case that there is indeed a planet within the 100 au gap (also in \citet{guzman18}). The best fit model also suggests that the disk $\alpha$ increases with radii in AS 209, which may have implications for studying disk accretion theory.
    \item We make synthetic observations for HR 8799 and Solar System analogs to show that DSHARP is capable of detecting giant planets in these systems. 
    \item We plot these potential young planets in the exoplanet mass-semimajor axis diagram (Figure \ref{fig:moneyplot}). We find that the occurrence rate for $>$ 5 $M_J$ planets beyond 5-10 au is $\sim$ 6\%, consistent with direction imaging constraints. Using disk features, we can probe a planet population which is not accessible by other planet searching techniques. These are Neptune to Jupiter mass planets beyond 10 au.  The occurrence rate is $\sim$ 50\%, suggesting a flat distribution beyond several au and planets with Neptune mass and above are common. On the other hand, we caution that there are large uncertainties for both the origin of these gaps and the inferred planet mass.
    
\end{itemize}

\acknowledgments S. Z. and Z. Z. thank Lee Hartmann for very useful discussions. S. Z and Z. Z. thank the referee for prompt and constructive comments. Z. Z. acknowledges support from the National Aeronautics and Space Administration through the Astrophysics Theory Program with Grant No. NNX17AK40G and Sloan Research Fellowship. Simulations are carried out with the support from the Texas Advanced Computing Center (TACC) at The University of Texas at Austin through XSEDE grant TG- AST130002. J.H. acknowledges support from the National Science Foundation Graduate Research Fellowship under Grant No. DGE-1144152. V.V.G. and J.C acknowledge support from the National Aeronautics and Space Administration under grant No.~15XRP15$\_$20140 issued through the Exoplanets Research Program. S.A. and J.H. acknowledge support from the National Aeronautics and Space Administration under grant No.~17-XRP17$\_$2-0012 issued through the Exoplanets Research Program. T.B. acknowledges funding from the European Research Council (ERC) under the European Union’s Horizon 2020 research and innovation programme under grant agreement No 714769. C.P.D. acknowledges support by the German Science Foundation (DFG) Research Unit FOR 2634, grants DU 414/22-1 and DU 414/23-1. A.I. acknowledges support from the National Aeronautics and Space Administration under grant No. NNX15AB06G issued through the Origins of Solar Systems program, and from the National Science Foundation under grant No. AST-1715719. L.P. acknowledges support from CONICYT project Basal AFB-170002 and from FCFM/U. de Chile Fondo de Instalaci\'on Acad\'emica. M.B. acknowledges funding from ANR of France under contract number ANR-16-CE31-0013 (Planet Forming disks). L.R. acknowledges support from the ngVLA Community Studies program, coordinated by the National Radio Astronomy Observatory, which is a facility of the National Science Foundation operated under cooperative agreement by Associated Universities, Inc. This paper makes use of ALMA data ADS/JAO.ALMA \#2016.1.00484.L.

\software{
{\tt Astropy} \citep{astropy},
{\tt CASA} \citep{mcmullin07},
{\tt Dusty FARGO-ADSG} \citep{baruteau2008a, baruteau2008b, baruteau2016},
{\tt Matplotlib} \citep{matplotlib},
{\tt Numpy} \citep{numpy},
{\tt Scipy} \citep{scipy}
}

\section*{Appendix}
The fitted gap widths and depths for all the models are listed in Table \ref{table:gapwidth} and Table \ref{table:gapdepth}. Column (4) shows the gap widths/depths of the gas; Column (5-11) show the gap widths/depths of the dust emission with increasing initial gas surface density $\Sigma_{g,0}$ (decreasing Stokes number $St_{max}$) under dust size distribution DSD1; similarly Column (12-16) show the gap widths/depths of the dust under DSD2. All widths/depths shown in Table \ref{table:gapwidth} and \ref{table:gapdepth} are derived from the images with a Gaussian convolution $\sigma$ = 0.06 $r_p$ (the larger kernel), except for the bottom of Table \ref{table:gapwidth} (below the horizontal line and above the double horizontal lines) where widths are derived using $\sigma$ = 0.025 $r_p$ (the smaller kernel). These widths with a smaller beam are listed only if the gap widths $\Delta$ $<$ 0.15 using the larger kernel ($\sigma$ = 0.06 $r_p$). Rows below the double lines show the individual widths of the gaps whose common gap is separated into two due to the horseshoe. The value on top the bar shows the width of the inner gap ($\Delta_1$), whereas the value under the bar shows the width of the outer gap ($\Delta_2$).

\startlongtable
\begin{deluxetable*}{ccrcccccccc|ccccc}
\tabletypesize{\scriptsize}
\tablecaption{Gap Widths for the Gas, DSD1 and DSD2 \label{table:gapwidth}}
\tablehead{
\colhead{h/r} & \colhead{$\alpha$} & \colhead{$q$} & \colhead{$\Delta_g$} & \colhead{$\Delta_{d,0p1}$} & \colhead{$\Delta_{d,0p3}$} & 
\colhead{$\Delta_{d,1}$} & \colhead{$\Delta_{d,3}$} &
\colhead{$\Delta_{d,10}$} & 
\colhead{$\Delta_{d,30}$} & \colhead{$\Delta_{d,100}$} & 
\colhead{$\Delta_{d,1}$} & \colhead{$\Delta_{d,3}$} &
\colhead{$\Delta_{d,10}$} & 
\colhead{$\Delta_{d,30}$} & \colhead{$\Delta_{d,100}$} 
}
\colnumbers
\startdata
 0.05 &  $10^{-4}$ &  3.3$\times 10^{-5}$ &       0.09 &             0.69 &             0.46 &           0.21 &           0.16 &            0.13 &            0.12 &             0.12 &               0.81 &               0.75 &                0.64 &                0.32 &                 0.19 \\
 0.05 &  $10^{-4}$ &    1$\times 10^{-4}$ &       0.24 &             0.80 &             0.61 &           0.48 &           0.25 &            0.24 &            0.22 &             0.21 &               1.00 &               0.87 &                0.76 &                0.57 &                 0.26 \\
 0.05 &  $10^{-4}$ &  3.3$\times 10^{-4}$ &       0.32 &             0.82 &             0.57 &           0.98 &           0.30 &            0.29 &            0.29 &             0.27 &               1.00 &               0.92 &                0.77 &                0.55 &                 0.48 \\
 0.05 &  $10^{-4}$ &    1$\times 10^{-3}$ &       0.42 &             0.56 &             0.56 &           0.55 &           0.42 &            0.40 &            0.38 &             0.37 &               1.00 &               0.56 &                0.56 &                0.56 &                 0.54 \\
 0.05 &  $10^{-4}$ &  3.3$\times 10^{-3}$ &       0.55 &             0.75 &             0.60 &           0.97 &           0.53 &            0.50 &            0.49 &             0.48 &               1.00 &               1.00 &                0.72 &                0.58 &                 0.98 \\
 0.05 &  $10^{-3}$ &  3.3$\times 10^{-5}$ &       0.00 &             0.00 &             0.13 &           0.00 &           0.00 &            0.00 &            0.08 &             0.00 &               0.00 &               0.00 &                0.00 &                0.13 &                 0.00 \\
 0.05 &  $10^{-3}$ &    1$\times 10^{-4}$ &       0.20 &             0.77 &             0.52 &           0.27 &           0.20 &            0.18 &            0.17 &             0.16 &               1.00 &               0.89 &                0.72 &                0.47 &                 0.24 \\
 0.05 &  $10^{-3}$ &  3.3$\times 10^{-4}$ &       0.27 &             0.77 &             0.54 &           0.38 &           0.30 &            0.27 &            0.26 &             0.24 &               1.00 &               0.96 &                0.72 &                0.51 &                 0.34 \\
 0.05 &  $10^{-3}$ &    1$\times 10^{-3}$ &       0.37 &             0.73 &             0.56 &           0.44 &           0.38 &            0.36 &            0.34 &             0.32 &               1.00 &               0.84 &                0.68 &                0.54 &                 0.41 \\
 0.05 &  $10^{-3}$ &  3.3$\times 10^{-3}$ &       0.57 &             0.94 &             0.62 &           0.54 &           0.50 &            0.47 &            0.46 &             0.45 &               0.99 &               0.98 &                0.94 &                0.62 &                 0.53 \\
 0.05 &  $10^{-2}$ &  3.3$\times 10^{-5}$ &       0.00 &             0.00 &             0.00 &           0.00 &           0.00 &            0.00 &            0.00 &             0.00 &               0.00 &               0.00 &                0.00 &                0.00 &                 0.00 \\
 0.05 &  $10^{-2}$ &    1$\times 10^{-4}$ &       0.00 &             0.00 &             0.00 &           0.00 &           0.00 &            0.00 &            0.00 &             0.00 &               0.00 &               0.00 &                0.00 &                0.00 &                 0.00 \\
 0.05 &  $10^{-2}$ &  3.3$\times 10^{-4}$ &       0.21 &             0.39 &             0.26 &           0.21 &           0.00 &            0.00 &            0.00 &             0.00 &               0.59 &               0.52 &                0.37 &                0.25 &                 0.19 \\
 0.05 &  $10^{-2}$ &    1$\times 10^{-3}$ &       0.30 &             1.00 &             0.62 &           0.38 &           0.33 &            0.33 &            0.31 &             0.30 &               1.00 &               1.00 &                0.98 &                0.58 &                 0.35 \\
 0.05 &  $10^{-2}$ &  3.3$\times 10^{-3}$ &       0.43 &             1.00 &             0.74 &           0.61 &           0.53 &            0.46 &            0.44 &             0.42 &               1.00 &               1.00 &                1.00 &                0.75 &                 0.60 \\
 0.07 &  $10^{-4}$ &  3.3$\times 10^{-5}$ &       0.00 &             0.00 &             0.00 &           0.11 &           0.00 &            0.00 &            0.00 &             0.00 &               0.00 &               0.00 &                0.00 &                0.00 &                 0.10 \\
 0.07 &  $10^{-4}$ &    1$\times 10^{-4}$ &       0.14 &             0.78 &             0.68 &           0.36 &           0.25 &            0.18 &            0.15 &             0.14 &               0.96 &               0.82 &                0.73 &                0.61 &                 0.32 \\
 0.07 &  $10^{-4}$ &  3.3$\times 10^{-4}$ &       0.33 &             1.00 &             0.75 &           0.63 &           0.36 &            0.33 &            0.31 &             0.30 &               1.00 &               1.00 &                0.81 &                0.68 &                 0.36 \\
 0.07 &  $10^{-4}$ &    1$\times 10^{-3}$ &       0.42 &             0.79 &             0.76 &           0.48 &           0.44 &            0.41 &            0.39 &             0.37 &               1.00 &               0.82 &                0.77 &                0.64 &                 0.44 \\
 0.07 &  $10^{-4}$ &  3.3$\times 10^{-3}$ &       0.56 &             1.00 &             1.00 &           0.84 &           0.58 &            0.52 &            0.50 &             0.48 &               1.00 &               1.00 &                1.00 &                0.63 &                 0.82 \\
 0.07 &  $10^{-3}$ &  3.3$\times 10^{-5}$ &       0.00 &             0.00 &             0.00 &           0.00 &           0.00 &            0.00 &            0.00 &             0.00 &               0.00 &               0.00 &                0.00 &                0.00 &                 0.00 \\
 0.07 &  $10^{-3}$ &    1$\times 10^{-4}$ &       0.00 &             0.00 &             0.18 &           0.15 &           0.00 &            0.00 &            0.00 &             0.00 &               0.00 &               0.00 &                0.00 &                0.17 &                 0.14 \\
 0.07 &  $10^{-3}$ &  3.3$\times 10^{-4}$ &       0.28 &             0.97 &             0.70 &           0.43 &           0.32 &            0.29 &            0.27 &             0.26 &               1.00 &               1.00 &                0.78 &                0.63 &                 0.38 \\
 0.07 &  $10^{-3}$ &    1$\times 10^{-3}$ &       0.36 &             0.82 &             0.71 &           0.53 &           0.43 &            0.37 &            0.35 &             0.33 &               1.00 &               1.00 &                0.78 &                0.66 &                 0.49 \\
 0.07 &  $10^{-3}$ &  3.3$\times 10^{-3}$ &       0.52 &             0.97 &             0.90 &           0.68 &           0.57 &            0.52 &            0.48 &             0.47 &               1.00 &               1.00 &                0.97 &                0.88 &                 0.65 \\
 0.07 &  $10^{-2}$ &  3.3$\times 10^{-5}$ &       0.00 &             0.00 &             0.00 &           0.00 &           0.00 &            0.00 &            0.00 &             0.00 &               0.00 &               0.00 &                0.00 &                0.00 &                 0.00 \\
 0.07 &  $10^{-2}$ &    1$\times 10^{-4}$ &       0.00 &             0.00 &             0.00 &           0.00 &           0.00 &            0.00 &            0.00 &             0.00 &               0.00 &               0.00 &                0.00 &                0.00 &                 0.00 \\
 0.07 &  $10^{-2}$ &  3.3$\times 10^{-4}$ &       0.00 &             0.00 &             0.00 &           0.00 &           0.00 &            0.00 &            0.00 &             0.00 &               0.00 &               0.00 &                0.00 &                0.00 &                 0.00 \\
 0.07 &  $10^{-2}$ &    1$\times 10^{-3}$ &       0.28 &             0.52 &             0.37 &           0.15 &           0.00 &            0.00 &            0.00 &             0.00 &               0.76 &               0.65 &                0.48 &                0.35 &                 0.13 \\
 0.07 &  $10^{-2}$ &  3.3$\times 10^{-3}$ &       0.40 &             1.00 &             0.97 &           0.58 &           0.46 &            0.43 &            0.41 &             0.41 &               1.00 &               1.00 &                1.00 &                0.99 &                 0.53 \\
  0.10&  $10^{-4}$ &  3.3$\times 10^{-5}$ &       0.00 &             0.00 &             0.00 &           0.00 &           0.00 &            0.00 &            0.00 &             0.00 &               0.00 &               0.00 &                0.00 &                0.00 &                 0.00 \\
  0.10&  $10^{-4}$ &    1$\times 10^{-4}$ &       0.00 &             0.00 &             0.00 &           0.00 &           0.12 &            0.12 &            0.11 &             0.11 &               0.00 &               0.00 &                0.00 &                0.00 &                 0.10 \\
  0.10&  $10^{-4}$ &  3.3$\times 10^{-4}$ &       0.21 &             0.83 &             0.79 &           0.55 &           0.43 &            0.32 &            0.22 &             0.19 &               1.00 &               0.89 &                0.80 &                0.71 &                 0.48 \\
  0.10&  $10^{-4}$ &    1$\times 10^{-3}$ &       0.41 &             1.00 &             0.81 &           0.75 &           0.50 &            0.44 &            0.42 &             0.40 &               1.00 &               1.00 &                0.84 &                0.78 &                 0.51 \\
  0.10 &  $10^{-4}$ &  3.3$\times 10^{-3}$ &       0.53 &             0.86 &             0.84 &           0.72 &           0.70 &            0.54 &            0.48 &             0.29 &               0.75 &               0.88 &                0.86 &                0.72 &                 0.71 \\
  0.10 &  $10^{-3}$ &  3.3$\times 10^{-5}$ &       0.00 &             0.00 &             0.00 &           0.00 &           0.00 &            0.00 &            0.00 &             0.00 &               0.00 &               0.00 &                0.00 &                0.00 &                 0.00 \\
  0.10 &  $10^{-3}$ &    1$\times 10^{-4}$ &       0.00 &             0.00 &             0.00 &           0.00 &           0.00 &            0.00 &            0.00 &             0.00 &               0.00 &               0.00 &                0.00 &                0.00 &                 0.00 \\
  0.10 &  $10^{-3}$ &  3.3$\times 10^{-4}$ &       0.00 &             0.00 &             0.00 &           0.18 &           0.15 &            0.15 &            0.00 &             0.00 &               0.00 &               0.00 &                0.00 &                0.00 &                 0.15 \\
  0.10 &  $10^{-3}$ &    1$\times 10^{-3}$ &       0.38 &             1.00 &             0.82 &           0.65 &           0.46 &            0.40 &            0.39 &             0.36 &               1.00 &               1.00 &                1.00 &                0.76 &                 0.56 \\
  0.10 &  $10^{-3}$ &  3.3$\times 10^{-3}$ &       0.49 &             1.00 &             0.98 &           0.74 &           0.59 &            0.52 &            0.49 &             0.47 &               1.00 &               1.00 &                1.00 &                0.81 &                 0.68 \\
  0.10 &  $10^{-2}$ &  3.3$\times 10^{-5}$ &       0.00 &             0.00 &             0.00 &           0.00 &           0.00 &            0.00 &            0.00 &             0.00 &               0.00 &               0.00 &                0.00 &                0.00 &                 0.00 \\
  0.10 &  $10^{-2}$ &    1$\times 10^{-4}$ &       0.00 &             0.00 &             0.00 &           0.00 &           0.00 &            0.00 &            0.00 &             0.00 &               0.31 &               0.00 &                0.00 &                0.00 &                 0.00 \\
  0.10 &  $10^{-2}$ &  3.3$\times 10^{-4}$ &       0.00 &             0.00 &             0.00 &           0.00 &           0.00 &            0.00 &            0.00 &             0.00 &               0.00 &               0.00 &                0.00 &                0.00 &                 0.00 \\
  0.10 &  $10^{-2}$ &    1$\times 10^{-3}$ &       0.11 &             0.00 &             0.00 &           0.00 &           0.00 &            0.00 &            0.00 &             0.00 &               0.20 &               0.00 &                0.00 &                0.00 &                 0.00 \\
  0.10 &  $10^{-2}$ &  3.3$\times 10^{-3}$ &       0.38 &             0.72 &             0.49 &           0.40 &           0.15 &            0.16 &            0.00 &             0.00 &               1.00 &               0.88 &                0.69 &                0.47 &                 0.38 \\
\hline
Kernel &  $\sigma$ &  $= 0.025 r_p$   &   &    &    &   &                    &                    &               &                &                   &                   &                   &                   &                   \\ 
 0.05 &  $10^{-4}$ &  3.3$\times 10^{-5}$ &       0.09 &               -- &               -- &             -- &             -- &            0.13 &            0.12 &             0.12 &                 -- &                 -- &                  -- &                  -- &                   -- \\
 0.05 &  $10^{-3}$ &  3.3$\times 10^{-5}$ &       0.00 &               -- &             0.13 &           0.11 &           0.11 &            0.09 &            0.08 &               -- &                 -- &                 -- &                  -- &                0.13 &                 0.11 \\
 0.05 &  $10^{-2}$ &  3.3$\times 10^{-4}$ &       0.21 &               -- &               -- &             -- &           0.20 &              -- &            0.11 &             0.13 &                 -- &                 -- &                  -- &                  -- &                   -- \\
 0.07 &  $10^{-4}$ &  3.3$\times 10^{-5}$ &       0.00 &               -- &               -- &           0.10 &           0.09 &            0.08 &            0.09 &             0.09 &                 -- &                 -- &                  -- &                0.08 &                 0.09 \\
 0.07 &  $10^{-3}$ &    1$\times 10^{-4}$ &       0.00 &               -- &               -- &           0.15 &           0.14 &            0.12 &            0.13 &               -- &                 -- &                 -- &                  -- &                0.17 &                 0.14 \\
 0.07 &  $10^{-2}$ &    1$\times 10^{-3}$ &       0.28 &               -- &               -- &             -- &           0.09 &            0.10 &            0.09 &             0.08 &                 -- &                 -- &                  -- &                  -- &                 0.12 \\
 0.10 &  $10^{-4}$ &  3.3$\times 10^{-5}$ &       0.00 &               -- &               -- &           0.07 &           0.08 &            0.08 &            0.08 &             0.08 &                 -- &                 -- &                  -- &                  -- &                 0.07 \\
 0.10 &  $10^{-4}$ &    1$\times 10^{-4}$ &       0.00 &               -- &               -- &           0.10 &           0.12 &            0.12 &            0.11 &             0.12 &                 -- &                 -- &                  -- &                0.09 &                 0.10 \\
 0.10 &  $10^{-3}$ &  3.3$\times 10^{-4}$ &       0.00 &               -- &               -- &             -- &             -- &              -- &            0.15 &             0.11 &                 -- &                 -- &                  -- &                0.16 &                   -- \\
 0.10 &  $10^{-2}$ &    1$\times 10^{-3}$ &       0.11 &               -- &               -- &             -- &           0.12 &            0.08 &            0.11 &               -- &                 -- &                 -- &                  -- &                  -- &                   -- \\
 0.10 &  $10^{-2}$ &  3.3$\times 10^{-3}$ &       0.38 &               -- &               -- &             -- &             -- &              -- &            0.40 &             0.13 &                 -- &                 -- &                  -- &                  -- &                   -- \\
 \hline
 \hline
 Common &  Gaps &  Separated  & by  &   Horseshoe       &   $\frac{\Delta_1}{\Delta_2}$     &   &                    &                    &               &                &                   &                   &                   &                   &                   \\ 
 0.05 &  $10^{-4}$ &    1$\times 10^{-4}$ &       0.24 &                   -- &                   -- &  $\frac{0.28}{0.10}$ &                   -- &                   -- &                   -- &                   -- &                   -- &                   -- &                   -- &                   -- &                   -- \\
 0.05 &  $10^{-4}$ &  3.3$\times 10^{-4}$ &       0.32 &                   -- &                   -- &  $\frac{0.32}{0.13}$ &                   -- &                   -- &                   -- &                   -- &                   -- &                   -- &                   -- &                   -- &  $\frac{0.29}{0.07}$ \\
 0.05 &  $10^{-4}$ &    1$\times 10^{-3}$ &       0.42 &  $\frac{0.10}{0.19}$ &  $\frac{0.11}{0.20}$ &  $\frac{0.10}{0.19}$ &  $\frac{0.25}{0.15}$ &  $\frac{0.26}{0.15}$ &                   -- &                   -- &                   -- &  $\frac{0.09}{0.17}$ &  $\frac{0.10}{0.19}$ &  $\frac{0.11}{0.20}$ &  $\frac{0.10}{0.18}$ \\
 0.05 &  $10^{-4}$ &  3.3$\times 10^{-3}$ &       0.55 &                   -- &                   -- &  $\frac{0.56}{0.81}$ &                   -- &                   -- &                   -- &                   -- &                   -- &                   -- &                   -- &                   -- &  $\frac{0.54}{0.86}$ \\
 0.05 &  $10^{-3}$ &  3.3$\times 10^{-3}$ &       0.57 &  $\frac{0.07}{0.93}$ &                   -- &                   -- &                   -- &                   -- &                   -- &                   -- &  $\frac{0.09}{0.99}$ &  $\frac{0.09}{0.97}$ &  $\frac{0.07}{0.93}$ &                   -- &                   -- \\
 0.07 &  $10^{-4}$ &  3.3$\times 10^{-4}$ &       0.33 &                   -- &                   -- &  $\frac{0.41}{0.19}$ &                   -- &                   -- &                   -- &                   -- &                   -- &                   -- &                   -- &                   -- &                   -- \\
 0.07 &  $10^{-4}$ &    1$\times 10^{-3}$ &       0.42 &                   -- &                   -- &                   -- &  $\frac{0.27}{0.19}$ &                   -- &                   -- &                   -- &                   -- &                   -- &                   -- &  $\frac{0.51}{0.19}$ &  $\frac{0.25}{0.17}$ \\
 0.07 &  $10^{-4}$ &  3.3$\times 10^{-3}$ &       0.56 &                   -- &                   -- &  $\frac{0.61}{0.39}$ &                   -- &                   -- &                   -- &                   -- &                   -- &                   -- &                   -- &                   -- &  $\frac{0.59}{0.27}$ \\
 0.07 &  $10^{-3}$ &  3.3$\times 10^{-4}$ &       0.28 &  $\frac{0.88}{0.74}$ &                   -- &                   -- &                   -- &                   -- &                   -- &                   -- &                   -- &                   -- &                   -- &                   -- &                   -- \\
 0.07 &  $10^{-2}$ &  3.3$\times 10^{-3}$ &       0.40 &                   -- &  $\frac{0.82}{0.62}$ &                   -- &                   -- &                   -- &                   -- &                   -- &                   -- &                   -- &                   -- &  $\frac{0.81}{0.88}$ &                   -- \\
 0.10 &  $10^{-4}$ &  3.3$\times 10^{-4}$ &       0.21 &                   -- &                   -- &                   -- &                   -- &                   -- &                   -- &                   -- &                   -- &                   -- &                   -- &  $\frac{0.62}{0.18}$ &                   -- \\
 0.10 &  $10^{-4}$ &    1$\times 10^{-3}$ &       0.41 &                   -- &                   -- &  $\frac{0.57}{0.18}$ &                   -- &                   -- &                   -- &                   -- &                   -- &                   -- &                   -- &  $\frac{0.58}{0.30}$ &  $\frac{0.27}{0.30}$ \\
 0.10 &  $10^{-4}$ &  3.3$\times 10^{-3}$ &       0.53 &  $\frac{0.16}{0.65}$ &  $\frac{0.19}{0.66}$ &  $\frac{0.19}{0.63}$ &  $\frac{0.16}{0.39}$ &  $\frac{0.35}{0.13}$ &  $\frac{0.33}{0.02}$ &                   -- &  $\frac{0.14}{0.67}$ &  $\frac{0.15}{0.65}$ &  $\frac{0.16}{0.65}$ &  $\frac{0.17}{0.64}$ &  $\frac{0.16}{0.42}$ \\
 0.10 &  $10^{-3}$ &    1$\times 10^{-3}$ &       0.38 &                   -- &                   -- &                   -- &                   -- &                   -- &  $\frac{0.18}{0.18}$ &  $\frac{0.17}{0.15}$ &                   -- &                   -- &                   -- &                   -- &                   -- \\
 0.10 &  $10^{-3}$ &  3.3$\times 10^{-3}$ &       0.49 &                   -- &  $\frac{0.85}{0.64}$ &                   -- &                   -- &                   -- &                   -- &                   -- &                   -- &                   -- &                   -- &                   -- &                   -- \\
 0.10 &  $10^{-2}$ &    1$\times 10^{-4}$ &       0.00 &                   -- &                   -- &                   -- &                   -- &                   -- &                   -- &                   -- &  $\frac{0.12}{0.12}$ &                   -- &                   -- &                   -- &                   -- \\
\enddata
\tablecomments{A summary of the gap widths of the gas surface density profile, and dust emission profile under dust size distribution DSD1 and DSD2. (1) aspect ratio $h/r$ (2) $\alpha$ viscosity (3) planet-stellar mass ratio $q$ (4) The width of the gas surface density (5-11) The gap width of the dust emission under DSD1, with initial gas surface density $\Sigma_{g,0}$ = 0.1, 0.3, 1, 3, 10, 30, 100 $\mathrm{g\,cm^{-2}}$ ($St_{max}$ = 1.57 $\times$ $10^{-1}$, 5.23 $\times$ $10^{-2}$, 1.57 $\times$ $10^{-2}$, 5.23 $\times$ $10^{-3}$, 1.57 $\times$ $10^{-3}$, 5.23 $\times$ $10^{-4}$, 1.57 $\times$ $10^{-4}$) (12-16) The gap width of the dust emission under DSD2, with initial gas surface density $\Sigma_{g,0}$ = 1, 3, 10, 30, 100 $\mathrm{g\,cm^{-2}}$ ($St_{max}$ = 1.57, 5.32 $\times$ $10^{-1}$, 1.57 $\times$ $10^{-1}$, 5.23 $\times$ $10^{-2}$, 1.57 $\times$ $10^{-2}$) While the gap widths $\Delta_g$ are found from unconvolved gas surface density profile, the rest of $\Delta_d$ are found from smoothed dust continuum intensity. The convolution beam for dust emission $\sigma$ = 0.06 $r_p$ for the top rows; $\sigma$ = 0.025 $r_p$ for 11 rows horizontal single and double lines. Bottom rows under the double lines are the gaps with the horseshoe that separates them into two gaps. The value on top the bar shows the width of the inner gap ($\Delta_1$), whereas the value under the bar shows the width of the out gap ($\Delta_2$).}
\end{deluxetable*}

\startlongtable
\begin{deluxetable*}{clrcccccccc|ccccc}
\setlength{\tabcolsep}{1.5pt}
\tabletypesize{\scriptsize}
\tablecaption{Gap Depths $\big(log_{10}(\delta - 1)\big)$ for the Gas, DSD1 and DSD2 \label{table:gapdepth}}
\tablehead{
\colhead{h/r} & \colhead{$\alpha$} & \colhead{$q$} & \colhead{$\delta_{g}-1$} & \colhead{$\delta_{d,0p1}-1$} & \colhead{$\delta_{d,0p3}-1$} & 
\colhead{$\delta_{d,1}-1$} & \colhead{$\delta_{d,3}-1$} &
\colhead{$\delta_{d,10}-1$} & 
\colhead{$\delta_{d,30}-1$} & \colhead{$\delta_{d,100}-1$} & 
\colhead{$\delta_{d,1}-1$} & \colhead{$\delta_{d,3}-1$} &
\colhead{$\delta_{d,10}-1$} & 
\colhead{$\delta_{d,30}-1$} & \colhead{$\delta_{d,100}-1$}
\\
\colhead{} & \colhead{} & \colhead{$(M_p/M_*)$} & \colhead{($log_{10}$)} 
& \colhead{($log_{10}$)} & \colhead{($log_{10}$)} & 
\colhead{($log_{10}$)} & \colhead{($log_{10}$)} &
\colhead{($log_{10}$)} & 
\colhead{($log_{10}$)} & \colhead{($log_{10}$)} & 
\colhead{($log_{10}$)} & \colhead{($log_{10}$)} &
\colhead{($log_{10}$)} & 
\colhead{($log_{10}$)} & \colhead{($log_{10}$)} 
}
\colnumbers
\startdata
 0.05 &  $10^{-4}$ &  3.3$\times 10^{-5}$ &      -1.03 &             1.94 &             1.25 &           0.62 &           0.24 &           -0.03 &           -0.21 &            -0.41 &               3.66 &               2.93 &                2.10 &                1.25 &                 0.50 \\
 0.05 &  $10^{-4}$ &    1$\times 10^{-4}$ &       0.34 &             1.66 &             1.05 &           0.57 &           0.39 &            0.29 &            0.23 &             0.10 &               3.67 &               2.62 &                1.60 &                0.87 &                 0.32 \\
 0.05 &  $10^{-4}$ &  3.3$\times 10^{-4}$ &       1.15 &             1.56 &             1.12 &           0.82 &           0.64 &            0.57 &            0.53 &             0.53 &               2.70 &               2.10 &                1.54 &                0.93 &                 0.54 \\
 0.05 &  $10^{-4}$ &    1$\times 10^{-3}$ &       3.43 &             3.12 &             2.56 &           2.14 &           1.90 &            1.72 &            1.65 &             1.56 &               3.66 &               3.47 &                3.13 &                2.64 &                 2.05 \\
 0.05 &  $10^{-4}$ &  3.3$\times 10^{-3}$ &       2.05 &             4.55 &             3.90 &           3.20 &           2.67 &            2.26 &            1.93 &             1.73 &               8.75 &               8.00 &                6.93 &                5.75 &                 4.10 \\
 0.05 &  $10^{-3}$ &  3.3$\times 10^{-5}$ &       -- &             -- &            -0.55 &          -0.73 &           -- &            -- &            -- &             -- &               -- &               -- &               -0.77 &               -0.61 &                -0.81 \\
 0.05 &  $10^{-3}$ &    1$\times 10^{-4}$ &      -0.27 &             2.24 &             1.46 &           0.72 &           0.24 &            0.01 &           -0.10 &            -0.26 &               4.94 &               3.83 &                2.55 &                1.46 &                 0.57 \\
 0.05 &  $10^{-3}$ &  3.3$\times 10^{-4}$ &       0.81 &             3.03 &             2.26 &           1.59 &           1.17 &            0.82 &            0.51 &             0.37 &               6.67 &               5.69 &                4.37 &                3.03 &                 1.84 \\
 0.05 &  $10^{-3}$ &    1$\times 10^{-3}$ &       2.35 &             2.54 &             3.08 &           2.86 &           2.49 &            2.17 &            1.89 &             1.49 &               3.40 &               3.08 &                2.72 &                3.23 &                 3.78 \\
 0.05 &  $10^{-3}$ &  3.3$\times 10^{-3}$ &       1.96 &             3.52 &             2.95 &           2.43 &           2.10 &            1.78 &            1.48 &             1.34 &               -- &               -- &                5.44 &                4.22 &                 3.13 \\
 0.05 &  $10^{-2}$ &  3.3$\times 10^{-5}$ &       -- &             -- &             -- &           -- &           -- &            -- &            -- &             -- &               -- &               -- &                -- &                -- &                 -- \\
 0.05 &  $10^{-2}$ &    1$\times 10^{-4}$ &       -- &             -- &             -- &           -- &           -- &            -- &            -- &             -- &               -- &               -- &                -- &                -- &                 -- \\
 0.05 &  $10^{-2}$ &  3.3$\times 10^{-4}$ &      -0.32 &             0.81 &             0.16 &          -0.25 &          -0.72 &            -- &           -0.72 &             -- &               1.45 &               1.20 &                0.75 &                0.11 &                -0.41 \\
 0.05 &  $10^{-2}$ &    1$\times 10^{-3}$ &       0.64 &             1.88 &             1.24 &           0.59 &           0.24 &            0.04 &           -0.01 &            -0.10 &               -- &               3.37 &                2.22 &                1.27 &                 0.44 \\
 0.05 &  $10^{-2}$ &  3.3$\times 10^{-3}$ &       2.69 &             3.05 &             2.50 &           2.01 &           1.69 &            1.35 &            1.05 &             0.74 &               4.87 &               4.12 &                5.00 &                3.85 &                 2.60 \\
 0.07 &  $10^{-4}$ &  3.3$\times 10^{-5}$ &       -- &             -- &             -- &          -0.49 &          -0.55 &           -0.58 &           -0.65 &             -- &               -- &               -- &                -- &                -- &                -0.61 \\
 0.07 &  $10^{-4}$ &    1$\times 10^{-4}$ &      -0.69 &             2.31 &             1.92 &           1.08 &           0.51 &            0.15 &           -0.04 &            -0.28 &               4.06 &               3.26 &                2.41 &                1.73 &                 0.88 \\
 0.07 &  $10^{-4}$ &  3.3$\times 10^{-4}$ &       0.59 &             2.13 &             1.88 &           1.32 &           0.93 &            0.60 &            0.39 &             0.25 &               3.09 &               2.32 &                1.94 &                1.54 &                 1.03 \\
 0.07 &  $10^{-4}$ &    1$\times 10^{-3}$ &       1.86 &             3.34 &             2.93 &           2.20 &           1.81 &            1.51 &            1.29 &             1.02 &               4.57 &               3.89 &                3.44 &                2.92 &                 2.20 \\
 0.07 &  $10^{-4}$ &  3.3$\times 10^{-3}$ &       2.60 &             5.10 &             4.63 &           3.79 &           3.24 &            2.75 &            2.42 &             2.04 &               9.89 &               9.11 &                8.10 &                7.02 &                 5.69 \\
 0.07 &  $10^{-3}$ &  3.3$\times 10^{-5}$ &       -- &             -- &             -- &           -- &           -- &            -- &            -- &             -- &               -- &               -- &                -- &                -- &                 -- \\
 0.07 &  $10^{-3}$ &    1$\times 10^{-4}$ &       -- &             -- &            -0.47 &          -0.39 &          -0.63 &           -0.65 &            -- &             -- &               -- &               -- &               -0.56 &               -0.55 &                -0.54 \\
 0.07 &  $10^{-3}$ &  3.3$\times 10^{-4}$ &       0.03 &             2.85 &             2.29 &           1.38 &           0.83 &            0.48 &            0.28 &             0.09 &               6.25 &               5.40 &                3.86 &                2.53 &                 1.33 \\
 0.07 &  $10^{-3}$ &    1$\times 10^{-3}$ &       1.04 &             3.50 &             2.96 &           2.12 &           1.64 &            1.26 &            0.93 &             0.72 &               7.71 &               6.91 &                5.78 &                4.49 &                 2.86 \\
 0.07 &  $10^{-3}$ &  3.3$\times 10^{-3}$ &       2.38 &             4.86 &             4.36 &           3.58 &           3.12 &            2.78 &            2.48 &             2.11 &               4.69 &               -- &                7.61 &                6.49 &                 4.89 \\
 0.07 &  $10^{-2}$ &  3.3$\times 10^{-5}$ &       -- &             -- &             -- &           -- &           -- &            -- &            -- &             -- &               -- &               -- &                -- &                -- &                 -- \\
 0.07 &  $10^{-2}$ &    1$\times 10^{-4}$ &       -- &             -- &             -- &           -- &           -- &            -- &            -- &             -- &               -- &               -- &                -- &                -- &                 -- \\
 0.07 &  $10^{-2}$ &  3.3$\times 10^{-4}$ &       -- &             -- &             -- &           -- &           -- &            -- &            -- &             -- &              -0.45 &               -- &                -- &                -- &                 -- \\
 0.07 &  $10^{-2}$ &    1$\times 10^{-3}$ &      -0.09 &             0.84 &             0.18 &          -0.37 &          -0.71 &            -- &            -- &             -- &               1.84 &               1.46 &                0.79 &                0.10 &                -0.49 \\
 0.07 &  $10^{-2}$ &  3.3$\times 10^{-3}$ &       1.08 &             2.09 &             1.63 &           1.01 &           0.58 &            0.28 &            0.16 &             0.11 &               -- &               -- &                2.99 &                1.84 &                 0.87 \\
  0.10 &  $10^{-4}$ &  3.3$\times 10^{-5}$ &       -- &             -- &             -- &           -- &           -- &            -- &            -- &             -- &               -- &               -- &                -- &                -- &                 -- \\
  0.10 &  $10^{-4}$ &    1$\times 10^{-4}$ &       -- &             -- &             -- &          -0.63 &          -0.29 &           -0.25 &           -0.38 &            -0.57 &               -- &               -- &                -- &                -- &                -0.80 \\
  0.10 &  $10^{-4}$ &  3.3$\times 10^{-4}$ &      -0.33 &             2.51 &             2.23 &           1.49 &           0.73 &            0.26 &            0.00 &            -0.27 &               4.10 &               3.50 &                3.00 &                2.20 &                 1.15 \\
  0.10 &  $10^{-4}$ &    1$\times 10^{-3}$ &       0.79 &             2.75 &             2.42 &           1.76 &           1.18 &            0.71 &            0.45 &             0.21 &               4.55 &               3.74 &                3.13 &                2.40 &                 1.46 \\
  0.10 &  $10^{-4}$ &  3.3$\times 10^{-3}$ &       1.84 &             3.71 &             3.08 &           2.81 &           2.36 &            1.87 &            1.52 &             1.14 &               5.84 &               4.47 &                3.77 &                3.37 &                 4.29 \\
  0.10 &  $10^{-3}$ &  3.3$\times 10^{-5}$ &       -- &             -- &             -- &           -- &           -- &            -- &            -- &             -- &               -- &               -- &                -- &                -- &                 -- \\
  0.10 &  $10^{-3}$ &    1$\times 10^{-4}$ &       -- &             -- &             -- &           -- &           -- &            -- &            -- &             -- &               -- &               -- &                -- &                -- &                 -- \\
  0.10 &  $10^{-3}$ &  3.3$\times 10^{-4}$ &       -- &             -- &             -- &          -0.41 &          -0.58 &           -0.57 &           -0.62 &             -- &               -- &               -- &                -- &               -0.67 &                -0.67 \\
  0.10 &  $10^{-3}$ &    1$\times 10^{-3}$ &       0.17 &             2.54 &             2.10 &           1.33 &           0.59 &            0.08 &           -0.18 &            -0.34 &               -- &               5.06 &                3.52 &                2.29 &                 1.14 \\
  0.10 &  $10^{-3}$ &  3.3$\times 10^{-3}$ &       1.33 &             3.19 &             2.74 &           2.04 &           1.40 &            0.97 &            0.63 &             0.26 &               -- &               6.64 &                5.63 &                4.52 &                 2.92 \\
  0.10 &  $10^{-2}$ &  3.3$\times 10^{-5}$ &       -- &             -- &             -- &           -- &           -- &            -- &            -- &             -- &               -- &               -- &                -- &                -- &                 -- \\
  0.10 &  $10^{-2}$ &    1$\times 10^{-4}$ &       -- &             -- &             -- &           -- &           -- &            -- &            -- &             -- &               -- &               -- &                -- &                -- &                 -- \\
  0.10 &  $10^{-2}$ &  3.3$\times 10^{-4}$ &       -- &             -- &             -- &           -- &           -- &            -- &            -- &             -- &               -- &               -- &                -- &                -- &                 -- \\
  0.10 &  $10^{-2}$ &    1$\times 10^{-3}$ &      -1.45 &             -- &             -- &           -- &           -- &            -- &            -- &             -- &               -- &               -- &                -- &                -- &                 -- \\
  0.10 &  $10^{-2}$ &  3.3$\times 10^{-3}$ &       0.17 &             1.24 &             0.55 &          -0.07 &          -0.61 &           -0.62 &           -0.62 &            -0.64 &               2.87 &               2.14 &                1.30 &                0.48 &                -0.25 \\
\enddata
\tablecomments{A summary of the gap depths of the gas surface density profiles and dust emission profiles under dust size distribution DSD1 and DSD2. The layout is similar to Table \ref{table:gapwidth}, except that the depths are listed in $log_{10}(\delta - 1)$ and only $\sigma$ = 0.06 $r_p$ kernel is applied to find the depths.}
\end{deluxetable*}

\clearpage

\end{document}